\documentclass[prb, aps, twocolumn, amsmath, amssymb, superscriptaddress, longbibliography]{revtex4-2}

\usepackage{CJK}
\usepackage{graphicx}
\usepackage{color}
\usepackage{braket}

\usepackage{xr-hyper}
\makeatletter
\newcommand*{\addFileDependency}[1]{
  \typeout{(#1)}
  \@addtofilelist{#1}
  \IfFileExists{#1}{}{\typeout{No file #1.}}
}
\makeatother

\newcommand*{\myexternaldocument}[1]{%
    \externaldocument{#1}%
    \addFileDependency{#1.tex}%
    \addFileDependency{#1.aux}%
}

\myexternaldocument{suppl}

\usepackage{dcolumn} 
\usepackage{bm}     
\usepackage{amsfonts}
\usepackage[bookmarks=false,colorlinks,citecolor=blue]{hyperref}
\hypersetup{colorlinks=true,linkcolor=blue,filecolor=blue,citecolor = blue, urlcolor=blue}
\usepackage{amsmath}

\usepackage[version=4]{mhchem}
\usepackage{siunitx}
\usepackage{glossaries}
\usepackage{todonotes}

\newcommand{\bfl}{\begin{flushleft}}
\newcommand{\efl}{\end{flushleft}}

\newacronym{RIXS}{RIXS}{resonant inelastic x-ray scattering}
\newacronym{FWHM}{FWHM}{full-width at half-maximum}
\newacronym{HWHM}{HWHM}{half-width at half-maximum}
\newacronym{2D}{2D}{two-dimensional}
\newacronym{XAS}{XAS}{x-ray absorption spectroscopy}
\newacronym{MBE}{MBE}{molecular beam epitaxy}
\newacronym{NSLS-II}{NSLS-II}{National Synchrotron Light Source II}
\newacronym{MIR}{MIR}{mid-infrared}

\begin{document}

\title{Magnetic excitations in Nd$_{n+1}$Ni$_{n}$O$_{3n+1}$ Ruddlesden-Popper nickelates observed via resonant inelastic x-ray scattering}

\date{\today}
\author{Sophia F. R. TenHuisen}
\affiliation{Department of Physics, Harvard University, Cambridge, MA, USA}
\affiliation{John A. Paulson School of Engineering and Applied Sciences, Harvard University, Cambridge, MA, USA}
\author{Grace A. Pan}
\affiliation{Department of Physics, Harvard University, Cambridge, MA, USA}
\author{Qi Song}
\affiliation{Department of Physics, Harvard University, Cambridge, MA, USA}
\author{Denitsa R. Baykusheva}
\affiliation{Department of Physics, Harvard University, Cambridge, MA, USA}
\email{Present address: Institute of Science and Technology Austria, Klosterneuburg, Austria}

\author{Dan Ferenc Segedin}
\affiliation{Department of Physics, Harvard University, Cambridge, MA, USA}

\author{Berit H. Goodge}
\affiliation{School of Applied and Engineering Physics, Cornell University, Ithaca, NY, USA}
\affiliation{Kavli Institute at Cornell for Nanoscale Science, Cornell University, Ithaca, NY, USA}

\author{Hanjong Paik}
\affiliation{Platform for the Accelerated Realization, Analysis and Discovery of Interface Materials (PARADIM), Cornell University, Ithaca, NY, USA}

\author{Jonathan Pelliciari}
\affiliation{National Synchrotron Light Source II, Brookhaven National Laboratory, Upton, New York 11973, USA}
\author{Valentina Bisogni}
\affiliation{National Synchrotron Light Source II, Brookhaven National Laboratory, Upton, New York 11973, USA}

\author{Yanhong Gu}
\affiliation{National Synchrotron Light Source II, Brookhaven National Laboratory, Upton, New York 11973, USA}

\author{Stefano Agrestini}
\affiliation{Diamond Light Source, Harwell Campus, Didcot OX11 0DE, UK}
\author{Abhishek Nag}
\affiliation{Diamond Light Source, Harwell Campus, Didcot OX11 0DE, UK}
\author{Mirian Garc\'{i}a-Fern\'{a}ndez}
\affiliation{Diamond Light Source, Harwell Campus, Didcot OX11 0DE, UK}
\author{Ke-Jin Zhou}
\affiliation{Diamond Light Source, Harwell Campus, Didcot OX11 0DE, UK}

\author{Lena F. Kourkoutis}
\thanks{Deceased}
\affiliation{School of Applied and Engineering Physics, Cornell University, Ithaca, NY, USA}
\affiliation{Kavli Institute at Cornell for Nanoscale Science, Cornell University, Ithaca, NY, USA}

\author{Charles M. Brooks}
\affiliation{Department of Physics, Harvard University, Cambridge, MA, USA}
\author{Julia A. Mundy}
\affiliation{Department of Physics, Harvard University, Cambridge, MA, USA}
\author{Mark P. M. Dean}
\thanks{mdean@bnl.gov}
\affiliation{Condensed Matter Physics and Materials Science Department, Brookhaven National Laboratory, Upton, New York 11973, USA}
\author{Matteo Mitrano}
\thanks{mmitrano@g.harvard.edu}
\affiliation{Department of Physics, Harvard University, Cambridge, MA, USA}

\begin{abstract}
Magnetic interactions are thought to play a key role in the properties of many unconventional superconductors, including cuprates, iron pnictides, and square-planar nickelates. Superconductivity was also recently observed in the bilayer and trilayer Ruddlesden-Popper nickelates, whose electronic structure is expected to differ from that of cuprates and square-planar nickelates. Here we study how electronic structure and magnetic interactions evolve with the number of layers, $n$, in thin film Ruddlesden-Popper nickelates Nd$_{n+1}$Ni$_{n}$O$_{3n+1}$ with $n=1,\:3$, and 5 using resonant inelastic x-ray scattering (RIXS). The RIXS spectra are consistent with a high-spin $|3d^8 \underline{L} \rangle$ electronic configuration, resembling that of La$_{2-x}$Sr$_x$NiO$_4$ and the parent perovskite, NdNiO$_3$. The magnetic excitations soften to lower energy in the structurally self-doped, higher-$n$ films. Our observations confirm that structural tuning is an effective route for altering electronic properties, such as magnetic superexchange, in this prominent family of materials.
\end{abstract}

\maketitle

\section{Introduction}
While much remains unknown about unconventional superconductivity, strong magnetic superexchange and reduced dimensionality likely play important roles in achieving high superconducting transition temperatures~\cite{Scalapino_2012, Keimer2015, Dean2015insights}. The square-planar family of nickelates, including the infinite-layer $R$NiO$_2$ ($R$ = La, Pr, Nd) and the quintuple-layer Nd$_6$Ni$_5$O$_{12}$ fit nicely into this picture in many ways, featuring \gls*{2D} transition-metal oxide planes and a $d^{9-\delta}$ electronic configuration~\cite{Li2019, zeng2020phase, pan2022superconductivity}. High energy magnetic excitations are observed throughout the nickelate phase diagram~\cite{Lu2021, dean2021strong,Chen2024, Fan2024capping,Rosa2024spin,Gao2024magnetic, Zhong2025epitaxial}, although much lower critical temperatures are found in the square-planar nickelates than in cuprates. Comparing these two seemingly similar families of materials can help uncover the origins of superconductivity and identify new strategies to optimize superconductivity.

\begin{figure}
    \centering
    \includegraphics[width=0.45\textwidth]{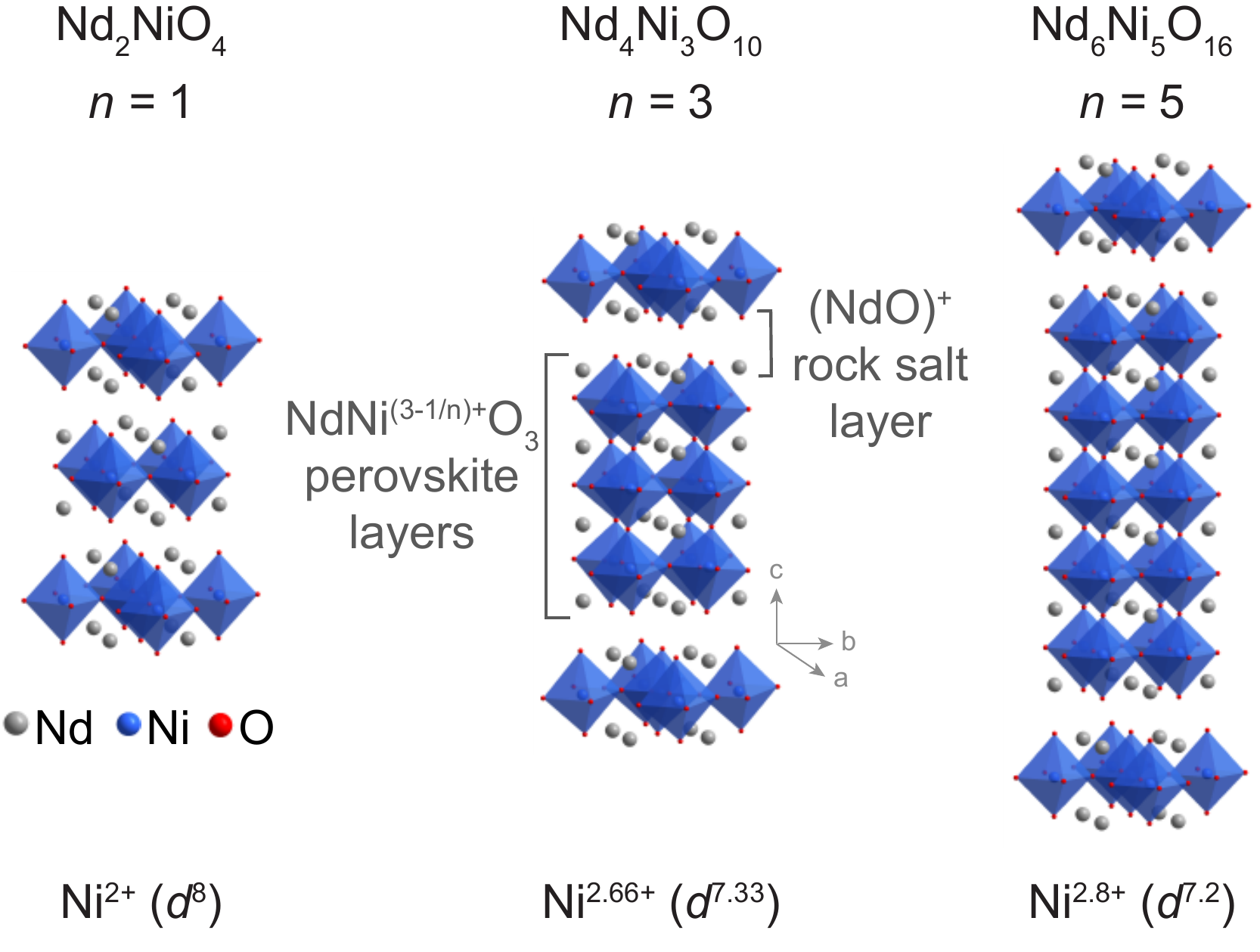}
    \caption{Structures of Ruddlesden-Popper Nd$_{n+1}$Ni$_{n}$O$_{3n+1}$ nickelates for $n=1$, $3$, and $5$. The unit cell hosts $n$ NdNiO$_3$ perovskite layers separated by NdO$^+$ rock-salt spacer layers. As $n$ varies, the nominal Ni valence changes as $d^{7 + 1/n}$.}
    \label{fig:crystals}
\end{figure}

Recently, superconductivity was also observed in bilayer and trilayer Ruddlesden-Popper nickelates $R_{3}$Ni$_{2}$O$_{7}$ ($n=2$) and $R_{4}$Ni$_{3}$O$_{10}$ ($n=3$)  for $R$ = La, Pr, under pressure~\cite{Sun2023,Zhang2024, chen2024nonbulk, Li2025signature, Huang2024signature, pei2024pressure} or epitaxial strain~\cite{Ko2024signatures, Liu2025superconductivity, Zhou2024ambient,bhatt2025resolving}. This discovery has expanded the variety of superconducting nickelates beyond the square-planar geometry~\cite{Li2019,zeng2020phase, pan2022superconductivity}, to octahedrally-coordinated nickelates and to substantially higher critical temperatures~\cite{Sun2023, Zhang2024}. More generally, the layered Ruddlesden-Popper nickelates, $R_{n+1}$Ni$_{n}$O$_{3n+1}$, provide a way to explore the nickelate phase diagram through structural tuning. As shown in Fig.~\ref{fig:crystals}, the Ruddlesden-Popper structure consists of $n$ layers of perovskite $R$NiO$_3$ separated by ($R$-O)$^+$ charged rock-salt layers. Increasing $n$ tunes the effective electron count of the NiO$_2$ planes by $1/n$ per nickel, from a $d^8$ configuration in $R_2$NiO$_4$ ($n = 1$) to a nominal $d^7$ configuration in $R$NiO$_3$ ($n = \infty$) compound, tuning the electronic behavior from semiconducting to metallic (see Supplemental Material Sec.~\ref{sec:transport}~\cite{suppl}) while in principle avoiding the disorder associated with chemical doping. This positions the layered Ruddlesden-Popper nickelates as a promising material family for exploring and tuning the superconducting ground state. 

Although these materials share the layered perovskite structure common to many cuprates, it is unclear whether they fit into the same `cuprate-like' picture which may be relevant to the square-planar nickelates~\cite{Lee2004infinite, Botana2020similarities, Hepting2021soft, Hepting2020electronic, Goodge2021doping, Rossi2021orbital}. The square-planar nickelates adopt a $d^{9-1/n}$ configuration, with dominant in-plane orbital polarization~\cite{Zhang2017large}. The key electronic interactions in the octahedrally-coordinated Ruddlesden-Popper nickelates may differ substantially from the square-planar nickelates, as Ruddlesden-Popper nickelates have a nominal $d^{7+1/n}$ configuration, with holes occupying both the in-plane $3d_{x^2-y^2}$ and out-of-plane $3d_{3z^2-r^2}$ orbitals~\cite{Norman2023, Fabbris2023resonant, li2017fermiology}. The character of the doped holes may also differ: while undoped $n=1$ cuprates and nickelates are both antiferromagnetic insulators, lightly hole-doped cuprates quickly become metallic and a superconducting dome emerges~\cite{Keimer2015}. Magnetic excitations are broadened due to increased metallicity, but remain at high energy~\cite{LeTacon2011intense}. In contrast, doped $n=1$ nickelates do not exhibit superconductivity, but instead display a wide array of insulating stripe phases~\cite{Boothroyd2004magnetic, Yoshizawa2000stripe, Kajimoto2003spontaneous, Ishizaka2003charge, Uchida2012pseudogap}. Holes and spins are bound via electron-electron and electron-phonon couplings, forming polarons~\cite{Chen1993charge}. This localizes spins in the lattice, resulting in a dramatic softening of the magnetic excitations with hole doping~\cite{Fabbris2017, Woo2005mapping}. 

Structural tuning across the Ruddlesden-Popper nickelate series offers access to a distinct subspace of material phases and may help identify properties that are inaccessible by doping alone. Here we study how electronic and magnetic interactions evolve with layer number $n$, performing high-resolution \gls*{RIXS} measurements on Nd$_{n+1}$Ni$_{n}$O$_{3n+1}$ thin films for $n=1$, $3$, and $5$. Orbital excitations reveal the expected $| d^8 \rangle$ orbital configuration in the $n=1$ compound. Upon increasing $n$, these features evolve towards a $| d^8 \underline{L} \rangle$ configuration, indicating holes are primarily added to the ligand bands rather than inducing a $|d^7\rangle$ state. Magnetic excitations with an energy scale of order 50-70~meV are prominent throughout the entire family of materials, with softening at higher $n$ due to increased doping. The observed softening is smaller than that expected based on effective doping alone, as dimensionality effects from increasing $n$ partially offset the doping-induced softening. Our observations demonstrate that the Ruddlesden-Popper nickelates share many key features with other perovskite nickelates, with structural tuning providing an effective and unique approach to modify electronic and magnetic properties.

\section{Experimental Methods}
\gls*{RIXS} has proven effective in determining the properties of nickelates \cite{Hepting2020electronic, Mitrano2024exploring}. We applied this technique to thin films of the $n=1$, $3$, and $5$ layer Ruddlesden-Popper Nd$_{n+1}$Ni$_{n}$O$_{3n+1}$ compounds, taken at the Ni $L_3$-edge and with $\pi$-polarized light in order to maximize the magnetic signal. \gls*{RIXS} measurements for the $n=1$ and $n=3$ compounds were performed at the \gls*{NSLS-II} beamline 2-ID (SIX)~\cite{Jarrige2018paving} at a temperature of 35~K, at a scattering angle of $2\theta = 150^\circ$ and with an experimental resolution of 31~meV. Measurements for the $n=5$ compound were performed at Diamond Light Source beamline ID21~\cite{Zhou2022i21} at a temperature of 20 K, at a scattering angle of $2\theta = 154^\circ$ and with an experimental resolution of 36~meV. To probe the dispersion of magnetic excitations, the scattering angle $2\theta$ was kept fixed while the sample angle $\theta$ was rotated to control the in-plane momentum transfer to the sample, along the $[H,0,L]$ direction of the pseudo-tetragonal unit cell. In order to compare spectra collected from different samples and at different beamlines, we present spectra normalized to the integrated intensity of the orbital excitations \cite{Braicovich2010momentum}. 

Nd$_{n+1}$Ni$_{n}$O$_{3n+1}$ films were synthesized using oxide \gls*{MBE} as described in Refs.~\cite{Pan2022synthesis,ferenc2023limits}. All films are epitaxially strained to the substrate. We focus on samples synthesized on (110)-oriented NdGaO$_3$, which has a pseudo-perovskite lattice constant of 3.86~\AA{} and thus provides a small amount of tensile strain, $\epsilon \approx +1.0\%$, to the Ruddlesden-Popper nickelate family \cite{Li2020contrasting}. Additional data for Nd$_6$Ni$_5$O$_{16}$ on (001)-oriented LaAlO$_3$ ($\epsilon \approx -0.9\%$) is shown in the Supplemental Material Sec.~\ref{sec:strain}~\cite{suppl}. Structural and electrical sample characterization is detailed in the Supplemental Material Sec.~\ref{sec:characterization}~\cite{suppl}. By studying compounds where the rare-earth site is occupied by neodymium rather than the more commonly studied lanthanum series, we avoid contamination on Ni $L_3$-edge \gls*{RIXS} from the nearby La $M_{4}$ absorption edge. We also emphasize that thin film techniques provide unique access to higher-$n$ Ruddlesden-Popper phases, as only compounds up to $n=3$ can be synthesized in bulk \cite{GREENBLATT1997174}.

\section{Orbital excitations and hole configuration}
\begin{figure}
    \centering
    \includegraphics[width=0.48\textwidth]{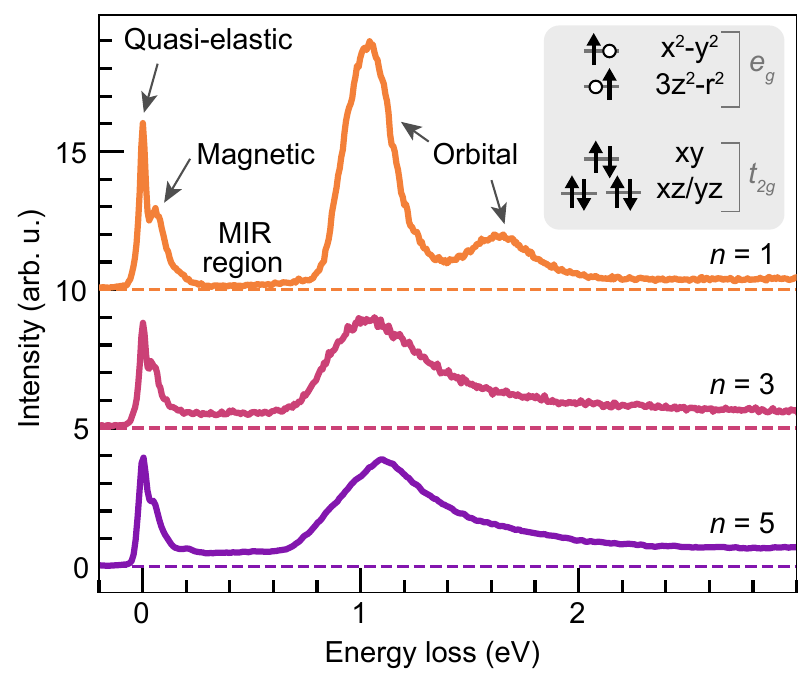}
    \caption{RIXS spectra of Ruddlesden-Popper nickelates for $n=1$ (orange), $3$ (dark pink) and $5$ (purple). The majority of the spectral weight observed around $\sim$1~eV and above corresponds to orbital excitations. Spectra are collected at $q = (-0.4,0)$ r.l.u. ($\theta = 24^\circ$) with $\pi$-incident polarization, probing primarily out-of-plane orbitals. Data for $n=1$ and $n=3$ were collected at 35~K, while data for $n=5$ were collected at 20~K. The inset depicts the ground state orbital configuration for $n=1$.}
    \label{fig:dds}
\end{figure}

To study the electronic structure of Nd$_{n+1}$Ni$_n$O$_{3n+1}$ we examine the orbital excitations observed in RIXS spectra, shown in Fig.~\ref{fig:dds} for films with $n=1$, $3$, and $5$. In Nd$_2$NiO$_4$ ($n=1$, $3d^8$) we observe two sharp orbital excitations at 1.04~eV and 1.6~eV, in good agreement with published \gls*{RIXS} data on bulk La$_2$NiO$_4$~\cite{Fabbris2017}. In Nd$_4$Ni$_3$O$_{10}$~($n=3$, nominal $d^{7.33}$) in Nd$_6$Ni$_5$O$_{16}$ ($n=5$, nominal $d^{7.2}$), the orbital excitations broaden in energy but remain centered at similar energies.

La$_2$NiO$_4$ is known to adopt a Ni $3d^8$ high-spin configuration in which the two holes reside in the Ni $3d$ $e_g$ orbitals and have parallel spin due to the Hund's exchange interaction (Fig.~\ref{fig:dds}). The orbital excitations were previously shown to be well-captured by excitations from this atomic configuration~\cite{Fabbris2017} and agree with the excitations we observe in Nd$_2$NiO$_4$ ($n=1$). The lower energy orbital excitation at 1.04~eV is attributed to a transition from the $B_{1g}$ symmetric ground state to an $E_g$ symmetric state with one of the holes moving into the $t_{2g}$ orbitals. The higher energy peak at 1.6~eV is composed of an $A_{1g}$ excitation from an $S=1$ to $S=0$ configuration, a $B_{2g}$ excitation with one of the holes transferred to the $t_{2g}$ manifold, and two $A_{2g}$ excitations, one with a single hole in the $t_{2g}$ manifold and one with both holes in the $t_{2g}$ manifold~\cite{Fabbris2017}. Thus the orbital excitations in  Nd$_2$NiO$_4$ ($n=1$) can be well-captured as excitations of the local $d$ multiple. 

As $n$ is increased, additional holes are doped into the perovskite layers, modifying the ground state and the orbital excitations. In Nd$_4$Ni$_3$O$_{10}$~($n=3$) the Ni sites have a nominal $d^{7.33}$ valence and in Nd$_6$Ni$_5$O$_{16}$~($n=5$) the Ni sites have a nominal $d^{7.2}$ valence. Nonetheless, the orbital excitations remain centered at approximately the same energies, and the polarization dependence indicates that holes remain roughly equally distributed between the in-plane and out-of-plane Ni $e_g$ orbitals (Fig.~\ref{fig:poldep}~\cite{suppl}). The orbital excitations broaden such that the two features can no longer be separately resolved, but remain centered at similar energies.  

The broadening of the orbital excitations in Nd$_4$Ni$_3$O$_{10}$~($n=3$) and Nd$_6$Ni$_5$O$_{16}$~($n=5$) are indicative of a $d^8 \underline{L}$ state rather than a $d^7$ state, sharing a strikingly similarity to the orbital excitations observed in hole-doped La$_{2-x}$Sr$_{x}$NiO$_4$~($n=1$)~\cite{Fabbris2017} and metallic NdNiO$_3$ ($n = \infty$) \cite{bisogni2016ground,Fursich2019resonant}, materials both known to adopt a $d^8 \underline{L}$ configuration. In addition to hole-doping, possible small variations in the crystal field environment between inequivalent Ni layers in Nd$_4$Ni$_3$O$_{10}$~($n=3$) and Nd$_6$Ni$_5$O$_{16}$~($n=5$) could additionally contribute to the observed broadening of the $d^8$ orbital excitations, with each inequivalent layer contributing excitations at slightly different energies. In contrast, the observed orbital excitations are inconsistent with a contribution from a $d^7$ configuration--such a contribution has been calculated in~\cite{bisogni2016ground, Fabbris2017} and was shown to yield additional peaks in the \gls*{RIXS} spectrum outside the energy range of these main $d^8$ excitations, above 2~eV and around ~0.25~eV respectively. 

We believe the spectral changes with increasing $n$ are driven primarily by changes in the effective doping rather than changes in hybridization. To leading order, the in-plane environment is expected to remain unchanged, as the Ni-O bond lengths are fixed for all $n$ by the substrate epitaxy. Changes in the apical Ni-O bonding with increasing $n$ may play a small role in the observed changes, but would manifest most strongly in changes to the orbital dichroism, which are not observed. We therefore consider doping to be the driving force behind the observed changes, consistent with the strong similarity of these data with prior measurements of hole-doped La$_{2-x}$Sr$_{x}$NiO$_4$~($n=1$) \cite{Fabbris2017}. 

In addition to hole-doping the perovskite layers, structural tuning introduces additional inter-layer couplings between adjacent Ni planes within each perovskite block, which might further modify the orbital excitations beyond a single-site picture. In bulk La$_3$Ni$_2$O$_7$ ($n=2$), an additional, Raman-like orbital excitation was observed at 0.4 eV, which is attributed to transitions between $d_{x^2-y^2}$ and $d_{3z^2-r^2}$ orbitals~\cite{Chen2024}. The energy of this excitation is primarily determined by the inter-layer hopping, which results in the formation of molecular subbands~\cite{Jung2022electronic}. If such features occurred in the higher $n$ films studied here, they would appear at substantially different energies, with differences of order ~$t$ \cite{Jung2022electronic}, which would be easily resolvable. We do not see evidence of any such additional peaks in the orbital excitations for Nd$_4$Ni$_3$O$_{10}$ ($n=3$) and  Nd$_6$Ni$_5$O$_{16}$ ($n=5$), which may be because for higher $n$ the distinct orbital subbands are strongly concentrated on distinct Ni layers within the perovskite blocks \cite{Jung2022electronic}, minimizing their cross-section for the highly local \gls*{RIXS} process. 

As shown in the Supplemental Material Fig.~\ref{fig:Emap} ~\cite{suppl}, Nd$_6$Ni$_5$O$_{16}$ ($n=5$) and Nd$_4$Ni$_3$O$_{10}$ ($n=3$) exhibit strong x-ray fluorescence features in which \gls*{RIXS} intensity appears at increasing energy loss as the incident energy increases. This is distinct from regular $dd$ excitations that appear at fixed energy loss independent of incident energy. These fluorescence features arise from and indicate the presence of hybridization between transition metal states and itinerant ligand states~\cite{Norman2023, Shen2023electronic, Fabbris2023resonant}. Similar hybridization features are commonly observed in cuprates~\cite{Lebert2017resonant} and nickelates~\cite{Norman2023, Shen2022role, bisogni2016ground, Fabbris2017, Rossi2021orbital, Hepting2020electronic, Shen2023electronic, Fabbris2023resonant, ren2024resolving}, including NdNiO$_3$ ($n=\infty$)~\cite{bisogni2016ground} and hole-doped La$_{2-x}$Sr$_x$NiO$_4$ ($n=1$)~\cite{Fabbris2017}, underscoring the role of hybridized itinerant ligand states as a unifying feature across these materials.

We also observe featureless, non-dispersing spectral weight in the \gls*{MIR} region, 0.3~-~0.7~eV, in the $n=3$, $5$ compounds, which is nearly absent in the $n=1$ compound (Fig.~\ref{fig:dds}). This energy scale is above that of magnons and multi-magnons. This feature appears only at the \gls*{XAS} resonance~(Fig. \ref{fig:Emap}), phenomenologically different from the clear Raman peak at 0.4 eV in La$_3$Ni$_2$O$_7$ ($n=2$) which is attributed to transitions between $d_{x^2-y^2}$ and $d_{3z^2-r^2}$ orbitals. Instead, the flat filling-in of the \gls*{MIR} spectral weight appears phenomenologically very similar to the behavior observed in NdNiO$_3$~($n=\infty$) upon heating through a metal-insulator transition~\cite{bisogni2016ground}, attributed to charge excitations within the $|d^8 \underline{L}\rangle$ states and associated with partial metallicity. This is consistent with the electronic transport properties of these films: the $n=1$ material is semiconducting with an exponentially increasing resistivity in the low-temperature limit, while the $n=3$ and $5$, on the other hand, are metallic at room temperature with modest resistivity upturns at low temperature (see Supplemental Material Sec.~\ref{sec:transport}~\cite{suppl}, see also references \cite{takeda1992crystal,gupta2024anisotropic,ren2024resolving,catalano2018rare} therein). 

Thus the high-energy RIXS features support a $|d^8 \underline{L}\rangle $ configuration for Ruddlesden-Popper nickelates, with mobile holes doped into hybridized ligand states as $n$ is increased. 

\section{magnetic excitations modified by structure and doping}
\begin{figure*}
    \centering
    \includegraphics[width=0.98\textwidth]{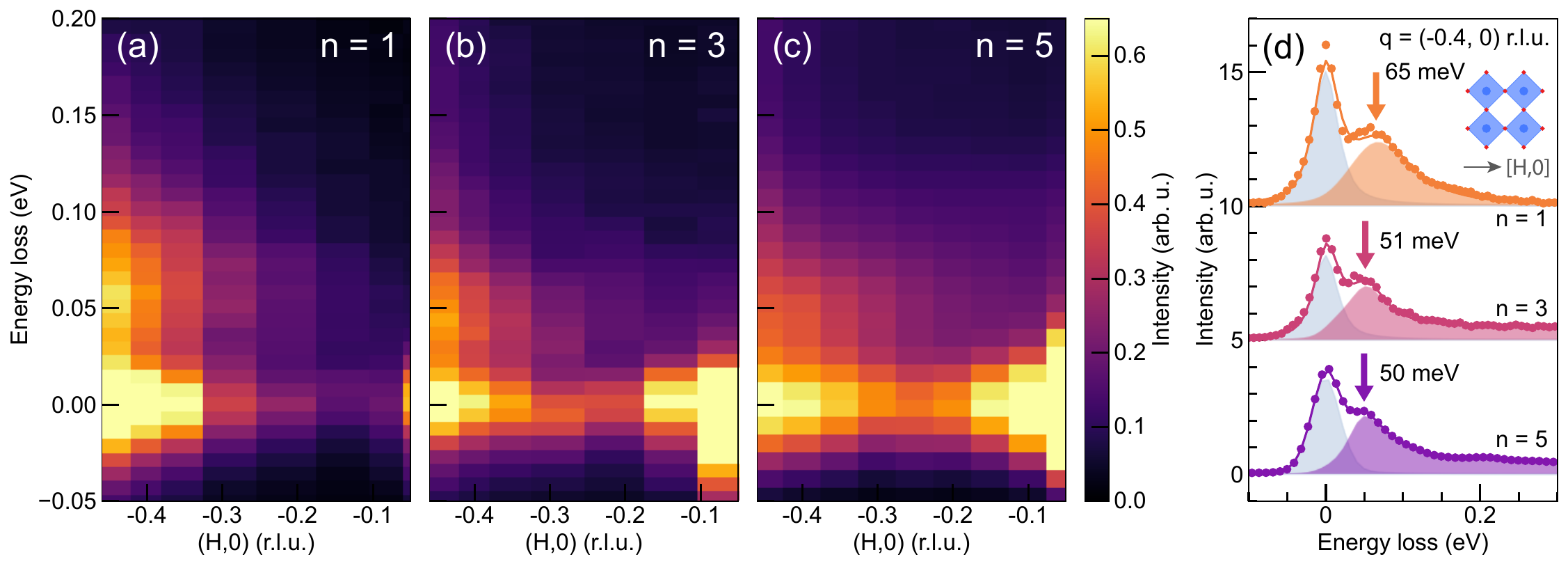}
    \caption{\gls*{RIXS} spectra along $[H,0]$ for $\pi$-incident polarization for (A) $n=1$, (B) $n=3$, and (C) $n=5$. The inelastic component of the \gls*{RIXS} spectra are dominated by magnetic excitations. (D) Linecuts at $q=(0.4,0)$ for $n=1$ (orange), $n=3$ (dark pink), and $n=5$ (purple) with the quasielastic scattering shown in gray and the inelastic (magnetic) scattering shown in color. Data for $n=1$ and $n=3$ were collected at 35~K, while data for $n=5$ were collected at 20~K.}
    \label{fig:rixsmaps}
\end{figure*}

To study the magnetic excitations in Nd$_{n+1}$Ni$_{n}$O$_{3n+1}$, we examine the momentum-dependence of the low energy-loss region of the \gls*{RIXS} spectra along the $[H, 0]$ direction for films with $n=1$, $3$, and $5$, shown in Fig.~\ref{fig:rixsmaps}. \gls*{RIXS} spectra show an elastic line and a broad inelastic feature extending out to 0.2~eV. In all three compounds these features are most prominent around the zone boundary, near $(0.4,0)$, and disperse to lower intensity and lower energy scales towards $(0,0)$. This behavior is typical for magnetic branches, so, consistent with prior work \cite{Fabbris2017, Bialo2023}, we assign these features to damped magnetic excitations. With increasing $n$, additional inelastic spectral weight appears approaching the zone center, $q = (0,0)$, and extends out to higher energy losses in $n=3$ and $5$ than in the $n=1$ sample. 

As shown in Fig.~\ref{fig:crystals}, the unit cells of Nd$_4$Ni$_3$O$_{10}$ ($n=3$) and Nd$_6$Ni$_5$O$_{16}$ ($n=5$) include sets of coupled nickel-oxide trilayers and quintuplelayers, so spin wave theory predicts a large number of distinct spin wave modes (Sec. \ref{sec:LSWT}). These modes are split by the interlayer magnetic exchange coupling, $J_z$, as has been observed in the bilayer systems La$_3$Ni$_2$O$_7$~\cite{Chen2024} and Sr$_3$Ir$_2$O$_7$~\cite{Kim2012large, mazzone2022antiferromagnetic}. Due to the large number of modes and mode-broadening effects arising from finite effective doping, resolving all the predicted modes individually is unlikely to be feasible. In fact, previous measurements of the trilayer square-planar nickelate La$_4$Ni$_3$O$_8$ observed only the average of the three modes expected based on spin wave theory \cite{dean2021strong}.

We attribute the enhanced spectral weight around 0.1~eV near $q=(0,0)$ in the higher $n$ compounds to a magnetic branch with partially optical character. The optical spectral weight in Nd$_4$Ni$_3$O$_{10}$ ($n=3$) appears weaker than in Nd$_6$Ni$_5$O$_{16}$ ($n=5$), which may be due to a reduced importance of interplane interactions in Nd$_4$Ni$_3$O$_{10}$. Neutron scattering measurements on bulk La$_4$Ni$_3$O$_{10}$ indeed reveal a concentration of spin density in the outer layers of the trilayer block~\cite{zhang2020intertwined}, reducing the magnetic coupling between planes. 

Since the individual spin-wave modes in the $n=3$ and $5$ materials cannot be resolved individually, we use a phenomenological model to separate the inelastic features from the elastic line and compare results for Nd$_{n+1}$Ni$_{n}$O$_{3n+1}$ compounds with $n=1$, $3$, and $5$. The phenomenological model includes a pseudovoigt peak for the quasi-elastic scattering, and an error function for the flat mid-infrared background in the $n=3$, $5$ compounds. For $n=1$ and $3$ compounds we use a single damped harmonic oscillator function to capture the remaining inelastic spectral weight. For $n=5$ we were better able to capture the inelastic features by instead using four pseudovoigt peaks, though our results are insensitive to the exact fitting model used. An example of the total fit is shown for $q = (0.4,0)$ in Fig.~\ref{fig:rixsmaps} D. The quasielastic scattering is indicated in light gray, while the sum of the inelastic components is indicated by the shaded colored region. The position of maximum intensity of the inelastic components is taken as the magnon bandwidth and is indicated by the arrows. At $q = (0.4,0)$ the magnetic excitations are peaked at $65$~meV in Nd$_2$NiO$_4$ ($n=1$) and soften to $51$~meV and $50$~meV in Nd$_4$Ni$_3$O$_{10}$ ($n=3$) and Nd$_6$Ni$_5$O$_{16}$ ($n=5$), respectively. 

The \gls*{RIXS} spectra for the Nd$_2$NiO$_4$ ($n=1$) sample qualitatively agree with published magnetic dispersions for La$_2$NiO$_4$ ($n=1$)~\cite{Fabbris2017,Bialo2023}, but the magnon bandwidth is reduced to about 75\% of that observed in the La-based material, consistent with reported differences between bulk Nd$_2$NiO$_4$ and La$_2$NiO$_4$~\cite{Batlle1991}. Larger perovskite distortions and rotations in Nd-based compounds can play a key role in driving this reduction in magnetic energy scales by bending the Ni-O-Ni bond angles~\cite{SaezPuche1989,RodriguezCarvajal1990,catalano2018rare}, weakening antiferromagnetic superexchange interactions. These rotations can also increase spin canting, introducing a ferromagnetic out-of-plane magnetization component which can cause an effective reduction in the magnon energy~\cite{Batlle1991, Batlle1992, RanjanMaiti2019, KhomskiiTMs}. Lastly, the large magnetic moment on the neodymium atoms can further interact with magnetic moments on Ni sites, competing with in-plane magnetic exchange interactions and introducing a large magnetic anisotropy~\cite{Batlle1991}. 

Magnetic energy scales are reduced by about 25\% in Nd$_4$Ni$_3$O$_{10}$ ($n=3$) and Nd$_6$Ni$_5$O$_{16}$ ($n=5$) relative to Nd$_2$NiO$_4$ ($n=1$) (Fig.~\ref{fig:rixsmaps}). We are unable to resolve a notable difference between Nd$_4$Ni$_3$O$_{10}$ ($n=3$) and Nd$_6$Ni$_5$O$_{16}$ ($n=5$). In fact, the magnon bandwidth in both Nd$_4$Ni$_3$O$_{10}$ ($n=3$) and Nd$_6$Ni$_5$O$_{16}$ ($n=5$) agrees well with that observed in NdNiO$_3$ ($n=\infty$)~\cite{Fursich2019resonant, Lu2018}, suggesting that as $n$ increases, magnetic interactions quickly approach the behavior seen in the end member of the series. For all $n$ studied, the magnetic bandwidth is similar to that of the Ruddlesden-Popper bilayer La$_3$Ni$_2$O$_7$ \cite{Chen2024} and the square-planar trilayer La$_4$Ni$_3$O$_8$~\cite{dean2021strong}, indicating common magnetic interaction strengths across these diverse layered nickelate materials. 

We interpret the softening of magnetic interactions with increasing $n$ to be dominantly driven by the increase in effective doping, which can disrupt the antiferromagnetic exchange by introducing states that act as spinless impurities. This suggests analogies between Ruddlesden-Popper nickelates and other materials such as square-planar nickelates~\cite{Lu2021} and La$_{2-x}$Sr$_x$NiO$_4$ ($n=1$), where magnon energies are reduced by a factor of two or more upon the addition of $0.5$ holes/site~\cite{Fabbris2017, Woo2005mapping, Freeman2005, Bourges2003}. Compared to these chemically-doped nickelates, however, the 25\% magnon softening in Nd$_4$Ni$_3$O$_{10}$ ($n=3$) and Nd$_6$Ni$_5$O$_{16}$ ($n=5$) (Fig.~\ref{fig:rixsmaps}) is less than that predicted solely on the basis of the effective doping of an additional $0.67$ and $0.8$ holes/site, respectively. The effective doping might be overestimated by simple electron counting arguments, as structurally-driven doping is predicted to introduce additional correlations that renormalize the effective doping amount in the layered square-planar nickelates~\cite{Labollita2021electronic, LaBollita2022correlated, Worm2022correlations}. Holes may also be unevenly distributed between planes within $n$-layer blocks, leading to contributions from inequivalent NiO$_2$ layers, further complicating the magnon mode structure probed by \gls*{RIXS}. However, both these effects would likely impart only small changes to the effective doping. 

Magnetic energy scales may be further modified by the increase of dimensionality with increasing $n$, in addition to changes due to the effective doping of $1/n$ holes per Ni$^{2+}$ ion. Increasing $n$ introduces an additional out-of-plane exchange pathway between neighboring NiO$_2$ layers which is absent in Nd$_2$NiO$_4$ ($n=1$). In Nd$_2$NiO$_4$ ($n=1$), the magnetic Ni sites can interact with four neighboring Ni sites. In higher $n$ compounds, Ni atoms in the inner layers can interact with six neighboring Ni sites, while Ni atoms in the two outer layers can interact with five neighboring Ni sites. Because each Ni atom now has more magnetic exchange pathways, we expect an overall increase in the magnetic bandwidth with increasing $n$ (Sec. \ref{sec:LSWT}), competing with the overall decrease due to increased doping. The out-of-plane exchange interaction $J_z$ can be quite strong due to the coupling of partially filled $d_{3z^2-r^2}$ orbitals by apical oxygen atoms. In fact, we see already for Nd$_4$Ni$_3$O$_{10}$ ($n=3$) that the magnetic bandwidth closely approaches the $\sim50$~meV bandwidth seen in NdNiO$_3$ ($n=\infty$)~\cite{Fursich2019resonant, Lu2018}. Thus the introduction of these out-of-plane exchange interactions may play a key role in setting the magnetic energy scales in higher order $n$ Ruddlesden-Popper nickelates, partially offsetting the magnetic softening due to increased doping. 

\section{Conclusion}
We used Ni $L_3$-edge \gls*{RIXS} to determine the evolution of the electronic and magnetic structure of layered Ruddlesden-Popper nickelates Nd$_{n+1}$Ni$_n$O$_{3n+1}$ with layer number $n$. We show that as $n$ is increased, holes are doped into itinerant ligand bands, moving the system from a $|d^8 \rangle$ configuration for $n=1$ to a $|d^8\underline{L}\rangle$ configuration. This finding agrees well with the behavior of the end members of the series, NdNiO$_3$ ($n = \infty$) and Sr-doped La$_{2-x}$Sr$_x$NiO$_4$ ($n = 1$), confirming systematic changes across this family of materials with doping and structural tuning. These doped holes cause a softening of the magnetic excitations in the higher $n$ compounds; however, the softening is partially mitigated due to the introduction of additional out-of-plane magnetic interactions with increasing $n$, which increase the energy scale of magnetic interactions. Thus, structural and dimensional control can tune both electronic and magnetic properties of materials in a complex and interdependent fashion, allowing access to a richer material space than that accessible by chemical doping alone. 

\section*{Data availability}
All data is available from authors upon reasonable request. 

\section*{Acknowledgments}
\noindent 
Work by S.F.R.T., D. R. B., J.P., V.B., M.P.M.D., and M.M.\ was supported by the U.S.\ Department of Energy (DOE), Division of Materials Science, under Contract DE-SC0012704. G.A.P.\ and D.F.S.\ are primarily supported by U.S. Department of Energy (DOE), Office of Basic Energy Sciences, Division of Materials Sciences and Engineering, under Award No.\ DE-SC0021925; and by NSF Graduate Research Fellowship Grant No. DGE-1745303. S.F.R.T. acknowledges additional support from the U.S. Department of Energy (DOE), Office of Science, Office of Workforce Development for Teachers and Scientists, Office of Science Graduate Student Research (SCGSR) program. The SCGSR program is administered by the Oak Ridge Institute for Science and Education for the DOE under contract number DE‐SC0014664. G.A.P.\ acknowledges additional support from the Paul \& Daisy Soros Fellowship for New Americans.  Q.S.\ was supported by the Science and Technology Center for Integrated Quantum Materials, NSF Grant No.~DMR-1231319. B.H.G and L.F.K. acknowledge support by PARADIM, NSF No. DMR-2039380. J.A.M. acknowledges support from the U.S.\ Department of Energy (DOE), Office of Basic Energy Sciences, Division of Materials Sciences and Engineering, under Award No.\ DE-SC0021925.  Materials growth and electron microscopy were supported by PARADIM under National Science Foundation (NSF) Cooperative Agreement No. DMR-2039380. Electron microscopy made use of the Cornell Center for Materials Research (CCMR) Shared Facilities. The Thermo Fisher Spectra 300 X-CFEG was acquired with support from PARADIM, an NSF MIP (DMR-2039380) and Cornell University. The FEI Titan Themis 300 was acquired through No. NSF-MRI-1429155, with additional support from Cornell University, the Weill Institute, and the Kavli Institute at Cornell. The Thermo Fisher Helios G4 UX FIB was acquired with support by NSF No. DMR-1539918. This research used beamline 2-ID of the National Synchrotron Light Source II, a U.S.\ DOE Office of Science User Facility operated for the DOE Office of Science by Brookhaven National Laboratory under Contract No.~DE-SC0012704. We acknowledge Diamond Light Source for time on Beamline I21 under Proposal MM27484.

\bibliography{refs.bib}

\begin{thebibliography}{77}%
\makeatletter
\providecommand \@ifxundefined [1]{%
 \@ifx{#1\undefined}
}%
\providecommand \@ifnum [1]{%
 \ifnum #1\expandafter \@firstoftwo
 \else \expandafter \@secondoftwo
 \fi
}%
\providecommand \@ifx [1]{%
 \ifx #1\expandafter \@firstoftwo
 \else \expandafter \@secondoftwo
 \fi
}%
\providecommand \natexlab [1]{#1}%
\providecommand \enquote  [1]{``#1''}%
\providecommand \bibnamefont  [1]{#1}%
\providecommand \bibfnamefont [1]{#1}%
\providecommand \citenamefont [1]{#1}%
\providecommand \href@noop [0]{\@secondoftwo}%
\providecommand \href [0]{\begingroup \@sanitize@url \@href}%
\providecommand \@href[1]{\@@startlink{#1}\@@href}%
\providecommand \@@href[1]{\endgroup#1\@@endlink}%
\providecommand \@sanitize@url [0]{\catcode `\\12\catcode `\$12\catcode `\&12\catcode `\#12\catcode `\^12\catcode `\_12\catcode `\%12\relax}%
\providecommand \@@startlink[1]{}%
\providecommand \@@endlink[0]{}%
\providecommand \url  [0]{\begingroup\@sanitize@url \@url }%
\providecommand \@url [1]{\endgroup\@href {#1}{\urlprefix }}%
\providecommand \urlprefix  [0]{URL }%
\providecommand \Eprint [0]{\href }%
\providecommand \doibase [0]{https://doi.org/}%
\providecommand \selectlanguage [0]{\@gobble}%
\providecommand \bibinfo  [0]{\@secondoftwo}%
\providecommand \bibfield  [0]{\@secondoftwo}%
\providecommand \translation [1]{[#1]}%
\providecommand \BibitemOpen [0]{}%
\providecommand \bibitemStop [0]{}%
\providecommand \bibitemNoStop [0]{.\EOS\space}%
\providecommand \EOS [0]{\spacefactor3000\relax}%
\providecommand \BibitemShut  [1]{\csname bibitem#1\endcsname}%
\let\auto@bib@innerbib\@empty
\bibitem [{\citenamefont {Scalapino}(2012)}]{Scalapino_2012}%
  \BibitemOpen
  \bibfield  {author} {\bibinfo {author} {\bibfnamefont {D.~J.}\ \bibnamefont {Scalapino}},\ }\bibfield  {title} {\bibinfo {title} {A common thread: The pairing interaction for unconventional superconductors},\ }\href {https://doi.org/10.1103/RevModPhys.84.1383} {\bibfield  {journal} {\bibinfo  {journal} {Rev. Mod. Phys.}\ }\textbf {\bibinfo {volume} {84}},\ \bibinfo {pages} {1383} (\bibinfo {year} {2012})}\BibitemShut {NoStop}%
\bibitem [{\citenamefont {Keimer}\ \emph {et~al.}(2015)\citenamefont {Keimer}, \citenamefont {Kivelson}, \citenamefont {Norman}, \citenamefont {Uchida},\ and\ \citenamefont {Zaanen}}]{Keimer2015}%
  \BibitemOpen
  \bibfield  {author} {\bibinfo {author} {\bibfnamefont {B.}~\bibnamefont {Keimer}}, \bibinfo {author} {\bibfnamefont {S.~A.}\ \bibnamefont {Kivelson}}, \bibinfo {author} {\bibfnamefont {M.~R.}\ \bibnamefont {Norman}}, \bibinfo {author} {\bibfnamefont {S.}~\bibnamefont {Uchida}},\ and\ \bibinfo {author} {\bibfnamefont {J.}~\bibnamefont {Zaanen}},\ }\bibfield  {title} {\bibinfo {title} {From quantum matter to high-temperature superconductivity in copper oxides},\ }\href {https://doi.org/10.1038/nature14165} {\bibfield  {journal} {\bibinfo  {journal} {Nature}\ }\textbf {\bibinfo {volume} {518}},\ \bibinfo {pages} {179} (\bibinfo {year} {2015})}\BibitemShut {NoStop}%
\bibitem [{\citenamefont {Dean}(2015)}]{Dean2015insights}%
  \BibitemOpen
  \bibfield  {author} {\bibinfo {author} {\bibfnamefont {M.}~\bibnamefont {Dean}},\ }\bibfield  {title} {\bibinfo {title} {Insights into the high temperature superconducting cuprates from resonant inelastic x-ray scattering},\ }\href {https://doi.org/https://doi.org/10.1016/j.jmmm.2014.03.057} {\bibfield  {journal} {\bibinfo  {journal} {J. Magn. Magn. Mater.}\ }\textbf {\bibinfo {volume} {376}},\ \bibinfo {pages} {3} (\bibinfo {year} {2015})}\BibitemShut {NoStop}%
\bibitem [{\citenamefont {Li}\ \emph {et~al.}(2019)\citenamefont {Li}, \citenamefont {Lee}, \citenamefont {Wang}, \citenamefont {Osada}, \citenamefont {Crossley}, \citenamefont {Lee}, \citenamefont {Cui}, \citenamefont {Hikita},\ and\ \citenamefont {Hwang}}]{Li2019}%
  \BibitemOpen
  \bibfield  {author} {\bibinfo {author} {\bibfnamefont {D.}~\bibnamefont {Li}}, \bibinfo {author} {\bibfnamefont {K.}~\bibnamefont {Lee}}, \bibinfo {author} {\bibfnamefont {B.~Y.}\ \bibnamefont {Wang}}, \bibinfo {author} {\bibfnamefont {M.}~\bibnamefont {Osada}}, \bibinfo {author} {\bibfnamefont {S.}~\bibnamefont {Crossley}}, \bibinfo {author} {\bibfnamefont {H.~R.}\ \bibnamefont {Lee}}, \bibinfo {author} {\bibfnamefont {Y.}~\bibnamefont {Cui}}, \bibinfo {author} {\bibfnamefont {Y.}~\bibnamefont {Hikita}},\ and\ \bibinfo {author} {\bibfnamefont {H.~Y.}\ \bibnamefont {Hwang}},\ }\bibfield  {title} {\bibinfo {title} {Superconductivity in an infinite-layer nickelate},\ }\href {https://doi.org/10.1038/s41586-019-1496-5} {\bibfield  {journal} {\bibinfo  {journal} {Nature}\ }\textbf {\bibinfo {volume} {572}},\ \bibinfo {pages} {624} (\bibinfo {year} {2019})}\BibitemShut {NoStop}%
\bibitem [{\citenamefont {Zeng}\ \emph {et~al.}(2020)\citenamefont {Zeng}, \citenamefont {Tang}, \citenamefont {Yin}, \citenamefont {Li}, \citenamefont {Li}, \citenamefont {Huang}, \citenamefont {Hu}, \citenamefont {Liu}, \citenamefont {Omar}, \citenamefont {Jani}, \citenamefont {Lim}, \citenamefont {Han}, \citenamefont {Wan}, \citenamefont {Yang}, \citenamefont {Pennycook}, \citenamefont {Wee},\ and\ \citenamefont {Ariando}}]{zeng2020phase}%
  \BibitemOpen
  \bibfield  {author} {\bibinfo {author} {\bibfnamefont {S.}~\bibnamefont {Zeng}}, \bibinfo {author} {\bibfnamefont {C.~S.}\ \bibnamefont {Tang}}, \bibinfo {author} {\bibfnamefont {X.}~\bibnamefont {Yin}}, \bibinfo {author} {\bibfnamefont {C.}~\bibnamefont {Li}}, \bibinfo {author} {\bibfnamefont {M.}~\bibnamefont {Li}}, \bibinfo {author} {\bibfnamefont {Z.}~\bibnamefont {Huang}}, \bibinfo {author} {\bibfnamefont {J.}~\bibnamefont {Hu}}, \bibinfo {author} {\bibfnamefont {W.}~\bibnamefont {Liu}}, \bibinfo {author} {\bibfnamefont {G.~J.}\ \bibnamefont {Omar}}, \bibinfo {author} {\bibfnamefont {H.}~\bibnamefont {Jani}}, \bibinfo {author} {\bibfnamefont {Z.~S.}\ \bibnamefont {Lim}}, \bibinfo {author} {\bibfnamefont {K.}~\bibnamefont {Han}}, \bibinfo {author} {\bibfnamefont {D.}~\bibnamefont {Wan}}, \bibinfo {author} {\bibfnamefont {P.}~\bibnamefont {Yang}}, \bibinfo {author} {\bibfnamefont {S.~J.}\ \bibnamefont {Pennycook}}, \bibinfo {author} {\bibfnamefont {A.~T.~S.}\ \bibnamefont {Wee}},\ and\ \bibinfo {author}
  {\bibfnamefont {A.}~\bibnamefont {Ariando}},\ }\bibfield  {title} {\bibinfo {title} {Phase diagram and superconducting dome of infinite-layer $\mathrm{Nd}_{1-x}\mathrm{Sr}_{x}\mathrm{NiO}_{2}$ thin films},\ }\href {https://doi.org/10.1103/PhysRevLett.125.147003} {\bibfield  {journal} {\bibinfo  {journal} {Phys. Rev. Lett.}\ }\textbf {\bibinfo {volume} {125}},\ \bibinfo {pages} {147003} (\bibinfo {year} {2020})}\BibitemShut {NoStop}%
\bibitem [{\citenamefont {Pan}\ \emph {et~al.}(2022{\natexlab{a}})\citenamefont {Pan}, \citenamefont {Ferenc~Segedin}, \citenamefont {LaBollita}, \citenamefont {Song}, \citenamefont {Nica}, \citenamefont {Goodge}, \citenamefont {Pierce}, \citenamefont {Doyle}, \citenamefont {Novakov}, \citenamefont {C{\'o}rdova~Carrizales} \emph {et~al.}}]{pan2022superconductivity}%
  \BibitemOpen
  \bibfield  {author} {\bibinfo {author} {\bibfnamefont {G.~A.}\ \bibnamefont {Pan}}, \bibinfo {author} {\bibfnamefont {D.}~\bibnamefont {Ferenc~Segedin}}, \bibinfo {author} {\bibfnamefont {H.}~\bibnamefont {LaBollita}}, \bibinfo {author} {\bibfnamefont {Q.}~\bibnamefont {Song}}, \bibinfo {author} {\bibfnamefont {E.~M.}\ \bibnamefont {Nica}}, \bibinfo {author} {\bibfnamefont {B.~H.}\ \bibnamefont {Goodge}}, \bibinfo {author} {\bibfnamefont {A.~T.}\ \bibnamefont {Pierce}}, \bibinfo {author} {\bibfnamefont {S.}~\bibnamefont {Doyle}}, \bibinfo {author} {\bibfnamefont {S.}~\bibnamefont {Novakov}}, \bibinfo {author} {\bibfnamefont {D.}~\bibnamefont {C{\'o}rdova~Carrizales}}, \emph {et~al.},\ }\bibfield  {title} {\bibinfo {title} {Superconductivity in a quintuple-layer square-planar nickelate},\ }\href {https://www.nature.com/articles/s41563-021-01142-9} {\bibfield  {journal} {\bibinfo  {journal} {Nat. Mater.}\ }\textbf {\bibinfo {volume} {21}},\ \bibinfo {pages} {160} (\bibinfo {year}
  {2022}{\natexlab{a}})}\BibitemShut {NoStop}%
\bibitem [{\citenamefont {Lu}\ \emph {et~al.}(2021)\citenamefont {Lu}, \citenamefont {Rossi}, \citenamefont {Nag}, \citenamefont {Osada}, \citenamefont {Li}, \citenamefont {Lee}, \citenamefont {Wang}, \citenamefont {Garcia-Fernandez}, \citenamefont {Agrestini}, \citenamefont {Shen}, \citenamefont {Been}, \citenamefont {Moritz}, \citenamefont {Devereaux}, \citenamefont {Zaanen}, \citenamefont {Hwang}, \citenamefont {Zhou},\ and\ \citenamefont {W.}}]{Lu2021}%
  \BibitemOpen
  \bibfield  {author} {\bibinfo {author} {\bibfnamefont {H.}~\bibnamefont {Lu}}, \bibinfo {author} {\bibfnamefont {M.}~\bibnamefont {Rossi}}, \bibinfo {author} {\bibfnamefont {A.}~\bibnamefont {Nag}}, \bibinfo {author} {\bibfnamefont {M.}~\bibnamefont {Osada}}, \bibinfo {author} {\bibfnamefont {D.}~\bibnamefont {Li}}, \bibinfo {author} {\bibfnamefont {K.}~\bibnamefont {Lee}}, \bibinfo {author} {\bibfnamefont {B.}~\bibnamefont {Wang}}, \bibinfo {author} {\bibfnamefont {M.}~\bibnamefont {Garcia-Fernandez}}, \bibinfo {author} {\bibfnamefont {S.}~\bibnamefont {Agrestini}}, \bibinfo {author} {\bibfnamefont {Z.}~\bibnamefont {Shen}}, \bibinfo {author} {\bibfnamefont {E.}~\bibnamefont {Been}}, \bibinfo {author} {\bibfnamefont {B.}~\bibnamefont {Moritz}}, \bibinfo {author} {\bibfnamefont {T.}~\bibnamefont {Devereaux}}, \bibinfo {author} {\bibfnamefont {J.}~\bibnamefont {Zaanen}}, \bibinfo {author} {\bibfnamefont {H.}~\bibnamefont {Hwang}}, \bibinfo {author} {\bibfnamefont {K.}~\bibnamefont {Zhou}},\ and\ \bibinfo
  {author} {\bibfnamefont {L.}~\bibnamefont {W.}},\ }\bibfield  {title} {\bibinfo {title} {Magnetic excitations in infinite-layer nickelates},\ }\href {https://doi.org/10.1126/science.abd7726} {\bibfield  {journal} {\bibinfo  {journal} {Science}\ }\textbf {\bibinfo {volume} {373}},\ \bibinfo {pages} {213} (\bibinfo {year} {2021})}\BibitemShut {NoStop}%
\bibitem [{\citenamefont {Lin}\ \emph {et~al.}(2021)\citenamefont {Lin}, \citenamefont {Villar~Arribi}, \citenamefont {Fabbris}, \citenamefont {Botana}, \citenamefont {Meyers}, \citenamefont {Miao}, \citenamefont {Shen}, \citenamefont {Mazzone}, \citenamefont {Feng}, \citenamefont {Chiuzb\ifmmode~\u{a}\else \u{a}\fi{}ian}, \citenamefont {Nag}, \citenamefont {Walters}, \citenamefont {Garc\'{\i}a-Fern\'andez}, \citenamefont {Zhou}, \citenamefont {Pelliciari}, \citenamefont {Jarrige}, \citenamefont {Freeland}, \citenamefont {Zhang}, \citenamefont {Mitchell}, \citenamefont {Bisogni}, \citenamefont {Liu}, \citenamefont {Norman},\ and\ \citenamefont {Dean}}]{dean2021strong}%
  \BibitemOpen
  \bibfield  {author} {\bibinfo {author} {\bibfnamefont {J.~Q.}\ \bibnamefont {Lin}}, \bibinfo {author} {\bibfnamefont {P.}~\bibnamefont {Villar~Arribi}}, \bibinfo {author} {\bibfnamefont {G.}~\bibnamefont {Fabbris}}, \bibinfo {author} {\bibfnamefont {A.~S.}\ \bibnamefont {Botana}}, \bibinfo {author} {\bibfnamefont {D.}~\bibnamefont {Meyers}}, \bibinfo {author} {\bibfnamefont {H.}~\bibnamefont {Miao}}, \bibinfo {author} {\bibfnamefont {Y.}~\bibnamefont {Shen}}, \bibinfo {author} {\bibfnamefont {D.~G.}\ \bibnamefont {Mazzone}}, \bibinfo {author} {\bibfnamefont {J.}~\bibnamefont {Feng}}, \bibinfo {author} {\bibfnamefont {S.~G.}\ \bibnamefont {Chiuzb\ifmmode~\u{a}\else \u{a}\fi{}ian}}, \bibinfo {author} {\bibfnamefont {A.}~\bibnamefont {Nag}}, \bibinfo {author} {\bibfnamefont {A.~C.}\ \bibnamefont {Walters}}, \bibinfo {author} {\bibfnamefont {M.}~\bibnamefont {Garc\'{\i}a-Fern\'andez}}, \bibinfo {author} {\bibfnamefont {K.-J.}\ \bibnamefont {Zhou}}, \bibinfo {author} {\bibfnamefont {J.}~\bibnamefont
  {Pelliciari}}, \bibinfo {author} {\bibfnamefont {I.}~\bibnamefont {Jarrige}}, \bibinfo {author} {\bibfnamefont {J.~W.}\ \bibnamefont {Freeland}}, \bibinfo {author} {\bibfnamefont {J.}~\bibnamefont {Zhang}}, \bibinfo {author} {\bibfnamefont {J.~F.}\ \bibnamefont {Mitchell}}, \bibinfo {author} {\bibfnamefont {V.}~\bibnamefont {Bisogni}}, \bibinfo {author} {\bibfnamefont {X.}~\bibnamefont {Liu}}, \bibinfo {author} {\bibfnamefont {M.~R.}\ \bibnamefont {Norman}},\ and\ \bibinfo {author} {\bibfnamefont {M.~P.~M.}\ \bibnamefont {Dean}},\ }\bibfield  {title} {\bibinfo {title} {Strong superexchange in a ${d}^{9\ensuremath{-}\ensuremath{\delta}}$ nickelate revealed by resonant inelastic x-ray scattering},\ }\href {https://doi.org/10.1103/PhysRevLett.126.087001} {\bibfield  {journal} {\bibinfo  {journal} {Phys. Rev. Lett.}\ }\textbf {\bibinfo {volume} {126}},\ \bibinfo {pages} {087001} (\bibinfo {year} {2021})}\BibitemShut {NoStop}%
\bibitem [{\citenamefont {Chen}\ \emph {et~al.}(2024{\natexlab{a}})\citenamefont {Chen}, \citenamefont {Choi}, \citenamefont {Jiang}, \citenamefont {Mei}, \citenamefont {Jiang}, \citenamefont {Li}, \citenamefont {Agrestini}, \citenamefont {Garcia-Fernandez}, \citenamefont {Sun}, \citenamefont {Huang}, \citenamefont {Shen}, \citenamefont {Wang}, \citenamefont {Hu}, \citenamefont {Lu}, \citenamefont {Zhou},\ and\ \citenamefont {Feng}}]{Chen2024}%
  \BibitemOpen
  \bibfield  {author} {\bibinfo {author} {\bibfnamefont {X.}~\bibnamefont {Chen}}, \bibinfo {author} {\bibfnamefont {J.}~\bibnamefont {Choi}}, \bibinfo {author} {\bibfnamefont {Z.}~\bibnamefont {Jiang}}, \bibinfo {author} {\bibfnamefont {J.}~\bibnamefont {Mei}}, \bibinfo {author} {\bibfnamefont {K.}~\bibnamefont {Jiang}}, \bibinfo {author} {\bibfnamefont {J.}~\bibnamefont {Li}}, \bibinfo {author} {\bibfnamefont {S.}~\bibnamefont {Agrestini}}, \bibinfo {author} {\bibfnamefont {M.}~\bibnamefont {Garcia-Fernandez}}, \bibinfo {author} {\bibfnamefont {H.}~\bibnamefont {Sun}}, \bibinfo {author} {\bibfnamefont {X.}~\bibnamefont {Huang}}, \bibinfo {author} {\bibfnamefont {D.}~\bibnamefont {Shen}}, \bibinfo {author} {\bibfnamefont {M.}~\bibnamefont {Wang}}, \bibinfo {author} {\bibfnamefont {J.}~\bibnamefont {Hu}}, \bibinfo {author} {\bibfnamefont {Y.}~\bibnamefont {Lu}}, \bibinfo {author} {\bibfnamefont {K.~J.}\ \bibnamefont {Zhou}},\ and\ \bibinfo {author} {\bibfnamefont {D.}~\bibnamefont {Feng}},\ }\bibfield
  {title} {\bibinfo {title} {Electronic and magnetic excitations in {La$_3$Ni$_2$O$_7$}},\ }\href {https://doi.org/10.1038/s41467-024-53863-5} {\bibfield  {journal} {\bibinfo  {journal} {Nat. Commun.}\ }\textbf {\bibinfo {volume} {15}},\ \bibinfo {pages} {9597} (\bibinfo {year} {2024}{\natexlab{a}})}\BibitemShut {NoStop}%
\bibitem [{\citenamefont {Fan}\ \emph {et~al.}(2024)\citenamefont {Fan}, \citenamefont {LaBollita}, \citenamefont {Gao}, \citenamefont {Khan}, \citenamefont {Gu}, \citenamefont {Kim}, \citenamefont {Li}, \citenamefont {Bhartiya}, \citenamefont {Li}, \citenamefont {Sun}, \citenamefont {Yang}, \citenamefont {Yan}, \citenamefont {Barbour}, \citenamefont {Zhou}, \citenamefont {Cano}, \citenamefont {Bernardini}, \citenamefont {Nie}, \citenamefont {Zhu}, \citenamefont {Bisogni}, \citenamefont {Mazzoli}, \citenamefont {Botana},\ and\ \citenamefont {Pelliciari}}]{Fan2024capping}%
  \BibitemOpen
  \bibfield  {author} {\bibinfo {author} {\bibfnamefont {S.}~\bibnamefont {Fan}}, \bibinfo {author} {\bibfnamefont {H.}~\bibnamefont {LaBollita}}, \bibinfo {author} {\bibfnamefont {Q.}~\bibnamefont {Gao}}, \bibinfo {author} {\bibfnamefont {N.}~\bibnamefont {Khan}}, \bibinfo {author} {\bibfnamefont {Y.}~\bibnamefont {Gu}}, \bibinfo {author} {\bibfnamefont {T.}~\bibnamefont {Kim}}, \bibinfo {author} {\bibfnamefont {J.}~\bibnamefont {Li}}, \bibinfo {author} {\bibfnamefont {V.}~\bibnamefont {Bhartiya}}, \bibinfo {author} {\bibfnamefont {Y.}~\bibnamefont {Li}}, \bibinfo {author} {\bibfnamefont {W.}~\bibnamefont {Sun}}, \bibinfo {author} {\bibfnamefont {J.}~\bibnamefont {Yang}}, \bibinfo {author} {\bibfnamefont {S.}~\bibnamefont {Yan}}, \bibinfo {author} {\bibfnamefont {A.}~\bibnamefont {Barbour}}, \bibinfo {author} {\bibfnamefont {X.}~\bibnamefont {Zhou}}, \bibinfo {author} {\bibfnamefont {A.}~\bibnamefont {Cano}}, \bibinfo {author} {\bibfnamefont {F.}~\bibnamefont {Bernardini}}, \bibinfo {author} {\bibfnamefont
  {Y.}~\bibnamefont {Nie}}, \bibinfo {author} {\bibfnamefont {Z.}~\bibnamefont {Zhu}}, \bibinfo {author} {\bibfnamefont {V.}~\bibnamefont {Bisogni}}, \bibinfo {author} {\bibfnamefont {C.}~\bibnamefont {Mazzoli}}, \bibinfo {author} {\bibfnamefont {A.~S.}\ \bibnamefont {Botana}},\ and\ \bibinfo {author} {\bibfnamefont {J.}~\bibnamefont {Pelliciari}},\ }\bibfield  {title} {\bibinfo {title} {Capping effects on spin and charge excitations in parent and superconducting {Nd$_{1-x}$Sr$_{x}$NiO$_{2}$}},\ }\href {https://doi.org/10.1103/PhysRevLett.133.206501} {\bibfield  {journal} {\bibinfo  {journal} {Phys. Rev. Lett.}\ }\textbf {\bibinfo {volume} {133}},\ \bibinfo {pages} {206501} (\bibinfo {year} {2024})}\BibitemShut {NoStop}%
\bibitem [{\citenamefont {Rosa}\ \emph {et~al.}(2024)\citenamefont {Rosa}, \citenamefont {Martinelli}, \citenamefont {Krieger}, \citenamefont {Braicovich}, \citenamefont {Brookes}, \citenamefont {Merzoni}, \citenamefont {Moretti~Sala}, \citenamefont {Yakhou-Harris}, \citenamefont {Arpaia}, \citenamefont {Preziosi}, \citenamefont {Salluzzo}, \citenamefont {Fidrysiak},\ and\ \citenamefont {Ghiringhelli}}]{Rosa2024spin}%
  \BibitemOpen
  \bibfield  {author} {\bibinfo {author} {\bibfnamefont {F.}~\bibnamefont {Rosa}}, \bibinfo {author} {\bibfnamefont {L.}~\bibnamefont {Martinelli}}, \bibinfo {author} {\bibfnamefont {G.}~\bibnamefont {Krieger}}, \bibinfo {author} {\bibfnamefont {L.}~\bibnamefont {Braicovich}}, \bibinfo {author} {\bibfnamefont {N.~B.}\ \bibnamefont {Brookes}}, \bibinfo {author} {\bibfnamefont {G.}~\bibnamefont {Merzoni}}, \bibinfo {author} {\bibfnamefont {M.}~\bibnamefont {Moretti~Sala}}, \bibinfo {author} {\bibfnamefont {F.}~\bibnamefont {Yakhou-Harris}}, \bibinfo {author} {\bibfnamefont {R.}~\bibnamefont {Arpaia}}, \bibinfo {author} {\bibfnamefont {D.}~\bibnamefont {Preziosi}}, \bibinfo {author} {\bibfnamefont {M.}~\bibnamefont {Salluzzo}}, \bibinfo {author} {\bibfnamefont {M.}~\bibnamefont {Fidrysiak}},\ and\ \bibinfo {author} {\bibfnamefont {G.}~\bibnamefont {Ghiringhelli}},\ }\bibfield  {title} {\bibinfo {title} {Spin excitations in {Nd$_{1-x}$Sr$_{x}$NiO$_{2}$} and {YBa$_{2}$Cu$_{3}$O$_{7-\delta}$}: The influence of
  hubbard $u$},\ }\href {https://doi.org/10.1103/PhysRevB.110.224431} {\bibfield  {journal} {\bibinfo  {journal} {Phys. Rev. B}\ }\textbf {\bibinfo {volume} {110}},\ \bibinfo {pages} {224431} (\bibinfo {year} {2024})}\BibitemShut {NoStop}%
\bibitem [{\citenamefont {Gao}\ \emph {et~al.}(2024)\citenamefont {Gao}, \citenamefont {Fan}, \citenamefont {Wang}, \citenamefont {Li}, \citenamefont {Ren}, \citenamefont {Biało}, \citenamefont {Drewanowski}, \citenamefont {Rothenbühler}, \citenamefont {Choi}, \citenamefont {Sutarto}, \citenamefont {Wang}, \citenamefont {Xiang}, \citenamefont {Hu}, \citenamefont {Zhou}, \citenamefont {Bisogni}, \citenamefont {Comin}, \citenamefont {Chang}, \citenamefont {Pelliciari}, \citenamefont {Zhou},\ and\ \citenamefont {Zhu}}]{Gao2024magnetic}%
  \BibitemOpen
  \bibfield  {author} {\bibinfo {author} {\bibfnamefont {Q.}~\bibnamefont {Gao}}, \bibinfo {author} {\bibfnamefont {S.}~\bibnamefont {Fan}}, \bibinfo {author} {\bibfnamefont {Q.}~\bibnamefont {Wang}}, \bibinfo {author} {\bibfnamefont {J.}~\bibnamefont {Li}}, \bibinfo {author} {\bibfnamefont {X.}~\bibnamefont {Ren}}, \bibinfo {author} {\bibfnamefont {I.}~\bibnamefont {Biało}}, \bibinfo {author} {\bibfnamefont {A.}~\bibnamefont {Drewanowski}}, \bibinfo {author} {\bibfnamefont {P.}~\bibnamefont {Rothenbühler}}, \bibinfo {author} {\bibfnamefont {J.}~\bibnamefont {Choi}}, \bibinfo {author} {\bibfnamefont {R.}~\bibnamefont {Sutarto}}, \bibinfo {author} {\bibfnamefont {Y.}~\bibnamefont {Wang}}, \bibinfo {author} {\bibfnamefont {T.}~\bibnamefont {Xiang}}, \bibinfo {author} {\bibfnamefont {J.}~\bibnamefont {Hu}}, \bibinfo {author} {\bibfnamefont {K.~J.}\ \bibnamefont {Zhou}}, \bibinfo {author} {\bibfnamefont {V.}~\bibnamefont {Bisogni}}, \bibinfo {author} {\bibfnamefont {R.}~\bibnamefont {Comin}}, \bibinfo {author}
  {\bibfnamefont {J.}~\bibnamefont {Chang}}, \bibinfo {author} {\bibfnamefont {J.}~\bibnamefont {Pelliciari}}, \bibinfo {author} {\bibfnamefont {X.~J.}\ \bibnamefont {Zhou}},\ and\ \bibinfo {author} {\bibfnamefont {Z.}~\bibnamefont {Zhu}},\ }\bibfield  {title} {\bibinfo {title} {Magnetic excitations in strained infinite-layer nickelate {PrNiO$_2$} films},\ }\href {https://doi.org/10.1038/s41467-024-49940-4} {\bibfield  {journal} {\bibinfo  {journal} {Nat. Commun.}\ }\textbf {\bibinfo {volume} {15}},\ \bibinfo {pages} {5576} (\bibinfo {year} {2024})}\BibitemShut {NoStop}%
\bibitem [{\citenamefont {Zhong}\ \emph {et~al.}(2025)\citenamefont {Zhong}, \citenamefont {Haobo}, \citenamefont {Wei}, \citenamefont {Zhang}, \citenamefont {Liu}, \citenamefont {Huang}, \citenamefont {Ni}, \citenamefont {dos Reis~Cantarino}, \citenamefont {Cao}, \citenamefont {Nie}, \citenamefont {Schmitt},\ and\ \citenamefont {Lu}}]{Zhong2025epitaxial}%
  \BibitemOpen
  \bibfield  {author} {\bibinfo {author} {\bibfnamefont {H.}~\bibnamefont {Zhong}}, \bibinfo {author} {\bibnamefont {Haobo}}, \bibinfo {author} {\bibfnamefont {Y.}~\bibnamefont {Wei}}, \bibinfo {author} {\bibfnamefont {Z.}~\bibnamefont {Zhang}}, \bibinfo {author} {\bibfnamefont {R.}~\bibnamefont {Liu}}, \bibinfo {author} {\bibfnamefont {X.}~\bibnamefont {Huang}}, \bibinfo {author} {\bibfnamefont {X.-S.}\ \bibnamefont {Ni}}, \bibinfo {author} {\bibfnamefont {M.}~\bibnamefont {dos Reis~Cantarino}}, \bibinfo {author} {\bibfnamefont {K.}~\bibnamefont {Cao}}, \bibinfo {author} {\bibfnamefont {Y.}~\bibnamefont {Nie}}, \bibinfo {author} {\bibfnamefont {T.}~\bibnamefont {Schmitt}},\ and\ \bibinfo {author} {\bibfnamefont {X.}~\bibnamefont {Lu}},\ }\href {https://arxiv.org/abs/2502.03178} {\bibinfo {title} {Epitaxial strain tuning of electronic and spin excitations in {La$_3$Ni$_2$O$_7$} thin films}} (\bibinfo {year} {2025}),\ \Eprint {https://arxiv.org/abs/2502.03178} {arXiv:2502.03178} \BibitemShut {NoStop}%
\bibitem [{\citenamefont {Sun}\ \emph {et~al.}(2023)\citenamefont {Sun}, \citenamefont {Huo}, \citenamefont {Hu}, \citenamefont {Li}, \citenamefont {Liu}, \citenamefont {Han}, \citenamefont {Tang}, \citenamefont {Mao}, \citenamefont {Yang}, \citenamefont {Wang} \emph {et~al.}}]{Sun2023}%
  \BibitemOpen
  \bibfield  {author} {\bibinfo {author} {\bibfnamefont {H.}~\bibnamefont {Sun}}, \bibinfo {author} {\bibfnamefont {M.}~\bibnamefont {Huo}}, \bibinfo {author} {\bibfnamefont {X.}~\bibnamefont {Hu}}, \bibinfo {author} {\bibfnamefont {J.}~\bibnamefont {Li}}, \bibinfo {author} {\bibfnamefont {Z.}~\bibnamefont {Liu}}, \bibinfo {author} {\bibfnamefont {Y.}~\bibnamefont {Han}}, \bibinfo {author} {\bibfnamefont {L.}~\bibnamefont {Tang}}, \bibinfo {author} {\bibfnamefont {Z.}~\bibnamefont {Mao}}, \bibinfo {author} {\bibfnamefont {P.}~\bibnamefont {Yang}}, \bibinfo {author} {\bibfnamefont {B.}~\bibnamefont {Wang}}, \emph {et~al.},\ }\bibfield  {title} {\bibinfo {title} {Signatures of superconductivity near 80 {K} in a nickelate under high pressure},\ }\href {https://doi.org/10.1038/s41586-023-06408-7} {\bibfield  {journal} {\bibinfo  {journal} {Nature}\ }\textbf {\bibinfo {volume} {621}},\ \bibinfo {pages} {493} (\bibinfo {year} {2023})}\BibitemShut {NoStop}%
\bibitem [{\citenamefont {Zhang}\ \emph {et~al.}(2024)\citenamefont {Zhang}, \citenamefont {Su}, \citenamefont {Huan}, \citenamefont {Shan}, \citenamefont {Sun}, \citenamefont {Huo}, \citenamefont {Ye}, \citenamefont {Zhang}, \citenamefont {Yang}, \citenamefont {Xu}, \citenamefont {Su}, \citenamefont {Li}, \citenamefont {Smidman}, \citenamefont {Wang}, \citenamefont {Jiao},\ and\ \citenamefont {Yuan}}]{Zhang2024}%
  \BibitemOpen
  \bibfield  {author} {\bibinfo {author} {\bibfnamefont {Y.}~\bibnamefont {Zhang}}, \bibinfo {author} {\bibfnamefont {D.}~\bibnamefont {Su}}, \bibinfo {author} {\bibfnamefont {Y.}~\bibnamefont {Huan}}, \bibinfo {author} {\bibfnamefont {Z.}~\bibnamefont {Shan}}, \bibinfo {author} {\bibfnamefont {H.}~\bibnamefont {Sun}}, \bibinfo {author} {\bibfnamefont {M.}~\bibnamefont {Huo}}, \bibinfo {author} {\bibfnamefont {K.}~\bibnamefont {Ye}}, \bibinfo {author} {\bibfnamefont {J.}~\bibnamefont {Zhang}}, \bibinfo {author} {\bibfnamefont {Z.}~\bibnamefont {Yang}}, \bibinfo {author} {\bibfnamefont {Y.}~\bibnamefont {Xu}}, \bibinfo {author} {\bibfnamefont {Y.}~\bibnamefont {Su}}, \bibinfo {author} {\bibfnamefont {R.}~\bibnamefont {Li}}, \bibinfo {author} {\bibfnamefont {M.}~\bibnamefont {Smidman}}, \bibinfo {author} {\bibfnamefont {M.}~\bibnamefont {Wang}}, \bibinfo {author} {\bibfnamefont {L.}~\bibnamefont {Jiao}},\ and\ \bibinfo {author} {\bibfnamefont {H.}~\bibnamefont {Yuan}},\ }\bibfield  {title} {\bibinfo {title}
  {High-temperature superconductivity with zero resistance and strange-metal behaviour in {La$_3$Ni$_2$O$_{7-\delta}$}},\ }\href {https://doi.org/10.1038/s41567-024-02515-y} {\bibfield  {journal} {\bibinfo  {journal} {Nat. Phys.}\ }\textbf {\bibinfo {volume} {20}},\ \bibinfo {pages} {1269–1273} (\bibinfo {year} {2024})}\BibitemShut {NoStop}%
\bibitem [{\citenamefont {Chen}\ \emph {et~al.}(2024{\natexlab{b}})\citenamefont {Chen}, \citenamefont {Poldi}, \citenamefont {Huyan}, \citenamefont {Chapai}, \citenamefont {Zheng}, \citenamefont {Budko}, \citenamefont {Welp}, \citenamefont {Canfield}, \citenamefont {Hemley}, \citenamefont {Mitchell},\ and\ \citenamefont {Phelan}}]{chen2024nonbulk}%
  \BibitemOpen
  \bibfield  {author} {\bibinfo {author} {\bibfnamefont {X.}~\bibnamefont {Chen}}, \bibinfo {author} {\bibfnamefont {E.~H.~T.}\ \bibnamefont {Poldi}}, \bibinfo {author} {\bibfnamefont {S.}~\bibnamefont {Huyan}}, \bibinfo {author} {\bibfnamefont {R.}~\bibnamefont {Chapai}}, \bibinfo {author} {\bibfnamefont {H.}~\bibnamefont {Zheng}}, \bibinfo {author} {\bibfnamefont {S.~L.}\ \bibnamefont {Budko}}, \bibinfo {author} {\bibfnamefont {U.}~\bibnamefont {Welp}}, \bibinfo {author} {\bibfnamefont {P.~C.}\ \bibnamefont {Canfield}}, \bibinfo {author} {\bibfnamefont {R.~J.}\ \bibnamefont {Hemley}}, \bibinfo {author} {\bibfnamefont {J.~F.}\ \bibnamefont {Mitchell}},\ and\ \bibinfo {author} {\bibfnamefont {D.}~\bibnamefont {Phelan}},\ }\href {https://arxiv.org/abs/2410.10666} {\bibinfo {title} {Non-bulk superconductivity in {Pr$_4$Ni$_3$O$_{10}$} single crystals under pressure}} (\bibinfo {year} {2024}{\natexlab{b}}),\ \Eprint {https://arxiv.org/abs/2410.10666} {arXiv:2410.10666} \BibitemShut {NoStop}%
\bibitem [{\citenamefont {Li}\ \emph {et~al.}(2025)\citenamefont {Li}, \citenamefont {Hao}, \citenamefont {Guo}, \citenamefont {Zhang}, \citenamefont {Zheng}, \citenamefont {Liu},\ and\ \citenamefont {Zhang}}]{Li2025signature}%
  \BibitemOpen
  \bibfield  {author} {\bibinfo {author} {\bibfnamefont {F.}~\bibnamefont {Li}}, \bibinfo {author} {\bibfnamefont {Y.}~\bibnamefont {Hao}}, \bibinfo {author} {\bibfnamefont {N.}~\bibnamefont {Guo}}, \bibinfo {author} {\bibfnamefont {J.}~\bibnamefont {Zhang}}, \bibinfo {author} {\bibfnamefont {Q.}~\bibnamefont {Zheng}}, \bibinfo {author} {\bibfnamefont {G.}~\bibnamefont {Liu}},\ and\ \bibinfo {author} {\bibfnamefont {J.}~\bibnamefont {Zhang}},\ }\href {https://arxiv.org/abs/2501.13511} {\bibinfo {title} {Signature of superconductivity in pressurized {La$_4$Ni$_3$O$_{10-x}$} single crystals grown at ambient pressure}} (\bibinfo {year} {2025}),\ \Eprint {https://arxiv.org/abs/2501.13511} {arXiv:2501.13511} \BibitemShut {NoStop}%
\bibitem [{\citenamefont {Huang}\ \emph {et~al.}(2024)\citenamefont {Huang}, \citenamefont {Zhang}, \citenamefont {Li}, \citenamefont {Huo}, \citenamefont {Chen}, \citenamefont {Qiu}, \citenamefont {Ma}, \citenamefont {Huang}, \citenamefont {Sun},\ and\ \citenamefont {Wang}}]{Huang2024signature}%
  \BibitemOpen
  \bibfield  {author} {\bibinfo {author} {\bibfnamefont {X.}~\bibnamefont {Huang}}, \bibinfo {author} {\bibfnamefont {H.}~\bibnamefont {Zhang}}, \bibinfo {author} {\bibfnamefont {J.}~\bibnamefont {Li}}, \bibinfo {author} {\bibfnamefont {M.}~\bibnamefont {Huo}}, \bibinfo {author} {\bibfnamefont {J.}~\bibnamefont {Chen}}, \bibinfo {author} {\bibfnamefont {Z.}~\bibnamefont {Qiu}}, \bibinfo {author} {\bibfnamefont {P.}~\bibnamefont {Ma}}, \bibinfo {author} {\bibfnamefont {C.}~\bibnamefont {Huang}}, \bibinfo {author} {\bibfnamefont {H.}~\bibnamefont {Sun}},\ and\ \bibinfo {author} {\bibfnamefont {M.}~\bibnamefont {Wang}},\ }\href {https://arxiv.org/abs/2410.07861} {\bibinfo {title} {Signature of superconductivity in pressurized trilayer-nickelate {Pr$_4$Ni$_3$O$_{10-\delta}$}}} (\bibinfo {year} {2024}),\ \Eprint {https://arxiv.org/abs/2410.07861} {arXiv:2410.07861} \BibitemShut {NoStop}%
\bibitem [{\citenamefont {Pei}\ \emph {et~al.}(2024)\citenamefont {Pei}, \citenamefont {Zhang}, \citenamefont {Peng}, \citenamefont {Huangfu}, \citenamefont {Zhu}, \citenamefont {Wang}, \citenamefont {Wu}, \citenamefont {Xing}, \citenamefont {Zhang}, \citenamefont {Chen}, \citenamefont {Zhao}, \citenamefont {Yang}, \citenamefont {Suo}, \citenamefont {Guo}, \citenamefont {Zeng},\ and\ \citenamefont {Qi}}]{pei2024pressure}%
  \BibitemOpen
  \bibfield  {author} {\bibinfo {author} {\bibfnamefont {C.}~\bibnamefont {Pei}}, \bibinfo {author} {\bibfnamefont {M.}~\bibnamefont {Zhang}}, \bibinfo {author} {\bibfnamefont {D.}~\bibnamefont {Peng}}, \bibinfo {author} {\bibfnamefont {S.}~\bibnamefont {Huangfu}}, \bibinfo {author} {\bibfnamefont {S.}~\bibnamefont {Zhu}}, \bibinfo {author} {\bibfnamefont {Q.}~\bibnamefont {Wang}}, \bibinfo {author} {\bibfnamefont {J.}~\bibnamefont {Wu}}, \bibinfo {author} {\bibfnamefont {Z.}~\bibnamefont {Xing}}, \bibinfo {author} {\bibfnamefont {L.}~\bibnamefont {Zhang}}, \bibinfo {author} {\bibfnamefont {Y.}~\bibnamefont {Chen}}, \bibinfo {author} {\bibfnamefont {J.}~\bibnamefont {Zhao}}, \bibinfo {author} {\bibfnamefont {W.}~\bibnamefont {Yang}}, \bibinfo {author} {\bibfnamefont {H.}~\bibnamefont {Suo}}, \bibinfo {author} {\bibfnamefont {H.}~\bibnamefont {Guo}}, \bibinfo {author} {\bibfnamefont {Q.}~\bibnamefont {Zeng}},\ and\ \bibinfo {author} {\bibfnamefont {Y.}~\bibnamefont {Qi}},\ }\href
  {https://arxiv.org/abs/2411.08677} {\bibinfo {title} {Pressure-induced superconductivity in {Pr$_4$Ni$_3$O$_{10}$} single crystals}} (\bibinfo {year} {2024}),\ \Eprint {https://arxiv.org/abs/2411.08677} {arXiv:2411.08677} \BibitemShut {NoStop}%
\bibitem [{\citenamefont {Ko}\ \emph {et~al.}(2024)\citenamefont {Ko}, \citenamefont {Yu}, \citenamefont {Liu}, \citenamefont {Bhatt}, \citenamefont {Li}, \citenamefont {Thampy}, \citenamefont {Kuo}, \citenamefont {Wang}, \citenamefont {Lee}, \citenamefont {Lee}, \citenamefont {Lee}, \citenamefont {Goodge}, \citenamefont {Muller},\ and\ \citenamefont {Hwang}}]{Ko2024signatures}%
  \BibitemOpen
  \bibfield  {author} {\bibinfo {author} {\bibfnamefont {E.~K.}\ \bibnamefont {Ko}}, \bibinfo {author} {\bibfnamefont {Y.}~\bibnamefont {Yu}}, \bibinfo {author} {\bibfnamefont {Y.}~\bibnamefont {Liu}}, \bibinfo {author} {\bibfnamefont {L.}~\bibnamefont {Bhatt}}, \bibinfo {author} {\bibfnamefont {J.}~\bibnamefont {Li}}, \bibinfo {author} {\bibfnamefont {V.}~\bibnamefont {Thampy}}, \bibinfo {author} {\bibfnamefont {C.-T.}\ \bibnamefont {Kuo}}, \bibinfo {author} {\bibfnamefont {B.~Y.}\ \bibnamefont {Wang}}, \bibinfo {author} {\bibfnamefont {Y.}~\bibnamefont {Lee}}, \bibinfo {author} {\bibfnamefont {K.}~\bibnamefont {Lee}}, \bibinfo {author} {\bibfnamefont {J.-S.}\ \bibnamefont {Lee}}, \bibinfo {author} {\bibfnamefont {B.~H.}\ \bibnamefont {Goodge}}, \bibinfo {author} {\bibfnamefont {D.~A.}\ \bibnamefont {Muller}},\ and\ \bibinfo {author} {\bibfnamefont {H.~Y.}\ \bibnamefont {Hwang}},\ }\bibfield  {title} {\bibinfo {title} {Signatures of ambient pressure superconductivity in thin film {La$_3$Ni$_2$O$_7$}},\
  }\bibfield  {journal} {\bibinfo  {journal} {Nature}\ }\href {https://doi.org/10.1038/s41586-024-08525-3} {10.1038/s41586-024-08525-3} (\bibinfo {year} {2024})\BibitemShut {NoStop}%
\bibitem [{\citenamefont {Liu}\ \emph {et~al.}(2025)\citenamefont {Liu}, \citenamefont {Ko}, \citenamefont {Tarn}, \citenamefont {Bhatt}, \citenamefont {Goodge}, \citenamefont {Muller}, \citenamefont {Raghu}, \citenamefont {Yu},\ and\ \citenamefont {Hwang}}]{Liu2025superconductivity}%
  \BibitemOpen
  \bibfield  {author} {\bibinfo {author} {\bibfnamefont {Y.}~\bibnamefont {Liu}}, \bibinfo {author} {\bibfnamefont {E.~K.}\ \bibnamefont {Ko}}, \bibinfo {author} {\bibfnamefont {Y.}~\bibnamefont {Tarn}}, \bibinfo {author} {\bibfnamefont {L.}~\bibnamefont {Bhatt}}, \bibinfo {author} {\bibfnamefont {B.~H.}\ \bibnamefont {Goodge}}, \bibinfo {author} {\bibfnamefont {D.~A.}\ \bibnamefont {Muller}}, \bibinfo {author} {\bibfnamefont {S.}~\bibnamefont {Raghu}}, \bibinfo {author} {\bibfnamefont {Y.}~\bibnamefont {Yu}},\ and\ \bibinfo {author} {\bibfnamefont {H.~Y.}\ \bibnamefont {Hwang}},\ }\href {https://arxiv.org/abs/2501.08022} {\bibinfo {title} {Superconductivity and normal-state transport in compressively strained {La$_2$PrNi$_2$O$_7$} thin films}} (\bibinfo {year} {2025}),\ \Eprint {https://arxiv.org/abs/2501.08022} {arXiv:2501.08022} \BibitemShut {NoStop}%
\bibitem [{\citenamefont {Zhou}\ \emph {et~al.}(2024)\citenamefont {Zhou}, \citenamefont {Lv}, \citenamefont {Wang}, \citenamefont {Nie}, \citenamefont {Chen}, \citenamefont {Li}, \citenamefont {Huang}, \citenamefont {Chen}, \citenamefont {Sun}, \citenamefont {Xue},\ and\ \citenamefont {Chen}}]{Zhou2024ambient}%
  \BibitemOpen
  \bibfield  {author} {\bibinfo {author} {\bibfnamefont {G.}~\bibnamefont {Zhou}}, \bibinfo {author} {\bibfnamefont {W.}~\bibnamefont {Lv}}, \bibinfo {author} {\bibfnamefont {H.}~\bibnamefont {Wang}}, \bibinfo {author} {\bibfnamefont {Z.}~\bibnamefont {Nie}}, \bibinfo {author} {\bibfnamefont {Y.}~\bibnamefont {Chen}}, \bibinfo {author} {\bibfnamefont {Y.}~\bibnamefont {Li}}, \bibinfo {author} {\bibfnamefont {H.}~\bibnamefont {Huang}}, \bibinfo {author} {\bibfnamefont {W.}~\bibnamefont {Chen}}, \bibinfo {author} {\bibfnamefont {Y.}~\bibnamefont {Sun}}, \bibinfo {author} {\bibfnamefont {Q.-K.}\ \bibnamefont {Xue}},\ and\ \bibinfo {author} {\bibfnamefont {Z.}~\bibnamefont {Chen}},\ }\href {https://arxiv.org/abs/2412.16622} {\bibinfo {title} {Ambient-pressure superconductivity onset above {40 K} in bilayer nickelate ultrathin films}} (\bibinfo {year} {2024}),\ \Eprint {https://arxiv.org/abs/2412.16622} {arXiv:2412.16622} \BibitemShut {NoStop}%
\bibitem [{\citenamefont {Bhatt}\ \emph {et~al.}(2025)\citenamefont {Bhatt}, \citenamefont {Jiang}, \citenamefont {Ko}, \citenamefont {Schnitzer}, \citenamefont {Pan}, \citenamefont {Segedin}, \citenamefont {Liu}, \citenamefont {Yu}, \citenamefont {Zhao}, \citenamefont {Morales} \emph {et~al.}}]{bhatt2025resolving}%
  \BibitemOpen
  \bibfield  {author} {\bibinfo {author} {\bibfnamefont {L.}~\bibnamefont {Bhatt}}, \bibinfo {author} {\bibfnamefont {A.~Y.}\ \bibnamefont {Jiang}}, \bibinfo {author} {\bibfnamefont {E.~K.}\ \bibnamefont {Ko}}, \bibinfo {author} {\bibfnamefont {N.}~\bibnamefont {Schnitzer}}, \bibinfo {author} {\bibfnamefont {G.~A.}\ \bibnamefont {Pan}}, \bibinfo {author} {\bibfnamefont {D.~F.}\ \bibnamefont {Segedin}}, \bibinfo {author} {\bibfnamefont {Y.}~\bibnamefont {Liu}}, \bibinfo {author} {\bibfnamefont {Y.}~\bibnamefont {Yu}}, \bibinfo {author} {\bibfnamefont {Y.-F.}\ \bibnamefont {Zhao}}, \bibinfo {author} {\bibfnamefont {E.~A.}\ \bibnamefont {Morales}}, \emph {et~al.},\ }\href {https://arxiv.org/abs/2501.08204} {\bibinfo {title} {Resolving structural origins for superconductivity in strain-engineered {La$_3$Ni$_2$O$_7$} thin films}} (\bibinfo {year} {2025})\BibitemShut {NoStop}%
\bibitem [{sup()}]{suppl}%
  \BibitemOpen
  \href@noop {} {}\bibinfo {note} {See Supplemental Material at [URL] for additional structural and electronic characterization, RIXS incident energy and polarization dependence, and $n=5$ strain dependence.}\BibitemShut {Stop}%
\bibitem [{\citenamefont {Lee}\ and\ \citenamefont {Pickett}(2004)}]{Lee2004infinite}%
  \BibitemOpen
  \bibfield  {author} {\bibinfo {author} {\bibfnamefont {K.~W.}\ \bibnamefont {Lee}}\ and\ \bibinfo {author} {\bibfnamefont {W.~E.}\ \bibnamefont {Pickett}},\ }\bibfield  {title} {\bibinfo {title} {Infinite-layer {LaNiO$_2$}: {Ni$^{1+}$} is not {Cu$^{2+}$}},\ }\href {https://doi.org/10.1103/PhysRevB.70.165109} {\bibfield  {journal} {\bibinfo  {journal} {Phys. Rev. B}\ }\textbf {\bibinfo {volume} {70}},\ \bibinfo {pages} {165109} (\bibinfo {year} {2004})}\BibitemShut {NoStop}%
\bibitem [{\citenamefont {Botana}\ and\ \citenamefont {Norman}(2020)}]{Botana2020similarities}%
  \BibitemOpen
  \bibfield  {author} {\bibinfo {author} {\bibfnamefont {A.~S.}\ \bibnamefont {Botana}}\ and\ \bibinfo {author} {\bibfnamefont {M.~R.}\ \bibnamefont {Norman}},\ }\bibfield  {title} {\bibinfo {title} {Similarities and differences between {LaNiO$_2$} and {CaCuO$_2$} and implications for superconductivity},\ }\href {https://doi.org/10.1103/PhysRevX.10.011024} {\bibfield  {journal} {\bibinfo  {journal} {Phys. Rev. X}\ }\textbf {\bibinfo {volume} {10}},\ \bibinfo {pages} {011024} (\bibinfo {year} {2020})}\BibitemShut {NoStop}%
\bibitem [{\citenamefont {Hepting}\ \emph {et~al.}(2021)\citenamefont {Hepting}, \citenamefont {Dean},\ and\ \citenamefont {Lee}}]{Hepting2021soft}%
  \BibitemOpen
  \bibfield  {author} {\bibinfo {author} {\bibfnamefont {M.}~\bibnamefont {Hepting}}, \bibinfo {author} {\bibfnamefont {M.~P.~M.}\ \bibnamefont {Dean}},\ and\ \bibinfo {author} {\bibfnamefont {W.-S.}\ \bibnamefont {Lee}},\ }\bibfield  {title} {\bibinfo {title} {Soft x-ray spectroscopy of low-valence nickelates},\ }\bibfield  {journal} {\bibinfo  {journal} {Front. Phys.}\ }\textbf {\bibinfo {volume} {9}},\ \href {https://doi.org/10.3389/fphy.2021.808683} {10.3389/fphy.2021.808683} (\bibinfo {year} {2021})\BibitemShut {NoStop}%
\bibitem [{\citenamefont {Hepting}\ \emph {et~al.}(2020)\citenamefont {Hepting}, \citenamefont {Li}, \citenamefont {Jia}, \citenamefont {Lu}, \citenamefont {Paris}, \citenamefont {Tseng}, \citenamefont {Feng}, \citenamefont {Osada}, \citenamefont {Been}, \citenamefont {Hikita}, \citenamefont {Chuang}, \citenamefont {Hussain}, \citenamefont {Zhou}, \citenamefont {Nag}, \citenamefont {García-Fernández}, \citenamefont {Rossi}, \citenamefont {Huang}, \citenamefont {Huang}, \citenamefont {Shen}, \citenamefont {Schmitt}, \citenamefont {Hwang}, \citenamefont {Moritz}, \citenamefont {Zaanen}, \citenamefont {Devereaux},\ and\ \citenamefont {Lee}}]{Hepting2020electronic}%
  \BibitemOpen
  \bibfield  {author} {\bibinfo {author} {\bibfnamefont {M.}~\bibnamefont {Hepting}}, \bibinfo {author} {\bibfnamefont {D.}~\bibnamefont {Li}}, \bibinfo {author} {\bibfnamefont {C.~J.}\ \bibnamefont {Jia}}, \bibinfo {author} {\bibfnamefont {H.}~\bibnamefont {Lu}}, \bibinfo {author} {\bibfnamefont {E.}~\bibnamefont {Paris}}, \bibinfo {author} {\bibfnamefont {Y.}~\bibnamefont {Tseng}}, \bibinfo {author} {\bibfnamefont {X.}~\bibnamefont {Feng}}, \bibinfo {author} {\bibfnamefont {M.}~\bibnamefont {Osada}}, \bibinfo {author} {\bibfnamefont {E.}~\bibnamefont {Been}}, \bibinfo {author} {\bibfnamefont {Y.}~\bibnamefont {Hikita}}, \bibinfo {author} {\bibfnamefont {Y.~D.}\ \bibnamefont {Chuang}}, \bibinfo {author} {\bibfnamefont {Z.}~\bibnamefont {Hussain}}, \bibinfo {author} {\bibfnamefont {K.~J.}\ \bibnamefont {Zhou}}, \bibinfo {author} {\bibfnamefont {A.}~\bibnamefont {Nag}}, \bibinfo {author} {\bibfnamefont {M.}~\bibnamefont {García-Fernández}}, \bibinfo {author} {\bibfnamefont {M.}~\bibnamefont {Rossi}},
  \bibinfo {author} {\bibfnamefont {H.~Y.}\ \bibnamefont {Huang}}, \bibinfo {author} {\bibfnamefont {D.~J.}\ \bibnamefont {Huang}}, \bibinfo {author} {\bibfnamefont {Z.~X.}\ \bibnamefont {Shen}}, \bibinfo {author} {\bibfnamefont {T.}~\bibnamefont {Schmitt}}, \bibinfo {author} {\bibfnamefont {H.~Y.}\ \bibnamefont {Hwang}}, \bibinfo {author} {\bibfnamefont {B.}~\bibnamefont {Moritz}}, \bibinfo {author} {\bibfnamefont {J.}~\bibnamefont {Zaanen}}, \bibinfo {author} {\bibfnamefont {T.~P.}\ \bibnamefont {Devereaux}},\ and\ \bibinfo {author} {\bibfnamefont {W.~S.}\ \bibnamefont {Lee}},\ }\bibfield  {title} {\bibinfo {title} {Electronic structure of the parent compound of superconducting infinite-layer nickelates},\ }\href {https://doi.org/10.1038/s41563-019-0585-z} {\bibfield  {journal} {\bibinfo  {journal} {Nat. Mater.}\ }\textbf {\bibinfo {volume} {19}},\ \bibinfo {pages} {381} (\bibinfo {year} {2020})}\BibitemShut {NoStop}%
\bibitem [{\citenamefont {Goodge}\ \emph {et~al.}(2021)\citenamefont {Goodge}, \citenamefont {Li}, \citenamefont {Lee}, \citenamefont {Osada}, \citenamefont {Wang}, \citenamefont {Sawatzky}, \citenamefont {Hwang},\ and\ \citenamefont {Kourkoutis}}]{Goodge2021doping}%
  \BibitemOpen
  \bibfield  {author} {\bibinfo {author} {\bibfnamefont {B.~H.}\ \bibnamefont {Goodge}}, \bibinfo {author} {\bibfnamefont {D.}~\bibnamefont {Li}}, \bibinfo {author} {\bibfnamefont {K.}~\bibnamefont {Lee}}, \bibinfo {author} {\bibfnamefont {M.}~\bibnamefont {Osada}}, \bibinfo {author} {\bibfnamefont {B.~Y.}\ \bibnamefont {Wang}}, \bibinfo {author} {\bibfnamefont {G.~A.}\ \bibnamefont {Sawatzky}}, \bibinfo {author} {\bibfnamefont {H.~Y.}\ \bibnamefont {Hwang}},\ and\ \bibinfo {author} {\bibfnamefont {L.~F.}\ \bibnamefont {Kourkoutis}},\ }\bibfield  {title} {\bibinfo {title} {Doping evolution of the mott–hubbard landscape in infinite-layer nickelates},\ }\href {https://doi.org/doi:10.1073/pnas.2007683118} {\bibfield  {journal} {\bibinfo  {journal} {Proc. Natl. Acad. Sci. U.S.A.}\ }\textbf {\bibinfo {volume} {118}},\ \bibinfo {pages} {e2007683118} (\bibinfo {year} {2021})}\BibitemShut {NoStop}%
\bibitem [{\citenamefont {Rossi}\ \emph {et~al.}(2021)\citenamefont {Rossi}, \citenamefont {Lu}, \citenamefont {Nag}, \citenamefont {Li}, \citenamefont {Osada}, \citenamefont {Lee}, \citenamefont {Wang}, \citenamefont {Agrestini}, \citenamefont {Garcia-Fernandez}, \citenamefont {Kas}, \citenamefont {Chuang}, \citenamefont {Shen}, \citenamefont {Hwang}, \citenamefont {Moritz}, \citenamefont {Zhou}, \citenamefont {Devereaux},\ and\ \citenamefont {Lee}}]{Rossi2021orbital}%
  \BibitemOpen
  \bibfield  {author} {\bibinfo {author} {\bibfnamefont {M.}~\bibnamefont {Rossi}}, \bibinfo {author} {\bibfnamefont {H.}~\bibnamefont {Lu}}, \bibinfo {author} {\bibfnamefont {A.}~\bibnamefont {Nag}}, \bibinfo {author} {\bibfnamefont {D.}~\bibnamefont {Li}}, \bibinfo {author} {\bibfnamefont {M.}~\bibnamefont {Osada}}, \bibinfo {author} {\bibfnamefont {K.}~\bibnamefont {Lee}}, \bibinfo {author} {\bibfnamefont {B.~Y.}\ \bibnamefont {Wang}}, \bibinfo {author} {\bibfnamefont {S.}~\bibnamefont {Agrestini}}, \bibinfo {author} {\bibfnamefont {M.}~\bibnamefont {Garcia-Fernandez}}, \bibinfo {author} {\bibfnamefont {J.~J.}\ \bibnamefont {Kas}}, \bibinfo {author} {\bibfnamefont {Y.-D.}\ \bibnamefont {Chuang}}, \bibinfo {author} {\bibfnamefont {Z.~X.}\ \bibnamefont {Shen}}, \bibinfo {author} {\bibfnamefont {H.~Y.}\ \bibnamefont {Hwang}}, \bibinfo {author} {\bibfnamefont {B.}~\bibnamefont {Moritz}}, \bibinfo {author} {\bibfnamefont {K.-J.}\ \bibnamefont {Zhou}}, \bibinfo {author} {\bibfnamefont {T.~P.}\ \bibnamefont
  {Devereaux}},\ and\ \bibinfo {author} {\bibfnamefont {W.~S.}\ \bibnamefont {Lee}},\ }\bibfield  {title} {\bibinfo {title} {Orbital and spin character of doped carriers in infinite-layer nickelates},\ }\href {https://doi.org/10.1103/PhysRevB.104.L220505} {\bibfield  {journal} {\bibinfo  {journal} {Phys. Rev. B}\ }\textbf {\bibinfo {volume} {104}},\ \bibinfo {pages} {L220505} (\bibinfo {year} {2021})}\BibitemShut {NoStop}%
\bibitem [{\citenamefont {Zhang}\ \emph {et~al.}(2017)\citenamefont {Zhang}, \citenamefont {Botana}, \citenamefont {Freeland}, \citenamefont {Phelan}, \citenamefont {Zheng}, \citenamefont {Pardo}, \citenamefont {Norman},\ and\ \citenamefont {Mitchell}}]{Zhang2017large}%
  \BibitemOpen
  \bibfield  {author} {\bibinfo {author} {\bibfnamefont {J.}~\bibnamefont {Zhang}}, \bibinfo {author} {\bibfnamefont {A.}~\bibnamefont {Botana}}, \bibinfo {author} {\bibfnamefont {J.}~\bibnamefont {Freeland}}, \bibinfo {author} {\bibfnamefont {D.}~\bibnamefont {Phelan}}, \bibinfo {author} {\bibfnamefont {H.}~\bibnamefont {Zheng}}, \bibinfo {author} {\bibfnamefont {V.}~\bibnamefont {Pardo}}, \bibinfo {author} {\bibfnamefont {M.}~\bibnamefont {Norman}},\ and\ \bibinfo {author} {\bibfnamefont {J.}~\bibnamefont {Mitchell}},\ }\bibfield  {title} {\bibinfo {title} {Large orbital polarization in a metallic square-planar nickelate},\ }\href {https://www.nature.com/articles/nphys4149} {\bibfield  {journal} {\bibinfo  {journal} {Nat. Phys.}\ }\textbf {\bibinfo {volume} {13}},\ \bibinfo {pages} {864} (\bibinfo {year} {2017})}\BibitemShut {NoStop}%
\bibitem [{\citenamefont {Norman}\ \emph {et~al.}(2023)\citenamefont {Norman}, \citenamefont {Botana}, \citenamefont {Karp}, \citenamefont {Hampel}, \citenamefont {Labollita}, \citenamefont {Millis}, \citenamefont {Fabbris}, \citenamefont {Shen},\ and\ \citenamefont {Dean}}]{Norman2023}%
  \BibitemOpen
  \bibfield  {author} {\bibinfo {author} {\bibfnamefont {M.~R.}\ \bibnamefont {Norman}}, \bibinfo {author} {\bibfnamefont {A.~S.}\ \bibnamefont {Botana}}, \bibinfo {author} {\bibfnamefont {J.}~\bibnamefont {Karp}}, \bibinfo {author} {\bibfnamefont {A.}~\bibnamefont {Hampel}}, \bibinfo {author} {\bibfnamefont {H.}~\bibnamefont {Labollita}}, \bibinfo {author} {\bibfnamefont {A.~J.}\ \bibnamefont {Millis}}, \bibinfo {author} {\bibfnamefont {G.}~\bibnamefont {Fabbris}}, \bibinfo {author} {\bibfnamefont {Y.}~\bibnamefont {Shen}},\ and\ \bibinfo {author} {\bibfnamefont {M.~P.~M.}\ \bibnamefont {Dean}},\ }\bibfield  {title} {\bibinfo {title} {Orbital polarization, charge transfer, and fluorescence in reduced-valence nickelates},\ }\href {https://doi.org/10.1103/PhysRevB.107.165124} {\bibfield  {journal} {\bibinfo  {journal} {Phys. Rev. B}\ }\textbf {\bibinfo {volume} {107}},\ \bibinfo {pages} {165124} (\bibinfo {year} {2023})}\BibitemShut {NoStop}%
\bibitem [{\citenamefont {Fabbris}\ \emph {et~al.}(2023)\citenamefont {Fabbris}, \citenamefont {Meyers}, \citenamefont {Shen}, \citenamefont {Bisogni}, \citenamefont {Zhang}, \citenamefont {Mitchell}, \citenamefont {Norman}, \citenamefont {Johnston}, \citenamefont {Feng}, \citenamefont {Chiuzb{\u{a}}ian}, \citenamefont {Nicolaou}, \citenamefont {Jaouen},\ and\ \citenamefont {Dean}}]{Fabbris2023resonant}%
  \BibitemOpen
  \bibfield  {author} {\bibinfo {author} {\bibfnamefont {G.}~\bibnamefont {Fabbris}}, \bibinfo {author} {\bibfnamefont {D.}~\bibnamefont {Meyers}}, \bibinfo {author} {\bibfnamefont {Y.}~\bibnamefont {Shen}}, \bibinfo {author} {\bibfnamefont {V.}~\bibnamefont {Bisogni}}, \bibinfo {author} {\bibfnamefont {J.}~\bibnamefont {Zhang}}, \bibinfo {author} {\bibfnamefont {J.~F.}\ \bibnamefont {Mitchell}}, \bibinfo {author} {\bibfnamefont {M.~R.}\ \bibnamefont {Norman}}, \bibinfo {author} {\bibfnamefont {S.}~\bibnamefont {Johnston}}, \bibinfo {author} {\bibfnamefont {J.}~\bibnamefont {Feng}}, \bibinfo {author} {\bibfnamefont {G.~S.}\ \bibnamefont {Chiuzb{\u{a}}ian}}, \bibinfo {author} {\bibfnamefont {A.}~\bibnamefont {Nicolaou}}, \bibinfo {author} {\bibfnamefont {N.}~\bibnamefont {Jaouen}},\ and\ \bibinfo {author} {\bibfnamefont {M.~P.~M.}\ \bibnamefont {Dean}},\ }\bibfield  {title} {\bibinfo {title} {Resonant inelastic x-ray scattering data for ruddlesden-popper and reduced ruddlesden-popper nickelates},\ }\href
  {https://doi.org/10.1038/s41597-023-02079-1} {\bibfield  {journal} {\bibinfo  {journal} {Scientific Data}\ }\textbf {\bibinfo {volume} {10}},\ \bibinfo {pages} {174} (\bibinfo {year} {2023})}\BibitemShut {NoStop}%
\bibitem [{\citenamefont {Li}\ \emph {et~al.}(2017)\citenamefont {Li}, \citenamefont {Zhou}, \citenamefont {Nummy}, \citenamefont {Zhang}, \citenamefont {Pardo}, \citenamefont {Pickett}, \citenamefont {Mitchell},\ and\ \citenamefont {Dessau}}]{li2017fermiology}%
  \BibitemOpen
  \bibfield  {author} {\bibinfo {author} {\bibfnamefont {H.}~\bibnamefont {Li}}, \bibinfo {author} {\bibfnamefont {X.}~\bibnamefont {Zhou}}, \bibinfo {author} {\bibfnamefont {T.}~\bibnamefont {Nummy}}, \bibinfo {author} {\bibfnamefont {J.}~\bibnamefont {Zhang}}, \bibinfo {author} {\bibfnamefont {V.}~\bibnamefont {Pardo}}, \bibinfo {author} {\bibfnamefont {W.~E.}\ \bibnamefont {Pickett}}, \bibinfo {author} {\bibfnamefont {J.~F.}\ \bibnamefont {Mitchell}},\ and\ \bibinfo {author} {\bibfnamefont {D.~S.}\ \bibnamefont {Dessau}},\ }\bibfield  {title} {\bibinfo {title} {Fermiology and electron dynamics of trilayer nickelate {La$_4$Ni$_3$O$_{10}$}},\ }\href {https://www.nature.com/articles/s41467-017-00777-0} {\bibfield  {journal} {\bibinfo  {journal} {Nat. Commun.}\ }\textbf {\bibinfo {volume} {8}},\ \bibinfo {pages} {704} (\bibinfo {year} {2017})}\BibitemShut {NoStop}%
\bibitem [{\citenamefont {Tacon}\ \emph {et~al.}(2011)\citenamefont {Tacon}, \citenamefont {Ghiringhelli}, \citenamefont {Chaloupka}, \citenamefont {Sala}, \citenamefont {Hinkov}, \citenamefont {Haverkort}, \citenamefont {Minola}, \citenamefont {Bakr}, \citenamefont {Zhou}, \citenamefont {Blanco-Canosa}, \citenamefont {Monney}, \citenamefont {Song}, \citenamefont {Sun}, \citenamefont {Lin}, \citenamefont {Luca}, \citenamefont {Salluzzo}, \citenamefont {Khaliullin}, \citenamefont {Schmitt}, \citenamefont {Braicovich},\ and\ \citenamefont {Keimer}}]{LeTacon2011intense}%
  \BibitemOpen
  \bibfield  {author} {\bibinfo {author} {\bibfnamefont {M.~L.}\ \bibnamefont {Tacon}}, \bibinfo {author} {\bibfnamefont {G.}~\bibnamefont {Ghiringhelli}}, \bibinfo {author} {\bibfnamefont {J.}~\bibnamefont {Chaloupka}}, \bibinfo {author} {\bibfnamefont {M.~M.}\ \bibnamefont {Sala}}, \bibinfo {author} {\bibfnamefont {V.}~\bibnamefont {Hinkov}}, \bibinfo {author} {\bibfnamefont {M.~W.}\ \bibnamefont {Haverkort}}, \bibinfo {author} {\bibfnamefont {M.}~\bibnamefont {Minola}}, \bibinfo {author} {\bibfnamefont {M.}~\bibnamefont {Bakr}}, \bibinfo {author} {\bibfnamefont {K.~J.}\ \bibnamefont {Zhou}}, \bibinfo {author} {\bibfnamefont {S.}~\bibnamefont {Blanco-Canosa}}, \bibinfo {author} {\bibfnamefont {C.}~\bibnamefont {Monney}}, \bibinfo {author} {\bibfnamefont {Y.~T.}\ \bibnamefont {Song}}, \bibinfo {author} {\bibfnamefont {G.~L.}\ \bibnamefont {Sun}}, \bibinfo {author} {\bibfnamefont {C.~T.}\ \bibnamefont {Lin}}, \bibinfo {author} {\bibfnamefont {G.~M.~D.}\ \bibnamefont {Luca}}, \bibinfo {author} {\bibfnamefont
  {M.}~\bibnamefont {Salluzzo}}, \bibinfo {author} {\bibfnamefont {G.}~\bibnamefont {Khaliullin}}, \bibinfo {author} {\bibfnamefont {T.}~\bibnamefont {Schmitt}}, \bibinfo {author} {\bibfnamefont {L.}~\bibnamefont {Braicovich}},\ and\ \bibinfo {author} {\bibfnamefont {B.}~\bibnamefont {Keimer}},\ }\bibfield  {title} {\bibinfo {title} {Intense paramagnon excitations in a large family of high-temperature superconductors},\ }\href {https://doi.org/10.1038/nphys2041} {\bibfield  {journal} {\bibinfo  {journal} {Nat. Phys.}\ }\textbf {\bibinfo {volume} {7}},\ \bibinfo {pages} {725} (\bibinfo {year} {2011})}\BibitemShut {NoStop}%
\bibitem [{\citenamefont {Boothroyd}\ \emph {et~al.}(2004)\citenamefont {Boothroyd}, \citenamefont {Freeman}, \citenamefont {Prabhakaran}, \citenamefont {Enderle},\ and\ \citenamefont {Kulda}}]{Boothroyd2004magnetic}%
  \BibitemOpen
  \bibfield  {author} {\bibinfo {author} {\bibfnamefont {A.~T.}\ \bibnamefont {Boothroyd}}, \bibinfo {author} {\bibfnamefont {P.~G.}\ \bibnamefont {Freeman}}, \bibinfo {author} {\bibfnamefont {D.}~\bibnamefont {Prabhakaran}}, \bibinfo {author} {\bibfnamefont {M.}~\bibnamefont {Enderle}},\ and\ \bibinfo {author} {\bibfnamefont {J.}~\bibnamefont {Kulda}},\ }\bibfield  {title} {\bibinfo {title} {Magnetic order and dynamics in stripe-ordered {La$_{2-x}$Sr$_x$NiO$_4$}},\ }\href {https://doi.org/10.1016/J.PHYSB.2003.11.007} {\bibfield  {journal} {\bibinfo  {journal} {Physica B}\ }\textbf {\bibinfo {volume} {345}},\ \bibinfo {pages} {1} (\bibinfo {year} {2004})}\BibitemShut {NoStop}%
\bibitem [{\citenamefont {Yoshizawa}\ \emph {et~al.}(2000)\citenamefont {Yoshizawa}, \citenamefont {Kakeshita}, \citenamefont {Kajimoto}, \citenamefont {Tanabe}, \citenamefont {Katsufuji},\ and\ \citenamefont {Tokura}}]{Yoshizawa2000stripe}%
  \BibitemOpen
  \bibfield  {author} {\bibinfo {author} {\bibfnamefont {H.}~\bibnamefont {Yoshizawa}}, \bibinfo {author} {\bibfnamefont {T.}~\bibnamefont {Kakeshita}}, \bibinfo {author} {\bibfnamefont {R.}~\bibnamefont {Kajimoto}}, \bibinfo {author} {\bibfnamefont {T.}~\bibnamefont {Tanabe}}, \bibinfo {author} {\bibfnamefont {T.}~\bibnamefont {Katsufuji}},\ and\ \bibinfo {author} {\bibfnamefont {Y.}~\bibnamefont {Tokura}},\ }\bibfield  {title} {\bibinfo {title} {Stripe order at low temperatures in {${\mathrm{La}}_{2\ensuremath{-}x}{\mathrm{Sr}}_{x}{\mathrm{NiO}}_{4}$} with $0.289\ensuremath{\lesssim}x\ensuremath{\lesssim}0.5$},\ }\href {https://doi.org/10.1103/PhysRevB.61.R854} {\bibfield  {journal} {\bibinfo  {journal} {Phys. Rev. B}\ }\textbf {\bibinfo {volume} {61}},\ \bibinfo {pages} {R854} (\bibinfo {year} {2000})}\BibitemShut {NoStop}%
\bibitem [{\citenamefont {Kajimoto}\ \emph {et~al.}(2003)\citenamefont {Kajimoto}, \citenamefont {Ishizaka}, \citenamefont {Yoshizawa},\ and\ \citenamefont {Tokura}}]{Kajimoto2003spontaneous}%
  \BibitemOpen
  \bibfield  {author} {\bibinfo {author} {\bibfnamefont {R.}~\bibnamefont {Kajimoto}}, \bibinfo {author} {\bibfnamefont {K.}~\bibnamefont {Ishizaka}}, \bibinfo {author} {\bibfnamefont {H.}~\bibnamefont {Yoshizawa}},\ and\ \bibinfo {author} {\bibfnamefont {Y.}~\bibnamefont {Tokura}},\ }\bibfield  {title} {\bibinfo {title} {Spontaneous rearrangement of the checkerboard charge order to stripe order in {La$_{1.5}$Sr$_{0.5}$NiO$_4$}},\ }\href {https://doi.org/10.1103/PhysRevB.67.014511} {\bibfield  {journal} {\bibinfo  {journal} {Phys. Rev. B}\ }\textbf {\bibinfo {volume} {67}},\ \bibinfo {pages} {014511} (\bibinfo {year} {2003})}\BibitemShut {NoStop}%
\bibitem [{\citenamefont {Ishizaka}\ \emph {et~al.}(2003)\citenamefont {Ishizaka}, \citenamefont {Taguchi}, \citenamefont {Kajimoto}, \citenamefont {Yoshizawa},\ and\ \citenamefont {Tokura}}]{Ishizaka2003charge}%
  \BibitemOpen
  \bibfield  {author} {\bibinfo {author} {\bibfnamefont {K.}~\bibnamefont {Ishizaka}}, \bibinfo {author} {\bibfnamefont {Y.}~\bibnamefont {Taguchi}}, \bibinfo {author} {\bibfnamefont {R.}~\bibnamefont {Kajimoto}}, \bibinfo {author} {\bibfnamefont {H.}~\bibnamefont {Yoshizawa}},\ and\ \bibinfo {author} {\bibfnamefont {Y.}~\bibnamefont {Tokura}},\ }\bibfield  {title} {\bibinfo {title} {Charge ordering and charge dynamics in {Nd$_{2-x}$Sr$_x$NiO$_4$} ($0.33 \geq x \geq 0.7$)},\ }\href {https://doi.org/10.1103/PhysRevB.67.184418} {\bibfield  {journal} {\bibinfo  {journal} {Phys. Rev. B}\ }\textbf {\bibinfo {volume} {67}},\ \bibinfo {pages} {184418} (\bibinfo {year} {2003})}\BibitemShut {NoStop}%
\bibitem [{\citenamefont {Uchida}\ \emph {et~al.}(2012)\citenamefont {Uchida}, \citenamefont {Yamasaki}, \citenamefont {Kaneko}, \citenamefont {Ishizaka}, \citenamefont {Okamoto}, \citenamefont {Nakao}, \citenamefont {Murakami},\ and\ \citenamefont {Tokura}}]{Uchida2012pseudogap}%
  \BibitemOpen
  \bibfield  {author} {\bibinfo {author} {\bibfnamefont {M.}~\bibnamefont {Uchida}}, \bibinfo {author} {\bibfnamefont {Y.}~\bibnamefont {Yamasaki}}, \bibinfo {author} {\bibfnamefont {Y.}~\bibnamefont {Kaneko}}, \bibinfo {author} {\bibfnamefont {K.}~\bibnamefont {Ishizaka}}, \bibinfo {author} {\bibfnamefont {J.}~\bibnamefont {Okamoto}}, \bibinfo {author} {\bibfnamefont {H.}~\bibnamefont {Nakao}}, \bibinfo {author} {\bibfnamefont {Y.}~\bibnamefont {Murakami}},\ and\ \bibinfo {author} {\bibfnamefont {Y.}~\bibnamefont {Tokura}},\ }\bibfield  {title} {\bibinfo {title} {Pseudogap-related charge dynamics in the layered nickelate {R$_{2-x}$Sr$_x$NiO$_4$} ($x \sim 1$)},\ }\href {https://doi.org/10.1103/PhysRevB.86.165126} {\bibfield  {journal} {\bibinfo  {journal} {Phys. Rev. B}\ }\textbf {\bibinfo {volume} {86}},\ \bibinfo {pages} {165126} (\bibinfo {year} {2012})}\BibitemShut {NoStop}%
\bibitem [{\citenamefont {Chen}\ \emph {et~al.}(1993)\citenamefont {Chen}, \citenamefont {Cheong},\ and\ \citenamefont {Cooper}}]{Chen1993charge}%
  \BibitemOpen
  \bibfield  {author} {\bibinfo {author} {\bibfnamefont {C.~H.}\ \bibnamefont {Chen}}, \bibinfo {author} {\bibfnamefont {S.-W.}\ \bibnamefont {Cheong}},\ and\ \bibinfo {author} {\bibfnamefont {A.~S.}\ \bibnamefont {Cooper}},\ }\bibfield  {title} {\bibinfo {title} {Charge modulations in {${\mathrm{La}}_{2\mathrm{\ensuremath{-}}\mathit{x}}$${\mathrm{Sr}}_{\mathit{x}}$${\mathrm{NiO}}_{4+\mathit{y}}$}: Ordering of polarons},\ }\href {https://doi.org/10.1103/PhysRevLett.71.2461} {\bibfield  {journal} {\bibinfo  {journal} {Phys. Rev. Lett.}\ }\textbf {\bibinfo {volume} {71}},\ \bibinfo {pages} {2461} (\bibinfo {year} {1993})}\BibitemShut {NoStop}%
\bibitem [{\citenamefont {Fabbris}\ \emph {et~al.}(2017)\citenamefont {Fabbris}, \citenamefont {Meyers}, \citenamefont {Xu}, \citenamefont {Katukuri}, \citenamefont {Hozoi}, \citenamefont {Liu}, \citenamefont {Chen}, \citenamefont {Okamoto}, \citenamefont {Schmitt}, \citenamefont {Uldry}, \citenamefont {Delley}, \citenamefont {Gu}, \citenamefont {Prabhakaran}, \citenamefont {Boothroyd}, \citenamefont {van~den Brink}, \citenamefont {Huang},\ and\ \citenamefont {Dean}}]{Fabbris2017}%
  \BibitemOpen
  \bibfield  {author} {\bibinfo {author} {\bibfnamefont {G.}~\bibnamefont {Fabbris}}, \bibinfo {author} {\bibfnamefont {D.}~\bibnamefont {Meyers}}, \bibinfo {author} {\bibfnamefont {L.}~\bibnamefont {Xu}}, \bibinfo {author} {\bibfnamefont {V.~M.}\ \bibnamefont {Katukuri}}, \bibinfo {author} {\bibfnamefont {L.}~\bibnamefont {Hozoi}}, \bibinfo {author} {\bibfnamefont {X.}~\bibnamefont {Liu}}, \bibinfo {author} {\bibfnamefont {Z.-Y.}\ \bibnamefont {Chen}}, \bibinfo {author} {\bibfnamefont {J.}~\bibnamefont {Okamoto}}, \bibinfo {author} {\bibfnamefont {T.}~\bibnamefont {Schmitt}}, \bibinfo {author} {\bibfnamefont {A.}~\bibnamefont {Uldry}}, \bibinfo {author} {\bibfnamefont {B.}~\bibnamefont {Delley}}, \bibinfo {author} {\bibfnamefont {G.~D.}\ \bibnamefont {Gu}}, \bibinfo {author} {\bibfnamefont {D.}~\bibnamefont {Prabhakaran}}, \bibinfo {author} {\bibfnamefont {A.~T.}\ \bibnamefont {Boothroyd}}, \bibinfo {author} {\bibfnamefont {J.}~\bibnamefont {van~den Brink}}, \bibinfo {author} {\bibfnamefont {D.~J.}\
  \bibnamefont {Huang}},\ and\ \bibinfo {author} {\bibfnamefont {M.~P.~M.}\ \bibnamefont {Dean}},\ }\bibfield  {title} {\bibinfo {title} {Doping dependence of collective spin and orbital excitations in the spin-1 quantum antiferromagnet {${\mathrm{La}}_{2\ensuremath{-}x}{\mathrm{Sr}}_{x}{\mathrm{NiO}}_{4}$ Observed by X Rays}},\ }\href {https://doi.org/10.1103/PhysRevLett.118.156402} {\bibfield  {journal} {\bibinfo  {journal} {Phys. Rev. Lett.}\ }\textbf {\bibinfo {volume} {118}},\ \bibinfo {pages} {156402} (\bibinfo {year} {2017})}\BibitemShut {NoStop}%
\bibitem [{\citenamefont {Woo}\ \emph {et~al.}(2005)\citenamefont {Woo}, \citenamefont {Boothroyd}, \citenamefont {Nakajima}, \citenamefont {Perring}, \citenamefont {Frost}, \citenamefont {Freeman}, \citenamefont {Prabhakaran}, \citenamefont {Yamada},\ and\ \citenamefont {Tranquada}}]{Woo2005mapping}%
  \BibitemOpen
  \bibfield  {author} {\bibinfo {author} {\bibfnamefont {H.}~\bibnamefont {Woo}}, \bibinfo {author} {\bibfnamefont {A.~T.}\ \bibnamefont {Boothroyd}}, \bibinfo {author} {\bibfnamefont {K.}~\bibnamefont {Nakajima}}, \bibinfo {author} {\bibfnamefont {T.~G.}\ \bibnamefont {Perring}}, \bibinfo {author} {\bibfnamefont {C.~D.}\ \bibnamefont {Frost}}, \bibinfo {author} {\bibfnamefont {P.~G.}\ \bibnamefont {Freeman}}, \bibinfo {author} {\bibfnamefont {D.}~\bibnamefont {Prabhakaran}}, \bibinfo {author} {\bibfnamefont {K.}~\bibnamefont {Yamada}},\ and\ \bibinfo {author} {\bibfnamefont {J.~M.}\ \bibnamefont {Tranquada}},\ }\bibfield  {title} {\bibinfo {title} {Mapping spin-wave dispersions in stripe-ordered {${\mathrm{La}}_{2\ensuremath{-}x}{\mathrm{Sr}}_{x}\mathrm{Ni}{\mathrm{O}}_{4}$} ($x=0.275$, 0.333)},\ }\href {https://doi.org/10.1103/PhysRevB.72.064437} {\bibfield  {journal} {\bibinfo  {journal} {Phys. Rev. B}\ }\textbf {\bibinfo {volume} {72}},\ \bibinfo {pages} {064437} (\bibinfo {year} {2005})}\BibitemShut
  {NoStop}%
\bibitem [{\citenamefont {Mitrano}\ \emph {et~al.}(2024)\citenamefont {Mitrano}, \citenamefont {Johnston}, \citenamefont {Kim},\ and\ \citenamefont {Dean}}]{Mitrano2024exploring}%
  \BibitemOpen
  \bibfield  {author} {\bibinfo {author} {\bibfnamefont {M.}~\bibnamefont {Mitrano}}, \bibinfo {author} {\bibfnamefont {S.}~\bibnamefont {Johnston}}, \bibinfo {author} {\bibfnamefont {Y.-J.}\ \bibnamefont {Kim}},\ and\ \bibinfo {author} {\bibfnamefont {M.~P.~M.}\ \bibnamefont {Dean}},\ }\bibfield  {title} {\bibinfo {title} {Exploring quantum materials with resonant inelastic x-ray scattering},\ }\href {https://doi.org/10.1103/PhysRevX.14.040501} {\bibfield  {journal} {\bibinfo  {journal} {Phys. Rev. X}\ }\textbf {\bibinfo {volume} {14}},\ \bibinfo {pages} {040501} (\bibinfo {year} {2024})}\BibitemShut {NoStop}%
\bibitem [{\citenamefont {Jarrige}\ \emph {et~al.}(2018)\citenamefont {Jarrige}, \citenamefont {Bisogni}, \citenamefont {Zhu}, \citenamefont {Leonhardt},\ and\ \citenamefont {Dvorak}}]{Jarrige2018paving}%
  \BibitemOpen
  \bibfield  {author} {\bibinfo {author} {\bibfnamefont {I.}~\bibnamefont {Jarrige}}, \bibinfo {author} {\bibfnamefont {V.}~\bibnamefont {Bisogni}}, \bibinfo {author} {\bibfnamefont {Y.}~\bibnamefont {Zhu}}, \bibinfo {author} {\bibfnamefont {W.}~\bibnamefont {Leonhardt}},\ and\ \bibinfo {author} {\bibfnamefont {J.}~\bibnamefont {Dvorak}},\ }\bibfield  {title} {\bibinfo {title} {Paving the way to ultra-high-resolution resonant inelastic x-ray scattering with the {SIX} beamline at {NSLS-II}},\ }\href {https://doi.org/10.1080/08940886.2018.1435949} {\bibfield  {journal} {\bibinfo  {journal} {Synchrotron Radiat. News}\ }\textbf {\bibinfo {volume} {31}},\ \bibinfo {pages} {7} (\bibinfo {year} {2018})}\BibitemShut {NoStop}%
\bibitem [{\citenamefont {Zhou}\ \emph {et~al.}(2022)\citenamefont {Zhou}, \citenamefont {Walters}, \citenamefont {Garcia-Fernandez}, \citenamefont {Rice}, \citenamefont {Hand}, \citenamefont {Nag}, \citenamefont {Li}, \citenamefont {Agrestini}, \citenamefont {Garland}, \citenamefont {Wang}, \citenamefont {Alcock}, \citenamefont {Nistea}, \citenamefont {Nutter}, \citenamefont {Rubies}, \citenamefont {Knap}, \citenamefont {Gaughran}, \citenamefont {Yuan}, \citenamefont {Chang}, \citenamefont {Emmins},\ and\ \citenamefont {Howell}}]{Zhou2022i21}%
  \BibitemOpen
  \bibfield  {author} {\bibinfo {author} {\bibfnamefont {K.~J.}\ \bibnamefont {Zhou}}, \bibinfo {author} {\bibfnamefont {A.}~\bibnamefont {Walters}}, \bibinfo {author} {\bibfnamefont {M.}~\bibnamefont {Garcia-Fernandez}}, \bibinfo {author} {\bibfnamefont {T.}~\bibnamefont {Rice}}, \bibinfo {author} {\bibfnamefont {M.}~\bibnamefont {Hand}}, \bibinfo {author} {\bibfnamefont {A.}~\bibnamefont {Nag}}, \bibinfo {author} {\bibfnamefont {J.}~\bibnamefont {Li}}, \bibinfo {author} {\bibfnamefont {S.}~\bibnamefont {Agrestini}}, \bibinfo {author} {\bibfnamefont {P.}~\bibnamefont {Garland}}, \bibinfo {author} {\bibfnamefont {H.}~\bibnamefont {Wang}}, \bibinfo {author} {\bibfnamefont {S.}~\bibnamefont {Alcock}}, \bibinfo {author} {\bibfnamefont {I.}~\bibnamefont {Nistea}}, \bibinfo {author} {\bibfnamefont {B.}~\bibnamefont {Nutter}}, \bibinfo {author} {\bibfnamefont {N.}~\bibnamefont {Rubies}}, \bibinfo {author} {\bibfnamefont {G.}~\bibnamefont {Knap}}, \bibinfo {author} {\bibfnamefont {M.}~\bibnamefont {Gaughran}},
  \bibinfo {author} {\bibfnamefont {F.}~\bibnamefont {Yuan}}, \bibinfo {author} {\bibfnamefont {P.}~\bibnamefont {Chang}}, \bibinfo {author} {\bibfnamefont {J.}~\bibnamefont {Emmins}},\ and\ \bibinfo {author} {\bibfnamefont {G.}~\bibnamefont {Howell}},\ }\bibfield  {title} {\bibinfo {title} {{{I21}: an advanced high-resolution resonant inelastic X-ray scattering beamline at {Diamond Light Source}}},\ }\href {https://doi.org/10.1107/S1600577522000601} {\bibfield  {journal} {\bibinfo  {journal} {J. Synchrotron Radiat.}\ }\textbf {\bibinfo {volume} {29}},\ \bibinfo {pages} {563} (\bibinfo {year} {2022})}\BibitemShut {NoStop}%
\bibitem [{\citenamefont {Braicovich}\ \emph {et~al.}(2010)\citenamefont {Braicovich}, \citenamefont {Moretti~Sala}, \citenamefont {Ament}, \citenamefont {Bisogni}, \citenamefont {Minola}, \citenamefont {Balestrino}, \citenamefont {Di~Castro}, \citenamefont {De~Luca}, \citenamefont {Salluzzo}, \citenamefont {Ghiringhelli},\ and\ \citenamefont {van~den Brink}}]{Braicovich2010momentum}%
  \BibitemOpen
  \bibfield  {author} {\bibinfo {author} {\bibfnamefont {L.}~\bibnamefont {Braicovich}}, \bibinfo {author} {\bibfnamefont {M.}~\bibnamefont {Moretti~Sala}}, \bibinfo {author} {\bibfnamefont {L.~J.~P.}\ \bibnamefont {Ament}}, \bibinfo {author} {\bibfnamefont {V.}~\bibnamefont {Bisogni}}, \bibinfo {author} {\bibfnamefont {M.}~\bibnamefont {Minola}}, \bibinfo {author} {\bibfnamefont {G.}~\bibnamefont {Balestrino}}, \bibinfo {author} {\bibfnamefont {D.}~\bibnamefont {Di~Castro}}, \bibinfo {author} {\bibfnamefont {G.~M.}\ \bibnamefont {De~Luca}}, \bibinfo {author} {\bibfnamefont {M.}~\bibnamefont {Salluzzo}}, \bibinfo {author} {\bibfnamefont {G.}~\bibnamefont {Ghiringhelli}},\ and\ \bibinfo {author} {\bibfnamefont {J.}~\bibnamefont {van~den Brink}},\ }\bibfield  {title} {\bibinfo {title} {Momentum and polarization dependence of single-magnon spectral weight for $\text{Cu}\text{ }{L}_{3}$-edge resonant inelastic x-ray scattering from layered cuprates},\ }\href {https://doi.org/10.1103/PhysRevB.81.174533} {\bibfield
   {journal} {\bibinfo  {journal} {Phys. Rev. B}\ }\textbf {\bibinfo {volume} {81}},\ \bibinfo {pages} {174533} (\bibinfo {year} {2010})}\BibitemShut {NoStop}%
\bibitem [{\citenamefont {Pan}\ \emph {et~al.}(2022{\natexlab{b}})\citenamefont {Pan}, \citenamefont {Song}, \citenamefont {Ferenc~Segedin}, \citenamefont {Jung}, \citenamefont {El-Sherif}, \citenamefont {Fleck}, \citenamefont {Goodge}, \citenamefont {Doyle}, \citenamefont {C\'ordova~Carrizales}, \citenamefont {N'Diaye}, \citenamefont {Shafer}, \citenamefont {Paik}, \citenamefont {Kourkoutis}, \citenamefont {El~Baggari}, \citenamefont {Botana}, \citenamefont {Brooks},\ and\ \citenamefont {Mundy}}]{Pan2022synthesis}%
  \BibitemOpen
  \bibfield  {author} {\bibinfo {author} {\bibfnamefont {G.~A.}\ \bibnamefont {Pan}}, \bibinfo {author} {\bibfnamefont {Q.}~\bibnamefont {Song}}, \bibinfo {author} {\bibfnamefont {D.}~\bibnamefont {Ferenc~Segedin}}, \bibinfo {author} {\bibfnamefont {M.-C.}\ \bibnamefont {Jung}}, \bibinfo {author} {\bibfnamefont {H.}~\bibnamefont {El-Sherif}}, \bibinfo {author} {\bibfnamefont {E.~E.}\ \bibnamefont {Fleck}}, \bibinfo {author} {\bibfnamefont {B.~H.}\ \bibnamefont {Goodge}}, \bibinfo {author} {\bibfnamefont {S.}~\bibnamefont {Doyle}}, \bibinfo {author} {\bibfnamefont {D.}~\bibnamefont {C\'ordova~Carrizales}}, \bibinfo {author} {\bibfnamefont {A.~T.}\ \bibnamefont {N'Diaye}}, \bibinfo {author} {\bibfnamefont {P.}~\bibnamefont {Shafer}}, \bibinfo {author} {\bibfnamefont {H.}~\bibnamefont {Paik}}, \bibinfo {author} {\bibfnamefont {L.~F.}\ \bibnamefont {Kourkoutis}}, \bibinfo {author} {\bibfnamefont {I.}~\bibnamefont {El~Baggari}}, \bibinfo {author} {\bibfnamefont {A.~S.}\ \bibnamefont {Botana}}, \bibinfo {author}
  {\bibfnamefont {C.~M.}\ \bibnamefont {Brooks}},\ and\ \bibinfo {author} {\bibfnamefont {J.~A.}\ \bibnamefont {Mundy}},\ }\bibfield  {title} {\bibinfo {title} {Synthesis and electronic properties of {${\mathrm{Nd}}_{n+1}{\mathrm{Ni}}_{n}{\mathrm{O}}_{3n+1}$ Ruddlesden-Popper} nickelate thin films},\ }\href {https://doi.org/10.1103/PhysRevMaterials.6.055003} {\bibfield  {journal} {\bibinfo  {journal} {Phys. Rev. Mater.}\ }\textbf {\bibinfo {volume} {6}},\ \bibinfo {pages} {055003} (\bibinfo {year} {2022}{\natexlab{b}})}\BibitemShut {NoStop}%
\bibitem [{\citenamefont {Ferenc~Segedin}\ \emph {et~al.}(2023)\citenamefont {Ferenc~Segedin}, \citenamefont {Goodge}, \citenamefont {Pan}, \citenamefont {Song}, \citenamefont {LaBollita}, \citenamefont {Jung}, \citenamefont {El-Sherif}, \citenamefont {Doyle}, \citenamefont {Turkiewicz}, \citenamefont {Taylor} \emph {et~al.}}]{ferenc2023limits}%
  \BibitemOpen
  \bibfield  {author} {\bibinfo {author} {\bibfnamefont {D.}~\bibnamefont {Ferenc~Segedin}}, \bibinfo {author} {\bibfnamefont {B.~H.}\ \bibnamefont {Goodge}}, \bibinfo {author} {\bibfnamefont {G.~A.}\ \bibnamefont {Pan}}, \bibinfo {author} {\bibfnamefont {Q.}~\bibnamefont {Song}}, \bibinfo {author} {\bibfnamefont {H.}~\bibnamefont {LaBollita}}, \bibinfo {author} {\bibfnamefont {M.-C.}\ \bibnamefont {Jung}}, \bibinfo {author} {\bibfnamefont {H.}~\bibnamefont {El-Sherif}}, \bibinfo {author} {\bibfnamefont {S.}~\bibnamefont {Doyle}}, \bibinfo {author} {\bibfnamefont {A.}~\bibnamefont {Turkiewicz}}, \bibinfo {author} {\bibfnamefont {N.~K.}\ \bibnamefont {Taylor}}, \emph {et~al.},\ }\bibfield  {title} {\bibinfo {title} {Limits to the strain engineering of layered square-planar nickelate thin films},\ }\href {https://doi.org/10.1038/s41467-023-37117-4} {\bibfield  {journal} {\bibinfo  {journal} {Nat. Commun.}\ }\textbf {\bibinfo {volume} {14}},\ \bibinfo {pages} {1468} (\bibinfo {year} {2023})}\BibitemShut
  {NoStop}%
\bibitem [{\citenamefont {Li}\ \emph {et~al.}(2021)\citenamefont {Li}, \citenamefont {He}, \citenamefont {Zhu}, \citenamefont {Si}, \citenamefont {Fan},\ and\ \citenamefont {Wen}}]{Li2020contrasting}%
  \BibitemOpen
  \bibfield  {author} {\bibinfo {author} {\bibfnamefont {Q.}~\bibnamefont {Li}}, \bibinfo {author} {\bibfnamefont {C.}~\bibnamefont {He}}, \bibinfo {author} {\bibfnamefont {X.}~\bibnamefont {Zhu}}, \bibinfo {author} {\bibfnamefont {J.}~\bibnamefont {Si}}, \bibinfo {author} {\bibfnamefont {X.}~\bibnamefont {Fan}},\ and\ \bibinfo {author} {\bibfnamefont {H.-H.}\ \bibnamefont {Wen}},\ }\bibfield  {title} {\bibinfo {title} {Contrasting physical properties of the trilayer nickelates {Nd$_4$Ni$_3$O$_{10}$} and {Nd$_4$Ni$_3$O$_8$}},\ }\href {https://doi.org/10.1007/s11433-020-1613-3} {\bibfield  {journal} {\bibinfo  {journal} {Sci. China Phys. Mech. Astron.}\ }\textbf {\bibinfo {volume} {64}},\ \bibinfo {pages} {227411} (\bibinfo {year} {2021})}\BibitemShut {NoStop}%
\bibitem [{\citenamefont {Greenblatt}(1997)}]{GREENBLATT1997174}%
  \BibitemOpen
  \bibfield  {author} {\bibinfo {author} {\bibfnamefont {M.}~\bibnamefont {Greenblatt}},\ }\bibfield  {title} {\bibinfo {title} {{Ruddlesden}-{Popper} $\mathrm{Ln}_{n+1}\mathrm{Ni}_{n}\mathrm{O}_{3n+1}$ nickelates: structure and properties},\ }\href {https://doi.org/https://doi.org/10.1016/S1359-0286(97)80062-9} {\bibfield  {journal} {\bibinfo  {journal} {Curr. Opin. Solid State Mater. Sci.}\ }\textbf {\bibinfo {volume} {2}},\ \bibinfo {pages} {174 } (\bibinfo {year} {1997})}\BibitemShut {NoStop}%
\bibitem [{\citenamefont {Bisogni}\ \emph {et~al.}(2016)\citenamefont {Bisogni}, \citenamefont {Catalano}, \citenamefont {Green}, \citenamefont {Gibert}, \citenamefont {Scherwitzl}, \citenamefont {Huang}, \citenamefont {Strocov}, \citenamefont {Zubko}, \citenamefont {Balandeh}, \citenamefont {Triscone}, \citenamefont {Sawatzky},\ and\ \citenamefont {Schmitt}}]{bisogni2016ground}%
  \BibitemOpen
  \bibfield  {author} {\bibinfo {author} {\bibfnamefont {V.}~\bibnamefont {Bisogni}}, \bibinfo {author} {\bibfnamefont {S.}~\bibnamefont {Catalano}}, \bibinfo {author} {\bibfnamefont {R.~J.}\ \bibnamefont {Green}}, \bibinfo {author} {\bibfnamefont {M.}~\bibnamefont {Gibert}}, \bibinfo {author} {\bibfnamefont {R.}~\bibnamefont {Scherwitzl}}, \bibinfo {author} {\bibfnamefont {Y.}~\bibnamefont {Huang}}, \bibinfo {author} {\bibfnamefont {V.~N.}\ \bibnamefont {Strocov}}, \bibinfo {author} {\bibfnamefont {P.}~\bibnamefont {Zubko}}, \bibinfo {author} {\bibfnamefont {S.}~\bibnamefont {Balandeh}}, \bibinfo {author} {\bibfnamefont {J.-M.}\ \bibnamefont {Triscone}}, \bibinfo {author} {\bibfnamefont {G.}~\bibnamefont {Sawatzky}},\ and\ \bibinfo {author} {\bibfnamefont {T.}~\bibnamefont {Schmitt}},\ }\bibfield  {title} {\bibinfo {title} {Ground-state oxygen holes and the metal--insulator transition in the negative charge-transfer rare-earth nickelates},\ }\href {https://doi.org/10.1038/ncomms13017} {\bibfield  {journal}
  {\bibinfo  {journal} {Nat. Commun.}\ }\textbf {\bibinfo {volume} {7}},\ \bibinfo {pages} {13017} (\bibinfo {year} {2016})}\BibitemShut {NoStop}%
\bibitem [{\citenamefont {F\"ursich}\ \emph {et~al.}(2019)\citenamefont {F\"ursich}, \citenamefont {Lu}, \citenamefont {Betto}, \citenamefont {Bluschke}, \citenamefont {Porras}, \citenamefont {Schierle}, \citenamefont {Ortiz}, \citenamefont {Suzuki}, \citenamefont {Cristiani}, \citenamefont {Logvenov}, \citenamefont {Brookes}, \citenamefont {Haverkort}, \citenamefont {Le~Tacon}, \citenamefont {Benckiser}, \citenamefont {Minola},\ and\ \citenamefont {Keimer}}]{Fursich2019resonant}%
  \BibitemOpen
  \bibfield  {author} {\bibinfo {author} {\bibfnamefont {K.}~\bibnamefont {F\"ursich}}, \bibinfo {author} {\bibfnamefont {Y.}~\bibnamefont {Lu}}, \bibinfo {author} {\bibfnamefont {D.}~\bibnamefont {Betto}}, \bibinfo {author} {\bibfnamefont {M.}~\bibnamefont {Bluschke}}, \bibinfo {author} {\bibfnamefont {J.}~\bibnamefont {Porras}}, \bibinfo {author} {\bibfnamefont {E.}~\bibnamefont {Schierle}}, \bibinfo {author} {\bibfnamefont {R.}~\bibnamefont {Ortiz}}, \bibinfo {author} {\bibfnamefont {H.}~\bibnamefont {Suzuki}}, \bibinfo {author} {\bibfnamefont {G.}~\bibnamefont {Cristiani}}, \bibinfo {author} {\bibfnamefont {G.}~\bibnamefont {Logvenov}}, \bibinfo {author} {\bibfnamefont {N.~B.}\ \bibnamefont {Brookes}}, \bibinfo {author} {\bibfnamefont {M.~W.}\ \bibnamefont {Haverkort}}, \bibinfo {author} {\bibfnamefont {M.}~\bibnamefont {Le~Tacon}}, \bibinfo {author} {\bibfnamefont {E.}~\bibnamefont {Benckiser}}, \bibinfo {author} {\bibfnamefont {M.}~\bibnamefont {Minola}},\ and\ \bibinfo {author} {\bibfnamefont
  {B.}~\bibnamefont {Keimer}},\ }\bibfield  {title} {\bibinfo {title} {Resonant inelastic x-ray scattering study of bond order and spin excitations in nickelate thin-film structures},\ }\href {https://doi.org/10.1103/PhysRevB.99.165124} {\bibfield  {journal} {\bibinfo  {journal} {Phys. Rev. B}\ }\textbf {\bibinfo {volume} {99}},\ \bibinfo {pages} {165124} (\bibinfo {year} {2019})}\BibitemShut {NoStop}%
\bibitem [{\citenamefont {Jung}\ \emph {et~al.}(2022)\citenamefont {Jung}, \citenamefont {Kapeghian}, \citenamefont {Hanson}, \citenamefont {Pamuk},\ and\ \citenamefont {Botana}}]{Jung2022electronic}%
  \BibitemOpen
  \bibfield  {author} {\bibinfo {author} {\bibfnamefont {M.-C.}\ \bibnamefont {Jung}}, \bibinfo {author} {\bibfnamefont {J.}~\bibnamefont {Kapeghian}}, \bibinfo {author} {\bibfnamefont {C.}~\bibnamefont {Hanson}}, \bibinfo {author} {\bibfnamefont {B.}~\bibnamefont {Pamuk}},\ and\ \bibinfo {author} {\bibfnamefont {A.~S.}\ \bibnamefont {Botana}},\ }\bibfield  {title} {\bibinfo {title} {Electronic structure of higher-order ruddlesden-popper nickelates},\ }\href {https://doi.org/10.1103/PhysRevB.105.085150} {\bibfield  {journal} {\bibinfo  {journal} {Phys. Rev. B}\ }\textbf {\bibinfo {volume} {105}},\ \bibinfo {pages} {085150} (\bibinfo {year} {2022})}\BibitemShut {NoStop}%
\bibitem [{\citenamefont {Shen}\ \emph {et~al.}(2023)\citenamefont {Shen}, \citenamefont {Sears}, \citenamefont {Fabbris}, \citenamefont {Li}, \citenamefont {Pelliciari}, \citenamefont {Mitrano}, \citenamefont {He}, \citenamefont {Zhang}, \citenamefont {Mitchell}, \citenamefont {Bisogni}, \citenamefont {Norman}, \citenamefont {Johnston},\ and\ \citenamefont {Dean}}]{Shen2023electronic}%
  \BibitemOpen
  \bibfield  {author} {\bibinfo {author} {\bibfnamefont {Y.}~\bibnamefont {Shen}}, \bibinfo {author} {\bibfnamefont {J.}~\bibnamefont {Sears}}, \bibinfo {author} {\bibfnamefont {G.}~\bibnamefont {Fabbris}}, \bibinfo {author} {\bibfnamefont {J.}~\bibnamefont {Li}}, \bibinfo {author} {\bibfnamefont {J.}~\bibnamefont {Pelliciari}}, \bibinfo {author} {\bibfnamefont {M.}~\bibnamefont {Mitrano}}, \bibinfo {author} {\bibfnamefont {W.}~\bibnamefont {He}}, \bibinfo {author} {\bibfnamefont {J.}~\bibnamefont {Zhang}}, \bibinfo {author} {\bibfnamefont {J.~F.}\ \bibnamefont {Mitchell}}, \bibinfo {author} {\bibfnamefont {V.}~\bibnamefont {Bisogni}}, \bibinfo {author} {\bibfnamefont {M.~R.}\ \bibnamefont {Norman}}, \bibinfo {author} {\bibfnamefont {S.}~\bibnamefont {Johnston}},\ and\ \bibinfo {author} {\bibfnamefont {M.~P.~M.}\ \bibnamefont {Dean}},\ }\bibfield  {title} {\bibinfo {title} {Electronic character of charge order in square-planar low-valence nickelates},\ }\href {https://doi.org/10.1103/PhysRevX.13.011021}
  {\bibfield  {journal} {\bibinfo  {journal} {Phys. Rev. X}\ }\textbf {\bibinfo {volume} {13}},\ \bibinfo {pages} {011021} (\bibinfo {year} {2023})}\BibitemShut {NoStop}%
\bibitem [{\citenamefont {Lebert}\ \emph {et~al.}(2017)\citenamefont {Lebert}, \citenamefont {Dean}, \citenamefont {Nicolaou}, \citenamefont {Pelliciari}, \citenamefont {Dantz}, \citenamefont {Schmitt}, \citenamefont {Yu}, \citenamefont {Azuma}, \citenamefont {Castellan}, \citenamefont {Miao}, \citenamefont {Gauzzi}, \citenamefont {Baptiste},\ and\ \citenamefont {D'astuto}}]{Lebert2017resonant}%
  \BibitemOpen
  \bibfield  {author} {\bibinfo {author} {\bibfnamefont {B.~W.}\ \bibnamefont {Lebert}}, \bibinfo {author} {\bibfnamefont {M.~P.~M.}\ \bibnamefont {Dean}}, \bibinfo {author} {\bibfnamefont {A.}~\bibnamefont {Nicolaou}}, \bibinfo {author} {\bibfnamefont {J.}~\bibnamefont {Pelliciari}}, \bibinfo {author} {\bibfnamefont {M.}~\bibnamefont {Dantz}}, \bibinfo {author} {\bibfnamefont {T.}~\bibnamefont {Schmitt}}, \bibinfo {author} {\bibfnamefont {R.}~\bibnamefont {Yu}}, \bibinfo {author} {\bibfnamefont {M.}~\bibnamefont {Azuma}}, \bibinfo {author} {\bibfnamefont {J.-P.}\ \bibnamefont {Castellan}}, \bibinfo {author} {\bibfnamefont {H.}~\bibnamefont {Miao}}, \bibinfo {author} {\bibfnamefont {A.}~\bibnamefont {Gauzzi}}, \bibinfo {author} {\bibfnamefont {B.}~\bibnamefont {Baptiste}},\ and\ \bibinfo {author} {\bibfnamefont {M.}~\bibnamefont {D'astuto}},\ }\bibfield  {title} {\bibinfo {title} {Resonant inelastic x-ray scattering study of spin-wave excitations in the cuprate parent compound {Ca$_2$CuO$_2$Cl$_2$}},\ }\href
  {https://doi.org/10.1103/PhysRevB.95.155110} {\bibfield  {journal} {\bibinfo  {journal} {Phys. Rev. B}\ }\textbf {\bibinfo {volume} {95}},\ \bibinfo {pages} {155110} (\bibinfo {year} {2017})}\BibitemShut {NoStop}%
\bibitem [{\citenamefont {Shen}\ \emph {et~al.}(2022)\citenamefont {Shen}, \citenamefont {Sears}, \citenamefont {Fabbris}, \citenamefont {Li}, \citenamefont {Pelliciari}, \citenamefont {Jarrige}, \citenamefont {He}, \citenamefont {Bo\ifmmode \check{z}\else \v{z}\fi{}ovi\ifmmode~\acute{c}\else \'{c}\fi{}}, \citenamefont {Mitrano}, \citenamefont {Zhang}, \citenamefont {Mitchell}, \citenamefont {Botana}, \citenamefont {Bisogni}, \citenamefont {Norman}, \citenamefont {Johnston},\ and\ \citenamefont {Dean}}]{Shen2022role}%
  \BibitemOpen
  \bibfield  {author} {\bibinfo {author} {\bibfnamefont {Y.}~\bibnamefont {Shen}}, \bibinfo {author} {\bibfnamefont {J.}~\bibnamefont {Sears}}, \bibinfo {author} {\bibfnamefont {G.}~\bibnamefont {Fabbris}}, \bibinfo {author} {\bibfnamefont {J.}~\bibnamefont {Li}}, \bibinfo {author} {\bibfnamefont {J.}~\bibnamefont {Pelliciari}}, \bibinfo {author} {\bibfnamefont {I.}~\bibnamefont {Jarrige}}, \bibinfo {author} {\bibfnamefont {X.}~\bibnamefont {He}}, \bibinfo {author} {\bibfnamefont {I.}~\bibnamefont {Bo\ifmmode \check{z}\else \v{z}\fi{}ovi\ifmmode~\acute{c}\else \'{c}\fi{}}}, \bibinfo {author} {\bibfnamefont {M.}~\bibnamefont {Mitrano}}, \bibinfo {author} {\bibfnamefont {J.}~\bibnamefont {Zhang}}, \bibinfo {author} {\bibfnamefont {J.~F.}\ \bibnamefont {Mitchell}}, \bibinfo {author} {\bibfnamefont {A.~S.}\ \bibnamefont {Botana}}, \bibinfo {author} {\bibfnamefont {V.}~\bibnamefont {Bisogni}}, \bibinfo {author} {\bibfnamefont {M.~R.}\ \bibnamefont {Norman}}, \bibinfo {author} {\bibfnamefont {S.}~\bibnamefont
  {Johnston}},\ and\ \bibinfo {author} {\bibfnamefont {M.~P.~M.}\ \bibnamefont {Dean}},\ }\bibfield  {title} {\bibinfo {title} {Role of oxygen states in the low valence nickelate {${\mathrm{La}}_{4}{\mathrm{Ni}}_{3}{\mathrm{O}}_{8}$}},\ }\href {https://doi.org/10.1103/PhysRevX.12.011055} {\bibfield  {journal} {\bibinfo  {journal} {Phys. Rev. X}\ }\textbf {\bibinfo {volume} {12}},\ \bibinfo {pages} {011055} (\bibinfo {year} {2022})}\BibitemShut {NoStop}%
\bibitem [{\citenamefont {Ren}\ \emph {et~al.}(2025)\citenamefont {Ren}, \citenamefont {Sutarto}, \citenamefont {Wu}, \citenamefont {Zhang}, \citenamefont {Huang}, \citenamefont {Xiang}, \citenamefont {Hu}, \citenamefont {Comin}, \citenamefont {Zhou},\ and\ \citenamefont {Zhu}}]{ren2024resolving}%
  \BibitemOpen
  \bibfield  {author} {\bibinfo {author} {\bibfnamefont {X.}~\bibnamefont {Ren}}, \bibinfo {author} {\bibfnamefont {R.}~\bibnamefont {Sutarto}}, \bibinfo {author} {\bibfnamefont {X.}~\bibnamefont {Wu}}, \bibinfo {author} {\bibfnamefont {J.}~\bibnamefont {Zhang}}, \bibinfo {author} {\bibfnamefont {H.}~\bibnamefont {Huang}}, \bibinfo {author} {\bibfnamefont {T.}~\bibnamefont {Xiang}}, \bibinfo {author} {\bibfnamefont {J.}~\bibnamefont {Hu}}, \bibinfo {author} {\bibfnamefont {R.}~\bibnamefont {Comin}}, \bibinfo {author} {\bibfnamefont {X.~J.}\ \bibnamefont {Zhou}},\ and\ \bibinfo {author} {\bibfnamefont {Z.}~\bibnamefont {Zhu}},\ }\bibfield  {title} {\bibinfo {title} {Resolving the electronic ground state of {La$_3$Ni$_2$O$_{7-\delta}$} films},\ }\href {https://doi.org/https://doi.org/10.1038/s42005-025-01971-z} {\bibfield  {journal} {\bibinfo  {journal} {Commun Phys}\ }\textbf {\bibinfo {volume} {8}},\ \bibinfo {pages} {52} (\bibinfo {year} {2025})}\BibitemShut {NoStop}%
\bibitem [{\citenamefont {Takeda}\ \emph {et~al.}(1992)\citenamefont {Takeda}, \citenamefont {Nishijima}, \citenamefont {Imanishi}, \citenamefont {Kanno}, \citenamefont {Yamamoto},\ and\ \citenamefont {Takano}}]{takeda1992crystal}%
  \BibitemOpen
  \bibfield  {author} {\bibinfo {author} {\bibfnamefont {Y.}~\bibnamefont {Takeda}}, \bibinfo {author} {\bibfnamefont {M.}~\bibnamefont {Nishijima}}, \bibinfo {author} {\bibfnamefont {N.}~\bibnamefont {Imanishi}}, \bibinfo {author} {\bibfnamefont {R.}~\bibnamefont {Kanno}}, \bibinfo {author} {\bibfnamefont {O.}~\bibnamefont {Yamamoto}},\ and\ \bibinfo {author} {\bibfnamefont {M.}~\bibnamefont {Takano}},\ }\bibfield  {title} {\bibinfo {title} {Crystal chemistry and transport properties of {Nd$_{2-x}A_x$NiO$_4$} {($A=$ Ca, Sr, or Ba, $0 \leq x \leq 1.4$)}},\ }\href {https://doi.org/https://doi.org/10.1016/S0022-4596(05)80299-3} {\bibfield  {journal} {\bibinfo  {journal} {J. Solid State Chem.}\ }\textbf {\bibinfo {volume} {96}},\ \bibinfo {pages} {72} (\bibinfo {year} {1992})}\BibitemShut {NoStop}%
\bibitem [{\citenamefont {Gupta}\ \emph {et~al.}(2024)\citenamefont {Gupta}, \citenamefont {Gong}, \citenamefont {Wu}, \citenamefont {Kang}, \citenamefont {Parzyck}, \citenamefont {Gregory}, \citenamefont {Costa}, \citenamefont {Sutarto}, \citenamefont {Sarker}, \citenamefont {Singer}, \citenamefont {Schlom}, \citenamefont {Shen},\ and\ \citenamefont {Hawthorn}}]{gupta2024anisotropic}%
  \BibitemOpen
  \bibfield  {author} {\bibinfo {author} {\bibfnamefont {N.~K.}\ \bibnamefont {Gupta}}, \bibinfo {author} {\bibfnamefont {R.}~\bibnamefont {Gong}}, \bibinfo {author} {\bibfnamefont {Y.}~\bibnamefont {Wu}}, \bibinfo {author} {\bibfnamefont {M.}~\bibnamefont {Kang}}, \bibinfo {author} {\bibfnamefont {C.~T.}\ \bibnamefont {Parzyck}}, \bibinfo {author} {\bibfnamefont {B.~Z.}\ \bibnamefont {Gregory}}, \bibinfo {author} {\bibfnamefont {N.}~\bibnamefont {Costa}}, \bibinfo {author} {\bibfnamefont {R.}~\bibnamefont {Sutarto}}, \bibinfo {author} {\bibfnamefont {S.}~\bibnamefont {Sarker}}, \bibinfo {author} {\bibfnamefont {A.}~\bibnamefont {Singer}}, \bibinfo {author} {\bibfnamefont {D.~G.}\ \bibnamefont {Schlom}}, \bibinfo {author} {\bibfnamefont {K.~M.}\ \bibnamefont {Shen}},\ and\ \bibinfo {author} {\bibfnamefont {D.~G.}\ \bibnamefont {Hawthorn}},\ }\href {https://arxiv.org/abs/2409.03210} {\bibinfo {title} {Anisotropic spin stripe domains in bilayer {La$_3$Ni$_2$O$_7$}}} (\bibinfo {year} {2024}),\ \Eprint
  {https://arxiv.org/abs/2409.03210} {arXiv:2409.03210} \BibitemShut {NoStop}%
\bibitem [{\citenamefont {Catalano}\ \emph {et~al.}(2018)\citenamefont {Catalano}, \citenamefont {Gibert}, \citenamefont {Fowlie}, \citenamefont {Iniguez}, \citenamefont {Triscone},\ and\ \citenamefont {Kreisel}}]{catalano2018rare}%
  \BibitemOpen
  \bibfield  {author} {\bibinfo {author} {\bibfnamefont {S.}~\bibnamefont {Catalano}}, \bibinfo {author} {\bibfnamefont {M.}~\bibnamefont {Gibert}}, \bibinfo {author} {\bibfnamefont {J.}~\bibnamefont {Fowlie}}, \bibinfo {author} {\bibfnamefont {J.}~\bibnamefont {Iniguez}}, \bibinfo {author} {\bibfnamefont {J.-M.}\ \bibnamefont {Triscone}},\ and\ \bibinfo {author} {\bibfnamefont {J.}~\bibnamefont {Kreisel}},\ }\bibfield  {title} {\bibinfo {title} {Rare-earth nickelates {$R$NiO$_3$}: thin films and heterostructures},\ }\href {https://iopscience.iop.org/article/10.1088/1361-6633/aaa37a} {\bibfield  {journal} {\bibinfo  {journal} {Rep. Prog. Phys.}\ }\textbf {\bibinfo {volume} {81}},\ \bibinfo {pages} {046501} (\bibinfo {year} {2018})}\BibitemShut {NoStop}%
\bibitem [{\citenamefont {Biało}\ \emph {et~al.}(2024)\citenamefont {Biało}, \citenamefont {Martinelli}, \citenamefont {Luca}, \citenamefont {Worm}, \citenamefont {Drewanowski}, \citenamefont {Jöhr}, \citenamefont {Choi}, \citenamefont {Garcia-Fernandez}, \citenamefont {Agrestini}, \citenamefont {Zhou}, \citenamefont {Kummer}, \citenamefont {Brookes}, \citenamefont {Guo}, \citenamefont {Edgeton}, \citenamefont {Eom}, \citenamefont {Tomczak}, \citenamefont {Held}, \citenamefont {Gibert}, \citenamefont {Wang},\ and\ \citenamefont {Chang}}]{Bialo2023}%
  \BibitemOpen
  \bibfield  {author} {\bibinfo {author} {\bibfnamefont {I.}~\bibnamefont {Biało}}, \bibinfo {author} {\bibfnamefont {L.}~\bibnamefont {Martinelli}}, \bibinfo {author} {\bibfnamefont {G.~D.}\ \bibnamefont {Luca}}, \bibinfo {author} {\bibfnamefont {P.}~\bibnamefont {Worm}}, \bibinfo {author} {\bibfnamefont {A.}~\bibnamefont {Drewanowski}}, \bibinfo {author} {\bibfnamefont {S.}~\bibnamefont {Jöhr}}, \bibinfo {author} {\bibfnamefont {J.}~\bibnamefont {Choi}}, \bibinfo {author} {\bibfnamefont {M.}~\bibnamefont {Garcia-Fernandez}}, \bibinfo {author} {\bibfnamefont {S.}~\bibnamefont {Agrestini}}, \bibinfo {author} {\bibfnamefont {K.~J.}\ \bibnamefont {Zhou}}, \bibinfo {author} {\bibfnamefont {K.}~\bibnamefont {Kummer}}, \bibinfo {author} {\bibfnamefont {N.~B.}\ \bibnamefont {Brookes}}, \bibinfo {author} {\bibfnamefont {L.}~\bibnamefont {Guo}}, \bibinfo {author} {\bibfnamefont {A.}~\bibnamefont {Edgeton}}, \bibinfo {author} {\bibfnamefont {C.~B.}\ \bibnamefont {Eom}}, \bibinfo {author} {\bibfnamefont {J.~M.}\
  \bibnamefont {Tomczak}}, \bibinfo {author} {\bibfnamefont {K.}~\bibnamefont {Held}}, \bibinfo {author} {\bibfnamefont {M.}~\bibnamefont {Gibert}}, \bibinfo {author} {\bibfnamefont {Q.}~\bibnamefont {Wang}},\ and\ \bibinfo {author} {\bibfnamefont {J.}~\bibnamefont {Chang}},\ }\bibfield  {title} {\bibinfo {title} {Strain-tuned incompatible magnetic exchange-interaction in {La$_2$NiO$_4$}},\ }\href {https://doi.org/10.1038/s42005-024-01701-x} {\bibfield  {journal} {\bibinfo  {journal} {Commun. Phys.}\ }\textbf {\bibinfo {volume} {7}},\ \bibinfo {pages} {230} (\bibinfo {year} {2024})}\BibitemShut {NoStop}%
\bibitem [{\citenamefont {Kim}\ \emph {et~al.}(2012)\citenamefont {Kim}, \citenamefont {Said}, \citenamefont {Casa}, \citenamefont {Upton}, \citenamefont {Gog}, \citenamefont {Daghofer}, \citenamefont {Jackeli}, \citenamefont {Brink}, \citenamefont {Khaliullin},\ and\ \citenamefont {Kim}}]{Kim2012large}%
  \BibitemOpen
  \bibfield  {author} {\bibinfo {author} {\bibfnamefont {J.}~\bibnamefont {Kim}}, \bibinfo {author} {\bibfnamefont {A.~H.}\ \bibnamefont {Said}}, \bibinfo {author} {\bibfnamefont {D.}~\bibnamefont {Casa}}, \bibinfo {author} {\bibfnamefont {M.~H.}\ \bibnamefont {Upton}}, \bibinfo {author} {\bibfnamefont {T.}~\bibnamefont {Gog}}, \bibinfo {author} {\bibfnamefont {M.}~\bibnamefont {Daghofer}}, \bibinfo {author} {\bibfnamefont {G.}~\bibnamefont {Jackeli}}, \bibinfo {author} {\bibfnamefont {J.~V.~D.}\ \bibnamefont {Brink}}, \bibinfo {author} {\bibfnamefont {G.}~\bibnamefont {Khaliullin}},\ and\ \bibinfo {author} {\bibfnamefont {B.~J.}\ \bibnamefont {Kim}},\ }\bibfield  {title} {\bibinfo {title} {Large spin-wave energy gap in the bilayer iridate {Sr$_3$Ir$_2$O$_7$}: Evidence for enhanced dipolar interactions near the {Mott} metal-insulator transition},\ }\href {https://doi.org/10.1103/PhysRevLett.109.157402} {\bibfield  {journal} {\bibinfo  {journal} {Phys. Rev. Lett.}\ }\textbf {\bibinfo {volume} {109}},\ \bibinfo
  {pages} {157402} (\bibinfo {year} {2012})}\BibitemShut {NoStop}%
\bibitem [{\citenamefont {Mazzone}\ \emph {et~al.}(2022)\citenamefont {Mazzone}, \citenamefont {Shen}, \citenamefont {Suwa}, \citenamefont {Fabbris}, \citenamefont {Yang}, \citenamefont {Zhang}, \citenamefont {Miao}, \citenamefont {Sears}, \citenamefont {Jia}, \citenamefont {Shi}, \citenamefont {Upton}, \citenamefont {Casa}, \citenamefont {Liu}, \citenamefont {Liu}, \citenamefont {Batista},\ and\ \citenamefont {Dean}}]{mazzone2022antiferromagnetic}%
  \BibitemOpen
  \bibfield  {author} {\bibinfo {author} {\bibfnamefont {D.~G.}\ \bibnamefont {Mazzone}}, \bibinfo {author} {\bibfnamefont {Y.}~\bibnamefont {Shen}}, \bibinfo {author} {\bibfnamefont {H.}~\bibnamefont {Suwa}}, \bibinfo {author} {\bibfnamefont {G.}~\bibnamefont {Fabbris}}, \bibinfo {author} {\bibfnamefont {J.}~\bibnamefont {Yang}}, \bibinfo {author} {\bibfnamefont {S.~S.}\ \bibnamefont {Zhang}}, \bibinfo {author} {\bibfnamefont {H.}~\bibnamefont {Miao}}, \bibinfo {author} {\bibfnamefont {J.}~\bibnamefont {Sears}}, \bibinfo {author} {\bibfnamefont {K.}~\bibnamefont {Jia}}, \bibinfo {author} {\bibfnamefont {Y.~G.}\ \bibnamefont {Shi}}, \bibinfo {author} {\bibfnamefont {M.~H.}\ \bibnamefont {Upton}}, \bibinfo {author} {\bibfnamefont {D.~M.}\ \bibnamefont {Casa}}, \bibinfo {author} {\bibfnamefont {X.}~\bibnamefont {Liu}}, \bibinfo {author} {\bibfnamefont {J.}~\bibnamefont {Liu}}, \bibinfo {author} {\bibfnamefont {C.~D.}\ \bibnamefont {Batista}},\ and\ \bibinfo {author} {\bibfnamefont {M.~P.}\ \bibnamefont
  {Dean}},\ }\bibfield  {title} {\bibinfo {title} {Antiferromagnetic excitonic insulator state in {Sr$_3$Ir$_2$O$_7$}},\ }\href {https://doi.org/10.1038/s41467-022-28207-w} {\bibfield  {journal} {\bibinfo  {journal} {Nat. Commun.}\ }\textbf {\bibinfo {volume} {13}},\ \bibinfo {pages} {913} (\bibinfo {year} {2022})}\BibitemShut {NoStop}%
\bibitem [{\citenamefont {Zhang}\ \emph {et~al.}(2020)\citenamefont {Zhang}, \citenamefont {Phelan}, \citenamefont {Botana}, \citenamefont {Chen}, \citenamefont {Zheng}, \citenamefont {Krogstad}, \citenamefont {Wang}, \citenamefont {Qiu}, \citenamefont {Rodriguez-Rivera}, \citenamefont {Osborn}, \citenamefont {Rosenkranz}, \citenamefont {Norman},\ and\ \citenamefont {Mitchell}}]{zhang2020intertwined}%
  \BibitemOpen
  \bibfield  {author} {\bibinfo {author} {\bibfnamefont {J.}~\bibnamefont {Zhang}}, \bibinfo {author} {\bibfnamefont {D.}~\bibnamefont {Phelan}}, \bibinfo {author} {\bibfnamefont {A.~S.}\ \bibnamefont {Botana}}, \bibinfo {author} {\bibfnamefont {Y.-S.}\ \bibnamefont {Chen}}, \bibinfo {author} {\bibfnamefont {H.}~\bibnamefont {Zheng}}, \bibinfo {author} {\bibfnamefont {M.}~\bibnamefont {Krogstad}}, \bibinfo {author} {\bibfnamefont {S.~G.}\ \bibnamefont {Wang}}, \bibinfo {author} {\bibfnamefont {Y.}~\bibnamefont {Qiu}}, \bibinfo {author} {\bibfnamefont {J.}~\bibnamefont {Rodriguez-Rivera}}, \bibinfo {author} {\bibfnamefont {R.}~\bibnamefont {Osborn}}, \bibinfo {author} {\bibfnamefont {S.}~\bibnamefont {Rosenkranz}}, \bibinfo {author} {\bibfnamefont {M.~R.}\ \bibnamefont {Norman}},\ and\ \bibinfo {author} {\bibfnamefont {J.~F.}\ \bibnamefont {Mitchell}},\ }\bibfield  {title} {\bibinfo {title} {Intertwined density waves in a metallic nickelate},\ }\href {https://www.nature.com/articles/s41467-020-19836-0}
  {\bibfield  {journal} {\bibinfo  {journal} {Nat. Commun.}\ }\textbf {\bibinfo {volume} {11}},\ \bibinfo {pages} {6003} (\bibinfo {year} {2020})}\BibitemShut {NoStop}%
\bibitem [{\citenamefont {Batlle}\ \emph {et~al.}(1992{\natexlab{a}})\citenamefont {Batlle}, \citenamefont {Obradors},\ and\ \citenamefont {Martnez}}]{Batlle1991}%
  \BibitemOpen
  \bibfield  {author} {\bibinfo {author} {\bibfnamefont {X.}~\bibnamefont {Batlle}}, \bibinfo {author} {\bibfnamefont {X.}~\bibnamefont {Obradors}},\ and\ \bibinfo {author} {\bibfnamefont {B.}~\bibnamefont {Martnez}},\ }\bibfield  {title} {\bibinfo {title} {Magnetic interactions, weak ferromagnetism, and field-induced transitions in {${\mathrm{Nd}}_{2}$${\mathrm{NiO}}_{4}$}},\ }\href {https://doi.org/10.1103/PhysRevB.45.2830} {\bibfield  {journal} {\bibinfo  {journal} {Phys. Rev. B}\ }\textbf {\bibinfo {volume} {45}},\ \bibinfo {pages} {2830} (\bibinfo {year} {1992}{\natexlab{a}})}\BibitemShut {NoStop}%
\bibitem [{\citenamefont {{Saez Puche}}\ \emph {et~al.}(1989)\citenamefont {{Saez Puche}}, \citenamefont {Fernández}, \citenamefont {{Rodríguez Carvajal}},\ and\ \citenamefont {Martínez}}]{SaezPuche1989}%
  \BibitemOpen
  \bibfield  {author} {\bibinfo {author} {\bibfnamefont {R.}~\bibnamefont {{Saez Puche}}}, \bibinfo {author} {\bibfnamefont {F.}~\bibnamefont {Fernández}}, \bibinfo {author} {\bibfnamefont {J.}~\bibnamefont {{Rodríguez Carvajal}}},\ and\ \bibinfo {author} {\bibfnamefont {J.~L.}\ \bibnamefont {Martínez}},\ }\bibfield  {title} {\bibinfo {title} {Magnetic and x-ray diffraction characterization of stoichiometric {Pr$_2$NiO$_4$} and {Nd$_2$NiO$_4$} oxides},\ }\href {https://doi.org/10.1016/0038-1098(89)90809-0} {\bibfield  {journal} {\bibinfo  {journal} {Solid State Commun.}\ }\textbf {\bibinfo {volume} {72}},\ \bibinfo {pages} {273} (\bibinfo {year} {1989})}\BibitemShut {NoStop}%
\bibitem [{\citenamefont {{Rodriguez-Carvajal}}\ \emph {et~al.}(1990)\citenamefont {{Rodriguez-Carvajal}}, \citenamefont {{Fernandez-Diaz}}, \citenamefont {Martinez},\ and\ \citenamefont {{Saez-Puche}}}]{RodriguezCarvajal1990}%
  \BibitemOpen
  \bibfield  {author} {\bibinfo {author} {\bibfnamefont {J.}~\bibnamefont {{Rodriguez-Carvajal}}}, \bibinfo {author} {\bibfnamefont {M.~T.}\ \bibnamefont {{Fernandez-Diaz}}}, \bibinfo {author} {\bibfnamefont {J.~L.}\ \bibnamefont {Martinez}},\ and\ \bibinfo {author} {\bibfnamefont {R.}~\bibnamefont {{Saez-Puche}}},\ }\bibfield  {title} {\bibinfo {title} {Structural phase transitions and three-dimensional magnetic ordering in the {Nd$_2$NiO$_4$} oxide},\ }\href {https://doi.org/10.1209/0295-5075/11/3/013} {\bibfield  {journal} {\bibinfo  {journal} {Europhys. Lett.}\ }\textbf {\bibinfo {volume} {11}},\ \bibinfo {pages} {261} (\bibinfo {year} {1990})}\BibitemShut {NoStop}%
\bibitem [{\citenamefont {Batlle}\ \emph {et~al.}(1992{\natexlab{b}})\citenamefont {Batlle}, \citenamefont {Martínez}, \citenamefont {Obradors}, \citenamefont {Pernet}, \citenamefont {Vallet}, \citenamefont {González-Calvet},\ and\ \citenamefont {Alonso}}]{Batlle1992}%
  \BibitemOpen
  \bibfield  {author} {\bibinfo {author} {\bibfnamefont {X.}~\bibnamefont {Batlle}}, \bibinfo {author} {\bibfnamefont {B.}~\bibnamefont {Martínez}}, \bibinfo {author} {\bibfnamefont {X.}~\bibnamefont {Obradors}}, \bibinfo {author} {\bibfnamefont {M.}~\bibnamefont {Pernet}}, \bibinfo {author} {\bibfnamefont {M.}~\bibnamefont {Vallet}}, \bibinfo {author} {\bibfnamefont {J.}~\bibnamefont {González-Calvet}},\ and\ \bibinfo {author} {\bibfnamefont {J.}~\bibnamefont {Alonso}},\ }\bibfield  {title} {\bibinfo {title} {Study of the magnetic properties of {Nd$_2$NiO$_4$}},\ }\href {https://doi.org/10.1016/0304-8853(92)90422-K} {\bibfield  {journal} {\bibinfo  {journal} {J. Magn. Magn. Mater.}\ }\textbf {\bibinfo {volume} {104-107}},\ \bibinfo {pages} {918} (\bibinfo {year} {1992}{\natexlab{b}})}\BibitemShut {NoStop}%
\bibitem [{\citenamefont {Maity}\ \emph {et~al.}(2019)\citenamefont {Maity}, \citenamefont {Ceretti}, \citenamefont {Keller}, \citenamefont {Schefer}, \citenamefont {Shang}, \citenamefont {Pomjakushina}, \citenamefont {Meven}, \citenamefont {Sheptyakov}, \citenamefont {Cervellino},\ and\ \citenamefont {Paulus}}]{RanjanMaiti2019}%
  \BibitemOpen
  \bibfield  {author} {\bibinfo {author} {\bibfnamefont {S.~R.}\ \bibnamefont {Maity}}, \bibinfo {author} {\bibfnamefont {M.}~\bibnamefont {Ceretti}}, \bibinfo {author} {\bibfnamefont {L.}~\bibnamefont {Keller}}, \bibinfo {author} {\bibfnamefont {J.}~\bibnamefont {Schefer}}, \bibinfo {author} {\bibfnamefont {T.}~\bibnamefont {Shang}}, \bibinfo {author} {\bibfnamefont {E.}~\bibnamefont {Pomjakushina}}, \bibinfo {author} {\bibfnamefont {M.}~\bibnamefont {Meven}}, \bibinfo {author} {\bibfnamefont {D.}~\bibnamefont {Sheptyakov}}, \bibinfo {author} {\bibfnamefont {A.}~\bibnamefont {Cervellino}},\ and\ \bibinfo {author} {\bibfnamefont {W.}~\bibnamefont {Paulus}},\ }\bibfield  {title} {\bibinfo {title} {Structural disorder and magnetic correlations driven by oxygen doping in {$\mathrm{N}{\mathrm{d}}_{2}\mathrm{Ni}{\mathrm{O}}_{4+\ensuremath{\delta}}$} ($\ensuremath{\delta}\ensuremath{\sim}0.11$)},\ }\href {https://doi.org/10.1103/PhysRevMaterials.3.083604} {\bibfield  {journal} {\bibinfo  {journal} {Phys. Rev.
  Mater.}\ }\textbf {\bibinfo {volume} {3}},\ \bibinfo {pages} {083604} (\bibinfo {year} {2019})}\BibitemShut {NoStop}%
\bibitem [{\citenamefont {Khomskii}(2014)}]{KhomskiiTMs}%
  \BibitemOpen
  \bibfield  {author} {\bibinfo {author} {\bibfnamefont {D.}~\bibnamefont {Khomskii}},\ }\href@noop {} {\emph {\bibinfo {title} {Transition Metal Compounds}}}\ (\bibinfo  {publisher} {Cambridge University Press},\ \bibinfo {year} {2014})\BibitemShut {NoStop}%
\bibitem [{\citenamefont {Lu}\ \emph {et~al.}(2018)\citenamefont {Lu}, \citenamefont {Betto}, \citenamefont {F\"ursich}, \citenamefont {Suzuki}, \citenamefont {Kim}, \citenamefont {Cristiani}, \citenamefont {Logvenov}, \citenamefont {Brookes}, \citenamefont {Benckiser}, \citenamefont {Haverkort}, \citenamefont {Khaliullin}, \citenamefont {Le~Tacon}, \citenamefont {Minola},\ and\ \citenamefont {Keimer}}]{Lu2018}%
  \BibitemOpen
  \bibfield  {author} {\bibinfo {author} {\bibfnamefont {Y.}~\bibnamefont {Lu}}, \bibinfo {author} {\bibfnamefont {D.}~\bibnamefont {Betto}}, \bibinfo {author} {\bibfnamefont {K.}~\bibnamefont {F\"ursich}}, \bibinfo {author} {\bibfnamefont {H.}~\bibnamefont {Suzuki}}, \bibinfo {author} {\bibfnamefont {H.-H.}\ \bibnamefont {Kim}}, \bibinfo {author} {\bibfnamefont {G.}~\bibnamefont {Cristiani}}, \bibinfo {author} {\bibfnamefont {G.}~\bibnamefont {Logvenov}}, \bibinfo {author} {\bibfnamefont {N.~B.}\ \bibnamefont {Brookes}}, \bibinfo {author} {\bibfnamefont {E.}~\bibnamefont {Benckiser}}, \bibinfo {author} {\bibfnamefont {M.~W.}\ \bibnamefont {Haverkort}}, \bibinfo {author} {\bibfnamefont {G.}~\bibnamefont {Khaliullin}}, \bibinfo {author} {\bibfnamefont {M.}~\bibnamefont {Le~Tacon}}, \bibinfo {author} {\bibfnamefont {M.}~\bibnamefont {Minola}},\ and\ \bibinfo {author} {\bibfnamefont {B.}~\bibnamefont {Keimer}},\ }\bibfield  {title} {\bibinfo {title} {Site-selective probe of magnetic excitations in rare-earth
  nickelates using resonant inelastic x-ray scattering},\ }\href {https://doi.org/10.1103/PhysRevX.8.031014} {\bibfield  {journal} {\bibinfo  {journal} {Phys. Rev. X}\ }\textbf {\bibinfo {volume} {8}},\ \bibinfo {pages} {031014} (\bibinfo {year} {2018})}\BibitemShut {NoStop}%
\bibitem [{\citenamefont {Freeman}\ \emph {et~al.}(2005)\citenamefont {Freeman}, \citenamefont {Boothroyd}, \citenamefont {Prabhakaran}, \citenamefont {Frost}, \citenamefont {Enderle},\ and\ \citenamefont {Hiess}}]{Freeman2005}%
  \BibitemOpen
  \bibfield  {author} {\bibinfo {author} {\bibfnamefont {P.~G.}\ \bibnamefont {Freeman}}, \bibinfo {author} {\bibfnamefont {A.~T.}\ \bibnamefont {Boothroyd}}, \bibinfo {author} {\bibfnamefont {D.}~\bibnamefont {Prabhakaran}}, \bibinfo {author} {\bibfnamefont {C.~D.}\ \bibnamefont {Frost}}, \bibinfo {author} {\bibfnamefont {M.}~\bibnamefont {Enderle}},\ and\ \bibinfo {author} {\bibfnamefont {A.}~\bibnamefont {Hiess}},\ }\bibfield  {title} {\bibinfo {title} {Spin dynamics of half-doped {${\mathrm{La}}_{3/2}{\mathrm{Sr}}_{1/2}\mathrm{Ni}{\mathrm{O}}_{4}$}},\ }\href {https://doi.org/10.1103/PhysRevB.71.174412} {\bibfield  {journal} {\bibinfo  {journal} {Phys. Rev. B}\ }\textbf {\bibinfo {volume} {71}},\ \bibinfo {pages} {174412} (\bibinfo {year} {2005})}\BibitemShut {NoStop}%
\bibitem [{\citenamefont {Bourges}\ \emph {et~al.}(2003)\citenamefont {Bourges}, \citenamefont {Sidis}, \citenamefont {Braden}, \citenamefont {Nakajima},\ and\ \citenamefont {Tranquada}}]{Bourges2003}%
  \BibitemOpen
  \bibfield  {author} {\bibinfo {author} {\bibfnamefont {P.}~\bibnamefont {Bourges}}, \bibinfo {author} {\bibfnamefont {Y.}~\bibnamefont {Sidis}}, \bibinfo {author} {\bibfnamefont {M.}~\bibnamefont {Braden}}, \bibinfo {author} {\bibfnamefont {K.}~\bibnamefont {Nakajima}},\ and\ \bibinfo {author} {\bibfnamefont {J.~M.}\ \bibnamefont {Tranquada}},\ }\bibfield  {title} {\bibinfo {title} {High-energy spin dynamics in {${\mathrm{L}\mathrm{a}}_{1.69}{\mathrm{S}\mathrm{r}}_{0.31}{\mathrm{N}\mathrm{i}\mathrm{O}}_{4}$}},\ }\href {https://doi.org/10.1103/PhysRevLett.90.147202} {\bibfield  {journal} {\bibinfo  {journal} {Phys. Rev. Lett.}\ }\textbf {\bibinfo {volume} {90}},\ \bibinfo {pages} {147202} (\bibinfo {year} {2003})}\BibitemShut {NoStop}%
\bibitem [{\citenamefont {LaBollita}\ and\ \citenamefont {Botana}(2021)}]{Labollita2021electronic}%
  \BibitemOpen
  \bibfield  {author} {\bibinfo {author} {\bibfnamefont {H.}~\bibnamefont {LaBollita}}\ and\ \bibinfo {author} {\bibfnamefont {A.~S.}\ \bibnamefont {Botana}},\ }\bibfield  {title} {\bibinfo {title} {Electronic structure and magnetic properties of higher-order layered nickelates: {${\mathrm{La}}_{n+1}{\mathrm{Ni}}_{n}{\mathrm{O}}_{2n+2}$} ($n=4-6$)},\ }\href {https://doi.org/10.1103/PhysRevB.104.035148} {\bibfield  {journal} {\bibinfo  {journal} {Phys. Rev. B}\ }\textbf {\bibinfo {volume} {104}},\ \bibinfo {pages} {035148} (\bibinfo {year} {2021})}\BibitemShut {NoStop}%
\bibitem [{\citenamefont {LaBollita}\ and\ \citenamefont {Botana}(2022)}]{LaBollita2022correlated}%
  \BibitemOpen
  \bibfield  {author} {\bibinfo {author} {\bibfnamefont {H.}~\bibnamefont {LaBollita}}\ and\ \bibinfo {author} {\bibfnamefont {A.~S.}\ \bibnamefont {Botana}},\ }\bibfield  {title} {\bibinfo {title} {Correlated electronic structure of a quintuple-layer nickelate},\ }\href {https://doi.org/10.1103/PhysRevB.105.085118} {\bibfield  {journal} {\bibinfo  {journal} {Phys. Rev. B}\ }\textbf {\bibinfo {volume} {105}},\ \bibinfo {pages} {085118} (\bibinfo {year} {2022})}\BibitemShut {NoStop}%
\bibitem [{\citenamefont {Worm}\ \emph {et~al.}(2022)\citenamefont {Worm}, \citenamefont {Si}, \citenamefont {Kitatani}, \citenamefont {Arita}, \citenamefont {Tomczak},\ and\ \citenamefont {Held}}]{Worm2022correlations}%
  \BibitemOpen
  \bibfield  {author} {\bibinfo {author} {\bibfnamefont {P.}~\bibnamefont {Worm}}, \bibinfo {author} {\bibfnamefont {L.}~\bibnamefont {Si}}, \bibinfo {author} {\bibfnamefont {M.}~\bibnamefont {Kitatani}}, \bibinfo {author} {\bibfnamefont {R.}~\bibnamefont {Arita}}, \bibinfo {author} {\bibfnamefont {J.~M.}\ \bibnamefont {Tomczak}},\ and\ \bibinfo {author} {\bibfnamefont {K.}~\bibnamefont {Held}},\ }\bibfield  {title} {\bibinfo {title} {Correlations tune the electronic structure of pentalayer nickelates into the superconducting regime},\ }\href {https://doi.org/10.1103/PhysRevMaterials.6.L091801} {\bibfield  {journal} {\bibinfo  {journal} {Phys. Rev. Mater.}\ }\textbf {\bibinfo {volume} {6}},\ \bibinfo {pages} {L091801} (\bibinfo {year} {2022})}\BibitemShut {NoStop}%
\end{thebibliography}%


\begin{thebibliography}{17}%
\makeatletter
\providecommand \@ifxundefined [1]{%
 \@ifx{#1\undefined}
}%
\providecommand \@ifnum [1]{%
 \ifnum #1\expandafter \@firstoftwo
 \else \expandafter \@secondoftwo
 \fi
}%
\providecommand \@ifx [1]{%
 \ifx #1\expandafter \@firstoftwo
 \else \expandafter \@secondoftwo
 \fi
}%
\providecommand \natexlab [1]{#1}%
\providecommand \enquote  [1]{``#1''}%
\providecommand \bibnamefont  [1]{#1}%
\providecommand \bibfnamefont [1]{#1}%
\providecommand \citenamefont [1]{#1}%
\providecommand \href@noop [0]{\@secondoftwo}%
\providecommand \href [0]{\begingroup \@sanitize@url \@href}%
\providecommand \@href[1]{\@@startlink{#1}\@@href}%
\providecommand \@@href[1]{\endgroup#1\@@endlink}%
\providecommand \@sanitize@url [0]{\catcode `\\12\catcode `\$12\catcode `\&12\catcode `\#12\catcode `\^12\catcode `\_12\catcode `\%12\relax}%
\providecommand \@@startlink[1]{}%
\providecommand \@@endlink[0]{}%
\providecommand \url  [0]{\begingroup\@sanitize@url \@url }%
\providecommand \@url [1]{\endgroup\@href {#1}{\urlprefix }}%
\providecommand \urlprefix  [0]{URL }%
\providecommand \Eprint [0]{\href }%
\providecommand \doibase [0]{https://doi.org/}%
\providecommand \selectlanguage [0]{\@gobble}%
\providecommand \bibinfo  [0]{\@secondoftwo}%
\providecommand \bibfield  [0]{\@secondoftwo}%
\providecommand \translation [1]{[#1]}%
\providecommand \BibitemOpen [0]{}%
\providecommand \bibitemStop [0]{}%
\providecommand \bibitemNoStop [0]{.\EOS\space}%
\providecommand \EOS [0]{\spacefactor3000\relax}%
\providecommand \BibitemShut  [1]{\csname bibitem#1\endcsname}%
\let\auto@bib@innerbib\@empty
\bibitem [{\citenamefont {Pan}\ \emph {et~al.}(2022{\natexlab{a}})\citenamefont {Pan}, \citenamefont {Ferenc~Segedin}, \citenamefont {LaBollita}, \citenamefont {Song}, \citenamefont {Nica}, \citenamefont {Goodge}, \citenamefont {Pierce}, \citenamefont {Doyle}, \citenamefont {Novakov}, \citenamefont {C{\'o}rdova~Carrizales} \emph {et~al.}}]{pan2022superconductivity}%
  \BibitemOpen
  \bibfield  {author} {\bibinfo {author} {\bibfnamefont {G.~A.}\ \bibnamefont {Pan}}, \bibinfo {author} {\bibfnamefont {D.}~\bibnamefont {Ferenc~Segedin}}, \bibinfo {author} {\bibfnamefont {H.}~\bibnamefont {LaBollita}}, \bibinfo {author} {\bibfnamefont {Q.}~\bibnamefont {Song}}, \bibinfo {author} {\bibfnamefont {E.~M.}\ \bibnamefont {Nica}}, \bibinfo {author} {\bibfnamefont {B.~H.}\ \bibnamefont {Goodge}}, \bibinfo {author} {\bibfnamefont {A.~T.}\ \bibnamefont {Pierce}}, \bibinfo {author} {\bibfnamefont {S.}~\bibnamefont {Doyle}}, \bibinfo {author} {\bibfnamefont {S.}~\bibnamefont {Novakov}}, \bibinfo {author} {\bibfnamefont {D.}~\bibnamefont {C{\'o}rdova~Carrizales}}, \emph {et~al.},\ }\bibfield  {title} {\bibinfo {title} {Superconductivity in a quintuple-layer square-planar nickelate},\ }\href {https://www.nature.com/articles/s41563-021-01142-9} {\bibfield  {journal} {\bibinfo  {journal} {Nat. Mater.}\ }\textbf {\bibinfo {volume} {21}},\ \bibinfo {pages} {160} (\bibinfo {year}
  {2022}{\natexlab{a}})}\BibitemShut {NoStop}%
\bibitem [{\citenamefont {Pan}\ \emph {et~al.}(2022{\natexlab{b}})\citenamefont {Pan}, \citenamefont {Song}, \citenamefont {Ferenc~Segedin}, \citenamefont {Jung}, \citenamefont {El-Sherif}, \citenamefont {Fleck}, \citenamefont {Goodge}, \citenamefont {Doyle}, \citenamefont {C\'ordova~Carrizales}, \citenamefont {N'Diaye}, \citenamefont {Shafer}, \citenamefont {Paik}, \citenamefont {Kourkoutis}, \citenamefont {El~Baggari}, \citenamefont {Botana}, \citenamefont {Brooks},\ and\ \citenamefont {Mundy}}]{Pan2022synthesis}%
  \BibitemOpen
  \bibfield  {author} {\bibinfo {author} {\bibfnamefont {G.~A.}\ \bibnamefont {Pan}}, \bibinfo {author} {\bibfnamefont {Q.}~\bibnamefont {Song}}, \bibinfo {author} {\bibfnamefont {D.}~\bibnamefont {Ferenc~Segedin}}, \bibinfo {author} {\bibfnamefont {M.-C.}\ \bibnamefont {Jung}}, \bibinfo {author} {\bibfnamefont {H.}~\bibnamefont {El-Sherif}}, \bibinfo {author} {\bibfnamefont {E.~E.}\ \bibnamefont {Fleck}}, \bibinfo {author} {\bibfnamefont {B.~H.}\ \bibnamefont {Goodge}}, \bibinfo {author} {\bibfnamefont {S.}~\bibnamefont {Doyle}}, \bibinfo {author} {\bibfnamefont {D.}~\bibnamefont {C\'ordova~Carrizales}}, \bibinfo {author} {\bibfnamefont {A.~T.}\ \bibnamefont {N'Diaye}}, \bibinfo {author} {\bibfnamefont {P.}~\bibnamefont {Shafer}}, \bibinfo {author} {\bibfnamefont {H.}~\bibnamefont {Paik}}, \bibinfo {author} {\bibfnamefont {L.~F.}\ \bibnamefont {Kourkoutis}}, \bibinfo {author} {\bibfnamefont {I.}~\bibnamefont {El~Baggari}}, \bibinfo {author} {\bibfnamefont {A.~S.}\ \bibnamefont {Botana}}, \bibinfo {author}
  {\bibfnamefont {C.~M.}\ \bibnamefont {Brooks}},\ and\ \bibinfo {author} {\bibfnamefont {J.~A.}\ \bibnamefont {Mundy}},\ }\bibfield  {title} {\bibinfo {title} {Synthesis and electronic properties of {${\mathrm{Nd}}_{n+1}{\mathrm{Ni}}_{n}{\mathrm{O}}_{3n+1}$ Ruddlesden-Popper} nickelate thin films},\ }\href {https://doi.org/10.1103/PhysRevMaterials.6.055003} {\bibfield  {journal} {\bibinfo  {journal} {Phys. Rev. Mater.}\ }\textbf {\bibinfo {volume} {6}},\ \bibinfo {pages} {055003} (\bibinfo {year} {2022}{\natexlab{b}})}\BibitemShut {NoStop}%
\bibitem [{\citenamefont {Ferenc~Segedin}\ \emph {et~al.}(2023)\citenamefont {Ferenc~Segedin}, \citenamefont {Goodge}, \citenamefont {Pan}, \citenamefont {Song}, \citenamefont {LaBollita}, \citenamefont {Jung}, \citenamefont {El-Sherif}, \citenamefont {Doyle}, \citenamefont {Turkiewicz}, \citenamefont {Taylor} \emph {et~al.}}]{ferenc2023limits}%
  \BibitemOpen
  \bibfield  {author} {\bibinfo {author} {\bibfnamefont {D.}~\bibnamefont {Ferenc~Segedin}}, \bibinfo {author} {\bibfnamefont {B.~H.}\ \bibnamefont {Goodge}}, \bibinfo {author} {\bibfnamefont {G.~A.}\ \bibnamefont {Pan}}, \bibinfo {author} {\bibfnamefont {Q.}~\bibnamefont {Song}}, \bibinfo {author} {\bibfnamefont {H.}~\bibnamefont {LaBollita}}, \bibinfo {author} {\bibfnamefont {M.-C.}\ \bibnamefont {Jung}}, \bibinfo {author} {\bibfnamefont {H.}~\bibnamefont {El-Sherif}}, \bibinfo {author} {\bibfnamefont {S.}~\bibnamefont {Doyle}}, \bibinfo {author} {\bibfnamefont {A.}~\bibnamefont {Turkiewicz}}, \bibinfo {author} {\bibfnamefont {N.~K.}\ \bibnamefont {Taylor}}, \emph {et~al.},\ }\bibfield  {title} {\bibinfo {title} {Limits to the strain engineering of layered square-planar nickelate thin films},\ }\href {https://doi.org/10.1038/s41467-023-37117-4} {\bibfield  {journal} {\bibinfo  {journal} {Nat. Commun.}\ }\textbf {\bibinfo {volume} {14}},\ \bibinfo {pages} {1468} (\bibinfo {year} {2023})}\BibitemShut
  {NoStop}%
\bibitem [{\citenamefont {Takeda}\ \emph {et~al.}(1992)\citenamefont {Takeda}, \citenamefont {Nishijima}, \citenamefont {Imanishi}, \citenamefont {Kanno}, \citenamefont {Yamamoto},\ and\ \citenamefont {Takano}}]{takeda1992crystal}%
  \BibitemOpen
  \bibfield  {author} {\bibinfo {author} {\bibfnamefont {Y.}~\bibnamefont {Takeda}}, \bibinfo {author} {\bibfnamefont {M.}~\bibnamefont {Nishijima}}, \bibinfo {author} {\bibfnamefont {N.}~\bibnamefont {Imanishi}}, \bibinfo {author} {\bibfnamefont {R.}~\bibnamefont {Kanno}}, \bibinfo {author} {\bibfnamefont {O.}~\bibnamefont {Yamamoto}},\ and\ \bibinfo {author} {\bibfnamefont {M.}~\bibnamefont {Takano}},\ }\bibfield  {title} {\bibinfo {title} {Crystal chemistry and transport properties of {Nd$_{2-x}A_x$NiO$_4$} {($A=$ Ca, Sr, or Ba, $0 \leq x \leq 1.4$)}},\ }\href {https://doi.org/https://doi.org/10.1016/S0022-4596(05)80299-3} {\bibfield  {journal} {\bibinfo  {journal} {J. Solid State Chem.}\ }\textbf {\bibinfo {volume} {96}},\ \bibinfo {pages} {72} (\bibinfo {year} {1992})}\BibitemShut {NoStop}%
\bibitem [{\citenamefont {Chen}\ \emph {et~al.}(2024)\citenamefont {Chen}, \citenamefont {Choi}, \citenamefont {Jiang}, \citenamefont {Mei}, \citenamefont {Jiang}, \citenamefont {Li}, \citenamefont {Agrestini}, \citenamefont {Garcia-Fernandez}, \citenamefont {Sun}, \citenamefont {Huang}, \citenamefont {Shen}, \citenamefont {Wang}, \citenamefont {Hu}, \citenamefont {Lu}, \citenamefont {Zhou},\ and\ \citenamefont {Feng}}]{Chen2024}%
  \BibitemOpen
  \bibfield  {author} {\bibinfo {author} {\bibfnamefont {X.}~\bibnamefont {Chen}}, \bibinfo {author} {\bibfnamefont {J.}~\bibnamefont {Choi}}, \bibinfo {author} {\bibfnamefont {Z.}~\bibnamefont {Jiang}}, \bibinfo {author} {\bibfnamefont {J.}~\bibnamefont {Mei}}, \bibinfo {author} {\bibfnamefont {K.}~\bibnamefont {Jiang}}, \bibinfo {author} {\bibfnamefont {J.}~\bibnamefont {Li}}, \bibinfo {author} {\bibfnamefont {S.}~\bibnamefont {Agrestini}}, \bibinfo {author} {\bibfnamefont {M.}~\bibnamefont {Garcia-Fernandez}}, \bibinfo {author} {\bibfnamefont {H.}~\bibnamefont {Sun}}, \bibinfo {author} {\bibfnamefont {X.}~\bibnamefont {Huang}}, \bibinfo {author} {\bibfnamefont {D.}~\bibnamefont {Shen}}, \bibinfo {author} {\bibfnamefont {M.}~\bibnamefont {Wang}}, \bibinfo {author} {\bibfnamefont {J.}~\bibnamefont {Hu}}, \bibinfo {author} {\bibfnamefont {Y.}~\bibnamefont {Lu}}, \bibinfo {author} {\bibfnamefont {K.~J.}\ \bibnamefont {Zhou}},\ and\ \bibinfo {author} {\bibfnamefont {D.}~\bibnamefont {Feng}},\ }\bibfield
  {title} {\bibinfo {title} {Electronic and magnetic excitations in {La$_3$Ni$_2$O$_7$}},\ }\href {https://doi.org/10.1038/s41467-024-53863-5} {\bibfield  {journal} {\bibinfo  {journal} {Nat. Commun.}\ }\textbf {\bibinfo {volume} {15}},\ \bibinfo {pages} {9597} (\bibinfo {year} {2024})}\BibitemShut {NoStop}%
\bibitem [{\citenamefont {Gupta}\ \emph {et~al.}(2024)\citenamefont {Gupta}, \citenamefont {Gong}, \citenamefont {Wu}, \citenamefont {Kang}, \citenamefont {Parzyck}, \citenamefont {Gregory}, \citenamefont {Costa}, \citenamefont {Sutarto}, \citenamefont {Sarker}, \citenamefont {Singer}, \citenamefont {Schlom}, \citenamefont {Shen},\ and\ \citenamefont {Hawthorn}}]{gupta2024anisotropic}%
  \BibitemOpen
  \bibfield  {author} {\bibinfo {author} {\bibfnamefont {N.~K.}\ \bibnamefont {Gupta}}, \bibinfo {author} {\bibfnamefont {R.}~\bibnamefont {Gong}}, \bibinfo {author} {\bibfnamefont {Y.}~\bibnamefont {Wu}}, \bibinfo {author} {\bibfnamefont {M.}~\bibnamefont {Kang}}, \bibinfo {author} {\bibfnamefont {C.~T.}\ \bibnamefont {Parzyck}}, \bibinfo {author} {\bibfnamefont {B.~Z.}\ \bibnamefont {Gregory}}, \bibinfo {author} {\bibfnamefont {N.}~\bibnamefont {Costa}}, \bibinfo {author} {\bibfnamefont {R.}~\bibnamefont {Sutarto}}, \bibinfo {author} {\bibfnamefont {S.}~\bibnamefont {Sarker}}, \bibinfo {author} {\bibfnamefont {A.}~\bibnamefont {Singer}}, \bibinfo {author} {\bibfnamefont {D.~G.}\ \bibnamefont {Schlom}}, \bibinfo {author} {\bibfnamefont {K.~M.}\ \bibnamefont {Shen}},\ and\ \bibinfo {author} {\bibfnamefont {D.~G.}\ \bibnamefont {Hawthorn}},\ }\href {https://arxiv.org/abs/2409.03210} {\bibinfo {title} {Anisotropic spin stripe domains in bilayer {La$_3$Ni$_2$O$_7$}}} (\bibinfo {year} {2024}),\ \Eprint
  {https://arxiv.org/abs/2409.03210} {arXiv:2409.03210} \BibitemShut {NoStop}%
\bibitem [{\citenamefont {Ren}\ \emph {et~al.}(2025)\citenamefont {Ren}, \citenamefont {Sutarto}, \citenamefont {Wu}, \citenamefont {Zhang}, \citenamefont {Huang}, \citenamefont {Xiang}, \citenamefont {Hu}, \citenamefont {Comin}, \citenamefont {Zhou},\ and\ \citenamefont {Zhu}}]{ren2024resolving}%
  \BibitemOpen
  \bibfield  {author} {\bibinfo {author} {\bibfnamefont {X.}~\bibnamefont {Ren}}, \bibinfo {author} {\bibfnamefont {R.}~\bibnamefont {Sutarto}}, \bibinfo {author} {\bibfnamefont {X.}~\bibnamefont {Wu}}, \bibinfo {author} {\bibfnamefont {J.}~\bibnamefont {Zhang}}, \bibinfo {author} {\bibfnamefont {H.}~\bibnamefont {Huang}}, \bibinfo {author} {\bibfnamefont {T.}~\bibnamefont {Xiang}}, \bibinfo {author} {\bibfnamefont {J.}~\bibnamefont {Hu}}, \bibinfo {author} {\bibfnamefont {R.}~\bibnamefont {Comin}}, \bibinfo {author} {\bibfnamefont {X.~J.}\ \bibnamefont {Zhou}},\ and\ \bibinfo {author} {\bibfnamefont {Z.}~\bibnamefont {Zhu}},\ }\bibfield  {title} {\bibinfo {title} {Resolving the electronic ground state of {La$_3$Ni$_2$O$_{7-\delta}$} films},\ }\href {https://doi.org/https://doi.org/10.1038/s42005-025-01971-z} {\bibfield  {journal} {\bibinfo  {journal} {Commun Phys}\ }\textbf {\bibinfo {volume} {8}},\ \bibinfo {pages} {52} (\bibinfo {year} {2025})}\BibitemShut {NoStop}%
\bibitem [{\citenamefont {Zhang}\ \emph {et~al.}(2020)\citenamefont {Zhang}, \citenamefont {Phelan}, \citenamefont {Botana}, \citenamefont {Chen}, \citenamefont {Zheng}, \citenamefont {Krogstad}, \citenamefont {Wang}, \citenamefont {Qiu}, \citenamefont {Rodriguez-Rivera}, \citenamefont {Osborn}, \citenamefont {Rosenkranz}, \citenamefont {Norman},\ and\ \citenamefont {Mitchell}}]{zhang2020intertwined}%
  \BibitemOpen
  \bibfield  {author} {\bibinfo {author} {\bibfnamefont {J.}~\bibnamefont {Zhang}}, \bibinfo {author} {\bibfnamefont {D.}~\bibnamefont {Phelan}}, \bibinfo {author} {\bibfnamefont {A.~S.}\ \bibnamefont {Botana}}, \bibinfo {author} {\bibfnamefont {Y.-S.}\ \bibnamefont {Chen}}, \bibinfo {author} {\bibfnamefont {H.}~\bibnamefont {Zheng}}, \bibinfo {author} {\bibfnamefont {M.}~\bibnamefont {Krogstad}}, \bibinfo {author} {\bibfnamefont {S.~G.}\ \bibnamefont {Wang}}, \bibinfo {author} {\bibfnamefont {Y.}~\bibnamefont {Qiu}}, \bibinfo {author} {\bibfnamefont {J.}~\bibnamefont {Rodriguez-Rivera}}, \bibinfo {author} {\bibfnamefont {R.}~\bibnamefont {Osborn}}, \bibinfo {author} {\bibfnamefont {S.}~\bibnamefont {Rosenkranz}}, \bibinfo {author} {\bibfnamefont {M.~R.}\ \bibnamefont {Norman}},\ and\ \bibinfo {author} {\bibfnamefont {J.~F.}\ \bibnamefont {Mitchell}},\ }\bibfield  {title} {\bibinfo {title} {Intertwined density waves in a metallic nickelate},\ }\href {https://www.nature.com/articles/s41467-020-19836-0}
  {\bibfield  {journal} {\bibinfo  {journal} {Nat. Commun.}\ }\textbf {\bibinfo {volume} {11}},\ \bibinfo {pages} {6003} (\bibinfo {year} {2020})}\BibitemShut {NoStop}%
\bibitem [{\citenamefont {Catalano}\ \emph {et~al.}(2018)\citenamefont {Catalano}, \citenamefont {Gibert}, \citenamefont {Fowlie}, \citenamefont {Iniguez}, \citenamefont {Triscone},\ and\ \citenamefont {Kreisel}}]{catalano2018rare}%
  \BibitemOpen
  \bibfield  {author} {\bibinfo {author} {\bibfnamefont {S.}~\bibnamefont {Catalano}}, \bibinfo {author} {\bibfnamefont {M.}~\bibnamefont {Gibert}}, \bibinfo {author} {\bibfnamefont {J.}~\bibnamefont {Fowlie}}, \bibinfo {author} {\bibfnamefont {J.}~\bibnamefont {Iniguez}}, \bibinfo {author} {\bibfnamefont {J.-M.}\ \bibnamefont {Triscone}},\ and\ \bibinfo {author} {\bibfnamefont {J.}~\bibnamefont {Kreisel}},\ }\bibfield  {title} {\bibinfo {title} {Rare-earth nickelates {$R$NiO$_3$}: thin films and heterostructures},\ }\href {https://iopscience.iop.org/article/10.1088/1361-6633/aaa37a} {\bibfield  {journal} {\bibinfo  {journal} {Rep. Prog. Phys.}\ }\textbf {\bibinfo {volume} {81}},\ \bibinfo {pages} {046501} (\bibinfo {year} {2018})}\BibitemShut {NoStop}%
\bibitem [{\citenamefont {Bhatt}\ \emph {et~al.}(2025)\citenamefont {Bhatt}, \citenamefont {Jiang}, \citenamefont {Ko}, \citenamefont {Schnitzer}, \citenamefont {Pan}, \citenamefont {Segedin}, \citenamefont {Liu}, \citenamefont {Yu}, \citenamefont {Zhao}, \citenamefont {Morales} \emph {et~al.}}]{bhatt2025resolving}%
  \BibitemOpen
  \bibfield  {author} {\bibinfo {author} {\bibfnamefont {L.}~\bibnamefont {Bhatt}}, \bibinfo {author} {\bibfnamefont {A.~Y.}\ \bibnamefont {Jiang}}, \bibinfo {author} {\bibfnamefont {E.~K.}\ \bibnamefont {Ko}}, \bibinfo {author} {\bibfnamefont {N.}~\bibnamefont {Schnitzer}}, \bibinfo {author} {\bibfnamefont {G.~A.}\ \bibnamefont {Pan}}, \bibinfo {author} {\bibfnamefont {D.~F.}\ \bibnamefont {Segedin}}, \bibinfo {author} {\bibfnamefont {Y.}~\bibnamefont {Liu}}, \bibinfo {author} {\bibfnamefont {Y.}~\bibnamefont {Yu}}, \bibinfo {author} {\bibfnamefont {Y.-F.}\ \bibnamefont {Zhao}}, \bibinfo {author} {\bibfnamefont {E.~A.}\ \bibnamefont {Morales}}, \emph {et~al.},\ }\href {https://arxiv.org/abs/2501.08204} {\bibinfo {title} {Resolving structural origins for superconductivity in strain-engineered {La$_3$Ni$_2$O$_7$} thin films}} (\bibinfo {year} {2025})\BibitemShut {NoStop}%
\bibitem [{\citenamefont {Mitrano}\ \emph {et~al.}(2024)\citenamefont {Mitrano}, \citenamefont {Johnston}, \citenamefont {Kim},\ and\ \citenamefont {Dean}}]{Mitrano2024exploring}%
  \BibitemOpen
  \bibfield  {author} {\bibinfo {author} {\bibfnamefont {M.}~\bibnamefont {Mitrano}}, \bibinfo {author} {\bibfnamefont {S.}~\bibnamefont {Johnston}}, \bibinfo {author} {\bibfnamefont {Y.-J.}\ \bibnamefont {Kim}},\ and\ \bibinfo {author} {\bibfnamefont {M.~P.~M.}\ \bibnamefont {Dean}},\ }\bibfield  {title} {\bibinfo {title} {Exploring quantum materials with resonant inelastic x-ray scattering},\ }\href {https://doi.org/10.1103/PhysRevX.14.040501} {\bibfield  {journal} {\bibinfo  {journal} {Phys. Rev. X}\ }\textbf {\bibinfo {volume} {14}},\ \bibinfo {pages} {040501} (\bibinfo {year} {2024})}\BibitemShut {NoStop}%
\bibitem [{\citenamefont {Norman}\ \emph {et~al.}(2023)\citenamefont {Norman}, \citenamefont {Botana}, \citenamefont {Karp}, \citenamefont {Hampel}, \citenamefont {Labollita}, \citenamefont {Millis}, \citenamefont {Fabbris}, \citenamefont {Shen},\ and\ \citenamefont {Dean}}]{Norman2023}%
  \BibitemOpen
  \bibfield  {author} {\bibinfo {author} {\bibfnamefont {M.~R.}\ \bibnamefont {Norman}}, \bibinfo {author} {\bibfnamefont {A.~S.}\ \bibnamefont {Botana}}, \bibinfo {author} {\bibfnamefont {J.}~\bibnamefont {Karp}}, \bibinfo {author} {\bibfnamefont {A.}~\bibnamefont {Hampel}}, \bibinfo {author} {\bibfnamefont {H.}~\bibnamefont {Labollita}}, \bibinfo {author} {\bibfnamefont {A.~J.}\ \bibnamefont {Millis}}, \bibinfo {author} {\bibfnamefont {G.}~\bibnamefont {Fabbris}}, \bibinfo {author} {\bibfnamefont {Y.}~\bibnamefont {Shen}},\ and\ \bibinfo {author} {\bibfnamefont {M.~P.~M.}\ \bibnamefont {Dean}},\ }\bibfield  {title} {\bibinfo {title} {Orbital polarization, charge transfer, and fluorescence in reduced-valence nickelates},\ }\href {https://doi.org/10.1103/PhysRevB.107.165124} {\bibfield  {journal} {\bibinfo  {journal} {Phys. Rev. B}\ }\textbf {\bibinfo {volume} {107}},\ \bibinfo {pages} {165124} (\bibinfo {year} {2023})}\BibitemShut {NoStop}%
\bibitem [{\citenamefont {Shen}\ \emph {et~al.}(2023)\citenamefont {Shen}, \citenamefont {Sears}, \citenamefont {Fabbris}, \citenamefont {Li}, \citenamefont {Pelliciari}, \citenamefont {Mitrano}, \citenamefont {He}, \citenamefont {Zhang}, \citenamefont {Mitchell}, \citenamefont {Bisogni}, \citenamefont {Norman}, \citenamefont {Johnston},\ and\ \citenamefont {Dean}}]{Shen2023electronic}%
  \BibitemOpen
  \bibfield  {author} {\bibinfo {author} {\bibfnamefont {Y.}~\bibnamefont {Shen}}, \bibinfo {author} {\bibfnamefont {J.}~\bibnamefont {Sears}}, \bibinfo {author} {\bibfnamefont {G.}~\bibnamefont {Fabbris}}, \bibinfo {author} {\bibfnamefont {J.}~\bibnamefont {Li}}, \bibinfo {author} {\bibfnamefont {J.}~\bibnamefont {Pelliciari}}, \bibinfo {author} {\bibfnamefont {M.}~\bibnamefont {Mitrano}}, \bibinfo {author} {\bibfnamefont {W.}~\bibnamefont {He}}, \bibinfo {author} {\bibfnamefont {J.}~\bibnamefont {Zhang}}, \bibinfo {author} {\bibfnamefont {J.~F.}\ \bibnamefont {Mitchell}}, \bibinfo {author} {\bibfnamefont {V.}~\bibnamefont {Bisogni}}, \bibinfo {author} {\bibfnamefont {M.~R.}\ \bibnamefont {Norman}}, \bibinfo {author} {\bibfnamefont {S.}~\bibnamefont {Johnston}},\ and\ \bibinfo {author} {\bibfnamefont {M.~P.~M.}\ \bibnamefont {Dean}},\ }\bibfield  {title} {\bibinfo {title} {Electronic character of charge order in square-planar low-valence nickelates},\ }\href {https://doi.org/10.1103/PhysRevX.13.011021}
  {\bibfield  {journal} {\bibinfo  {journal} {Phys. Rev. X}\ }\textbf {\bibinfo {volume} {13}},\ \bibinfo {pages} {011021} (\bibinfo {year} {2023})}\BibitemShut {NoStop}%
\bibitem [{\citenamefont {Dahlbom}\ \emph {et~al.}(2025)\citenamefont {Dahlbom}, \citenamefont {Zhang}, \citenamefont {Miles}, \citenamefont {Quinn}, \citenamefont {Niraula}, \citenamefont {Thipe}, \citenamefont {Wilson}, \citenamefont {Matin}, \citenamefont {Mankad}, \citenamefont {Hahn}, \citenamefont {Pajerowski}, \citenamefont {Johnston}, \citenamefont {Wang}, \citenamefont {Lane}, \citenamefont {Li}, \citenamefont {Bai}, \citenamefont {Mourigal}, \citenamefont {Batista},\ and\ \citenamefont {Barros}}]{Sunny}%
  \BibitemOpen
  \bibfield  {author} {\bibinfo {author} {\bibfnamefont {D.}~\bibnamefont {Dahlbom}}, \bibinfo {author} {\bibfnamefont {H.}~\bibnamefont {Zhang}}, \bibinfo {author} {\bibfnamefont {C.}~\bibnamefont {Miles}}, \bibinfo {author} {\bibfnamefont {S.}~\bibnamefont {Quinn}}, \bibinfo {author} {\bibfnamefont {A.}~\bibnamefont {Niraula}}, \bibinfo {author} {\bibfnamefont {B.}~\bibnamefont {Thipe}}, \bibinfo {author} {\bibfnamefont {M.}~\bibnamefont {Wilson}}, \bibinfo {author} {\bibfnamefont {S.}~\bibnamefont {Matin}}, \bibinfo {author} {\bibfnamefont {H.}~\bibnamefont {Mankad}}, \bibinfo {author} {\bibfnamefont {S.}~\bibnamefont {Hahn}}, \bibinfo {author} {\bibfnamefont {D.}~\bibnamefont {Pajerowski}}, \bibinfo {author} {\bibfnamefont {S.}~\bibnamefont {Johnston}}, \bibinfo {author} {\bibfnamefont {Z.}~\bibnamefont {Wang}}, \bibinfo {author} {\bibfnamefont {H.}~\bibnamefont {Lane}}, \bibinfo {author} {\bibfnamefont {Y.~W.}\ \bibnamefont {Li}}, \bibinfo {author} {\bibfnamefont {X.}~\bibnamefont {Bai}}, \bibinfo
  {author} {\bibfnamefont {M.}~\bibnamefont {Mourigal}}, \bibinfo {author} {\bibfnamefont {C.~D.}\ \bibnamefont {Batista}},\ and\ \bibinfo {author} {\bibfnamefont {K.}~\bibnamefont {Barros}},\ }\href {https://arxiv.org/abs/2501.13095} {\bibinfo {title} {Sunny.jl: A julia package for spin dynamics}} (\bibinfo {year} {2025}),\ \Eprint {https://arxiv.org/abs/2501.13095} {arXiv:2501.13095} \BibitemShut {NoStop}%
\bibitem [{\citenamefont {Li}\ \emph {et~al.}(2021)\citenamefont {Li}, \citenamefont {He}, \citenamefont {Zhu}, \citenamefont {Si}, \citenamefont {Fan},\ and\ \citenamefont {Wen}}]{Li2020contrasting}%
  \BibitemOpen
  \bibfield  {author} {\bibinfo {author} {\bibfnamefont {Q.}~\bibnamefont {Li}}, \bibinfo {author} {\bibfnamefont {C.}~\bibnamefont {He}}, \bibinfo {author} {\bibfnamefont {X.}~\bibnamefont {Zhu}}, \bibinfo {author} {\bibfnamefont {J.}~\bibnamefont {Si}}, \bibinfo {author} {\bibfnamefont {X.}~\bibnamefont {Fan}},\ and\ \bibinfo {author} {\bibfnamefont {H.-H.}\ \bibnamefont {Wen}},\ }\bibfield  {title} {\bibinfo {title} {Contrasting physical properties of the trilayer nickelates {Nd$_4$Ni$_3$O$_{10}$} and {Nd$_4$Ni$_3$O$_8$}},\ }\href {https://doi.org/10.1007/s11433-020-1613-3} {\bibfield  {journal} {\bibinfo  {journal} {Sci. China Phys. Mech. Astron.}\ }\textbf {\bibinfo {volume} {64}},\ \bibinfo {pages} {227411} (\bibinfo {year} {2021})}\BibitemShut {NoStop}%
\bibitem [{\citenamefont {F\"ursich}\ \emph {et~al.}(2019)\citenamefont {F\"ursich}, \citenamefont {Lu}, \citenamefont {Betto}, \citenamefont {Bluschke}, \citenamefont {Porras}, \citenamefont {Schierle}, \citenamefont {Ortiz}, \citenamefont {Suzuki}, \citenamefont {Cristiani}, \citenamefont {Logvenov}, \citenamefont {Brookes}, \citenamefont {Haverkort}, \citenamefont {Le~Tacon}, \citenamefont {Benckiser}, \citenamefont {Minola},\ and\ \citenamefont {Keimer}}]{Fursich2019resonant}%
  \BibitemOpen
  \bibfield  {author} {\bibinfo {author} {\bibfnamefont {K.}~\bibnamefont {F\"ursich}}, \bibinfo {author} {\bibfnamefont {Y.}~\bibnamefont {Lu}}, \bibinfo {author} {\bibfnamefont {D.}~\bibnamefont {Betto}}, \bibinfo {author} {\bibfnamefont {M.}~\bibnamefont {Bluschke}}, \bibinfo {author} {\bibfnamefont {J.}~\bibnamefont {Porras}}, \bibinfo {author} {\bibfnamefont {E.}~\bibnamefont {Schierle}}, \bibinfo {author} {\bibfnamefont {R.}~\bibnamefont {Ortiz}}, \bibinfo {author} {\bibfnamefont {H.}~\bibnamefont {Suzuki}}, \bibinfo {author} {\bibfnamefont {G.}~\bibnamefont {Cristiani}}, \bibinfo {author} {\bibfnamefont {G.}~\bibnamefont {Logvenov}}, \bibinfo {author} {\bibfnamefont {N.~B.}\ \bibnamefont {Brookes}}, \bibinfo {author} {\bibfnamefont {M.~W.}\ \bibnamefont {Haverkort}}, \bibinfo {author} {\bibfnamefont {M.}~\bibnamefont {Le~Tacon}}, \bibinfo {author} {\bibfnamefont {E.}~\bibnamefont {Benckiser}}, \bibinfo {author} {\bibfnamefont {M.}~\bibnamefont {Minola}},\ and\ \bibinfo {author} {\bibfnamefont
  {B.}~\bibnamefont {Keimer}},\ }\bibfield  {title} {\bibinfo {title} {Resonant inelastic x-ray scattering study of bond order and spin excitations in nickelate thin-film structures},\ }\href {https://doi.org/10.1103/PhysRevB.99.165124} {\bibfield  {journal} {\bibinfo  {journal} {Phys. Rev. B}\ }\textbf {\bibinfo {volume} {99}},\ \bibinfo {pages} {165124} (\bibinfo {year} {2019})}\BibitemShut {NoStop}%
\bibitem [{\citenamefont {Gao}\ \emph {et~al.}(2024)\citenamefont {Gao}, \citenamefont {Fan}, \citenamefont {Wang}, \citenamefont {Li}, \citenamefont {Ren}, \citenamefont {Biało}, \citenamefont {Drewanowski}, \citenamefont {Rothenbühler}, \citenamefont {Choi}, \citenamefont {Sutarto}, \citenamefont {Wang}, \citenamefont {Xiang}, \citenamefont {Hu}, \citenamefont {Zhou}, \citenamefont {Bisogni}, \citenamefont {Comin}, \citenamefont {Chang}, \citenamefont {Pelliciari}, \citenamefont {Zhou},\ and\ \citenamefont {Zhu}}]{Gao2024magnetic}%
  \BibitemOpen
  \bibfield  {author} {\bibinfo {author} {\bibfnamefont {Q.}~\bibnamefont {Gao}}, \bibinfo {author} {\bibfnamefont {S.}~\bibnamefont {Fan}}, \bibinfo {author} {\bibfnamefont {Q.}~\bibnamefont {Wang}}, \bibinfo {author} {\bibfnamefont {J.}~\bibnamefont {Li}}, \bibinfo {author} {\bibfnamefont {X.}~\bibnamefont {Ren}}, \bibinfo {author} {\bibfnamefont {I.}~\bibnamefont {Biało}}, \bibinfo {author} {\bibfnamefont {A.}~\bibnamefont {Drewanowski}}, \bibinfo {author} {\bibfnamefont {P.}~\bibnamefont {Rothenbühler}}, \bibinfo {author} {\bibfnamefont {J.}~\bibnamefont {Choi}}, \bibinfo {author} {\bibfnamefont {R.}~\bibnamefont {Sutarto}}, \bibinfo {author} {\bibfnamefont {Y.}~\bibnamefont {Wang}}, \bibinfo {author} {\bibfnamefont {T.}~\bibnamefont {Xiang}}, \bibinfo {author} {\bibfnamefont {J.}~\bibnamefont {Hu}}, \bibinfo {author} {\bibfnamefont {K.~J.}\ \bibnamefont {Zhou}}, \bibinfo {author} {\bibfnamefont {V.}~\bibnamefont {Bisogni}}, \bibinfo {author} {\bibfnamefont {R.}~\bibnamefont {Comin}}, \bibinfo {author}
  {\bibfnamefont {J.}~\bibnamefont {Chang}}, \bibinfo {author} {\bibfnamefont {J.}~\bibnamefont {Pelliciari}}, \bibinfo {author} {\bibfnamefont {X.~J.}\ \bibnamefont {Zhou}},\ and\ \bibinfo {author} {\bibfnamefont {Z.}~\bibnamefont {Zhu}},\ }\bibfield  {title} {\bibinfo {title} {Magnetic excitations in strained infinite-layer nickelate {PrNiO$_2$} films},\ }\href {https://doi.org/10.1038/s41467-024-49940-4} {\bibfield  {journal} {\bibinfo  {journal} {Nat. Commun.}\ }\textbf {\bibinfo {volume} {15}},\ \bibinfo {pages} {5576} (\bibinfo {year} {2024})}\BibitemShut {NoStop}%
\end{thebibliography}%
\end{document}


\title{Supplemental Material for: Magnetic excitations in Nd$_{n+1}$Ni$_{n}$O$_{3n+1}$ Ruddlesden-Popper nickelates observed via resonant inelastic x-ray scattering}

\date{\today}
\author{Sophia F. R. TenHuisen}
\affiliation{Department of Physics, Harvard University, Cambridge, MA, USA}
\affiliation{John A. Paulson School of Engineering and Applied Sciences, Harvard University, Cambridge, MA, USA}
\author{Grace A. Pan}
\affiliation{Department of Physics, Harvard University, Cambridge, MA, USA}
\author{Qi Song}
\affiliation{Department of Physics, Harvard University, Cambridge, MA, USA}
\author{Denitsa R. Baykusheva}
\affiliation{Department of Physics, Harvard University, Cambridge, MA, USA}
\email{Present address: Institute of Science and Technology Austria, Klosterneuburg, Austria}

\author{Dan Ferenc Segedin}
\affiliation{Department of Physics, Harvard University, Cambridge, MA, USA}

\author{Berit H. Goodge}
\affiliation{School of Applied and Engineering Physics, Cornell University, Ithaca, NY, USA}
\affiliation{Kavli Institute at Cornell for Nanoscale Science, Cornell University, Ithaca, NY, USA}

\author{Hanjong Paik}
\affiliation{Platform for the Accelerated Realization, Analysis and Discovery of Interface Materials (PARADIM), Cornell University, Ithaca, NY, USA}

\author{Jonathan Pelliciari}
\affiliation{National Synchrotron Light Source II, Brookhaven National Laboratory, Upton, New York 11973, USA}
\author{Valentina Bisogni}
\affiliation{National Synchrotron Light Source II, Brookhaven National Laboratory, Upton, New York 11973, USA}

\author{Yanhong Gu}
\affiliation{National Synchrotron Light Source II, Brookhaven National Laboratory, Upton, New York 11973, USA}

\author{Stefano Agrestini}
\affiliation{Diamond Light Source, Harwell Campus, Didcot OX11 0DE, UK}
\author{Abhishek Nag}
\affiliation{Diamond Light Source, Harwell Campus, Didcot OX11 0DE, UK}
\author{Mirian Garc\'{i}a-Fern\'{a}ndez}
\affiliation{Diamond Light Source, Harwell Campus, Didcot OX11 0DE, UK}
\author{Ke-Jin Zhou}
\affiliation{Diamond Light Source, Harwell Campus, Didcot OX11 0DE, UK}

\author{Lena F. Kourkoutis}
\thanks{Deceased}
\affiliation{School of Applied and Engineering Physics, Cornell University, Ithaca, NY, USA}
\affiliation{Kavli Institute at Cornell for Nanoscale Science, Cornell University, Ithaca, NY, USA}
\author{Charles M. Brooks}
\affiliation{Department of Physics, Harvard University, Cambridge, MA, USA}
\author{Julia A. Mundy}
\affiliation{Department of Physics, Harvard University, Cambridge, MA, USA}
\author{Mark P. M. Dean}
\thanks{mdean@bnl.gov}
\affiliation{Condensed Matter Physics and Materials Science Department, Brookhaven National Laboratory, Upton, New York 11973, USA}
\author{Matteo Mitrano}
\thanks{mmitrano@g.harvard.edu}
\affiliation{Department of Physics, Harvard University, Cambridge, MA, USA}

\maketitle

\section{Sample characterization}\label{sec:characterization}

\noindent Thin film Nd$_{n+1}$Ni$_n$O$_{3n+1}$ samples on (001)-oriented LaAlO$_3$ and (110)-oriented NdGaO$_3$ were synthesized using reactive-oxide molecular-beam epitaxy according to the procedures described in Refs.~\cite{pan2022superconductivity,Pan2022synthesis}. The synthesis process uses distilled ozone to reach a total chamber pressure of $\sim$1-3 $\times 10^{-6}$ and source materials of elemental neodymium and nickel evaporating with flux rates of $\sim$1 $\times 10^{13}$ atoms/cm$^2 \cdot$ s. The $n = 3, 5$ compounds were synthesized with a substrate temperature of 660 - 690 \textdegree C and the $n = 1$ compound at 1000 \textdegree C.

\subsection{X-ray diffraction}

\begin{figure}
    \centering
    \includegraphics[width=0.75\linewidth]{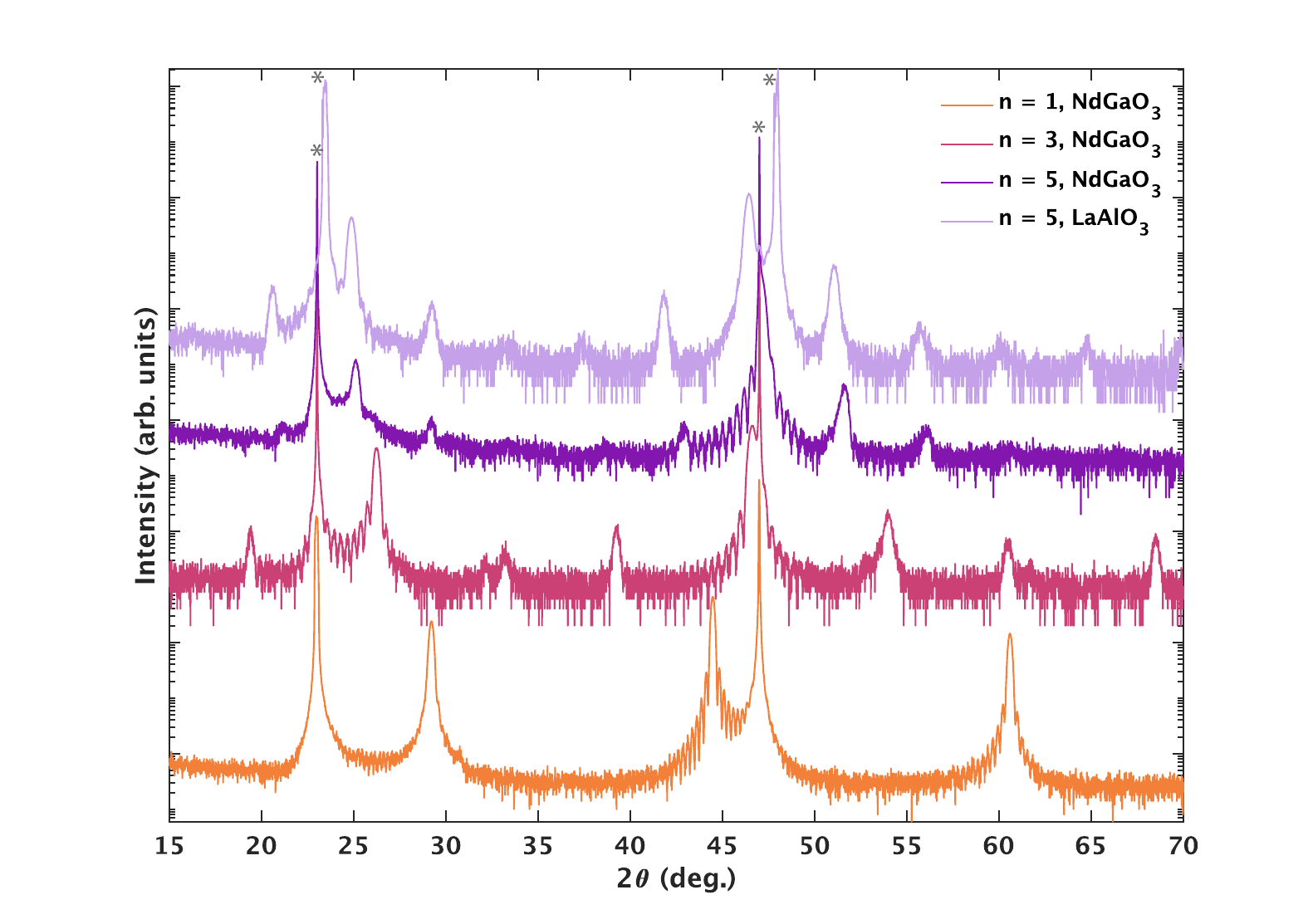}
\caption{X-ray diffraction patterns of the four Nd$_{n+1}$Ni$_n$O$_{3n+1}$ thin film samples: $n=1$ on NdGaO$_3$ (orange), $n=3$ on NdGaO$_3$ (dark pink), $n=5$ on NdGaO$_3$ (dark purple), and $n=5$ on LaAlO$_3$ (light purple). Grey asterisks indicate substrate peaks. The $n=3$ on NdGaO$_3$ and $n=5$ on NdGaO$_3$ patterns are reproduced from Ref.~\cite{pan2022superconductivity}.}
    \label{fig:xrd}
\end{figure}

\begin{figure}
    \centering
    \includegraphics[width=1\linewidth]{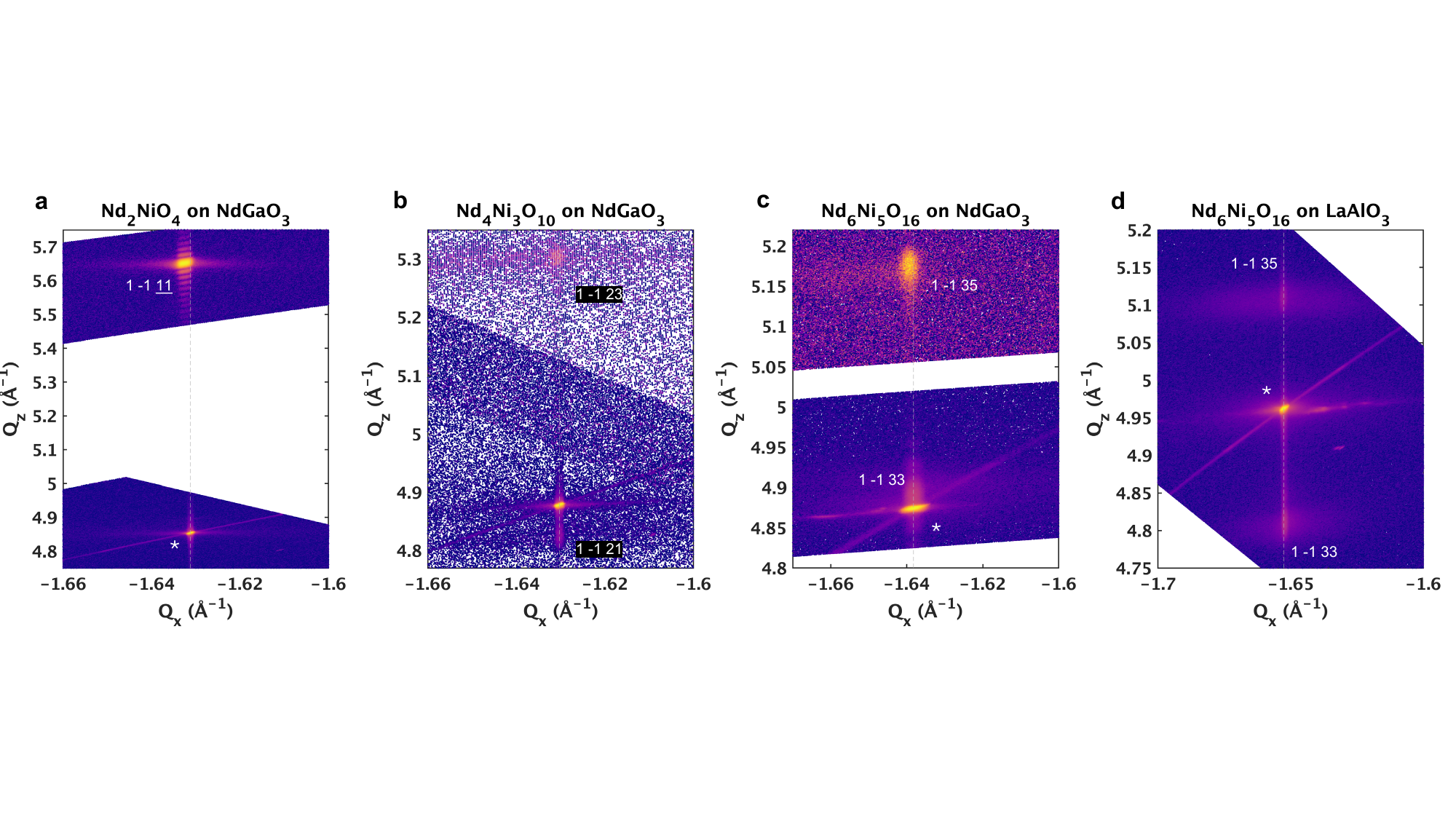}
    \caption{Reciprocal space mapping of the four Nd$_{n+1}$Ni$_n$O$_{3n+1}$ thin film samples.  White asterisks indicate the 3 3 -2 peaks of NdGaO$_3$ and -1 0 3 peak of LaAlO$_3$.  Film peaks are labeled directly on the maps.  Grey dashed lines intersect the substrate peaks and are guides to the eye.}
    \label{fig:rsm}
\end{figure}

\noindent Thin film x-ray diffraction was performed on a Malvern Panalytical Empyrean diffractometer using Cu K$\alpha_1$ radiation.  Figure~\ref{fig:xrd} displays the x-ray diffraction patterns for the four films considered in this study.  All films exhibit sharp superlattice peaks, demonstrating the long-ranged structural order of the Ruddlesden-Popper motif.  Reciprocal space maps (RSMs)  of the compounds on LaAlO$_3$ ($a_{pc} = 3.79$ \AA, $\epsilon \approx -0.9\%$) and NdGaO$_3$ ($a_{pc} = 3.86$ \AA, $\epsilon \approx +1.0\%$), shown in Fig.~\ref{fig:rsm}, indicate that the samples in this study are largely strained to their substrates.  The $n = 1$ compound has partially relaxed to an intermediate strain state with $a_{pc} \approx 3.835$ \AA.  This could arise from the increased presence of rock salt layers, which may relieve the strain generated by external epitaxy \cite{ferenc2023limits}.

\subsection{Electron microscopy}

\noindent Specimens for cross-sectional scanning transmission electron microscopy (STEM) analysis were prepared using either a Thermo Fisher Scientific Helios G4 UX or an FEI Helios 660 focused ion beam (FIB), with a final thinning process performed at an energy of 2 keV. Aberration-corrected STEM imaging was conducted on a Thermo Fisher Scientific Spectra 300 X-FEG or Titan Themis CryoS/TEM microscope operating at 300 kV, using a probe convergence semi-angle of 30~mrad. High-angle annular dark-field (HAADF) STEM images were captured using collection angles ranging from 54 to 200~mrad.

\begin{figure}
    \centering
    \includegraphics[width=.55\linewidth]{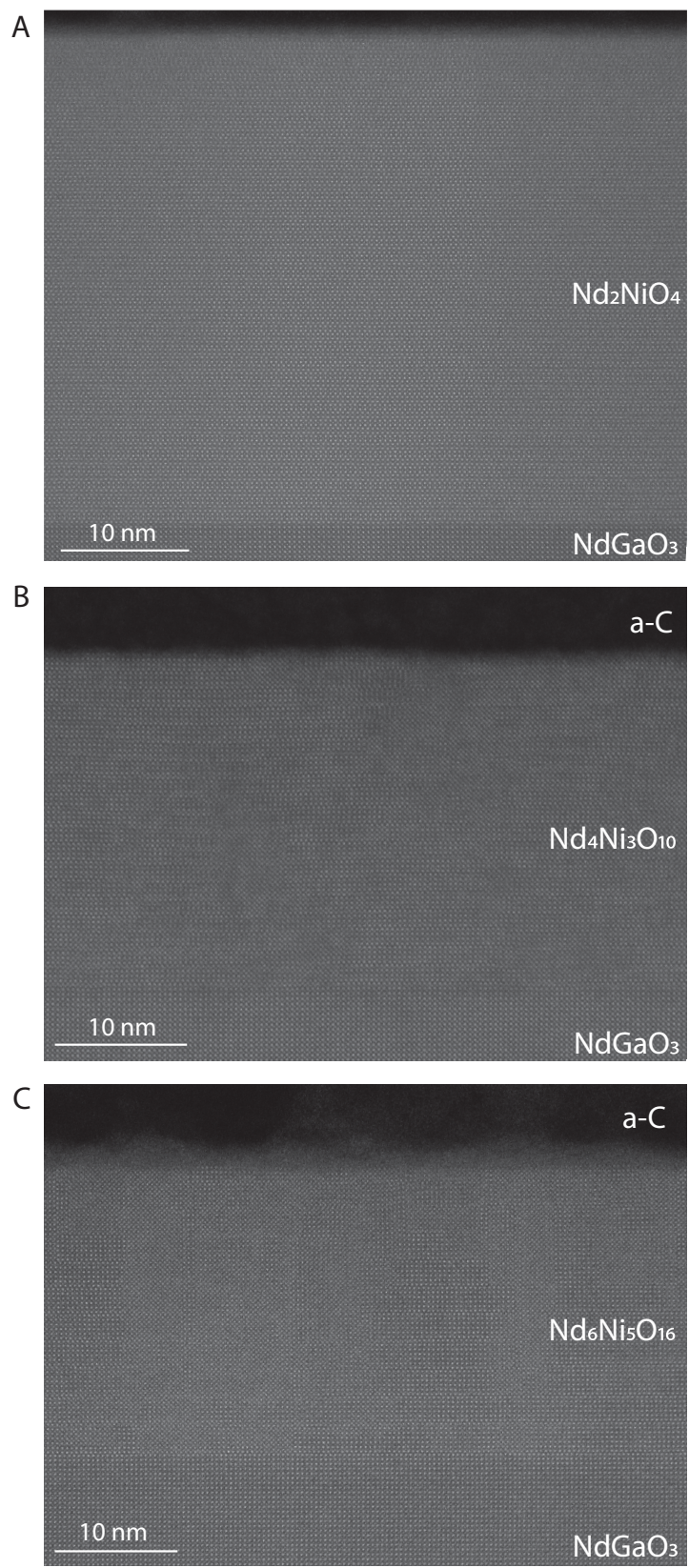}
\caption{Representative HAADF-STEM images of the (A) $n = 1$, (B) $n = 3$, and (C) $n = 5$ Nd$_{n+1}$Ni$_n$O$_{3n+1}$ thin films synthesized on NdGaO$_3$ substrates. a-C, amorphous carbon.}
    \label{fig:stem}
\end{figure}

Fig. \ref{fig:stem} shows regions of the Nd$_{n+1}$Ni$_n$O$_{3n+1}$ ($n = 1, 3, 5$) thin films synthesized on NdGaO$_3$.  The films have well-ordered, horizontally-stacked rock salt layers characteristic of the layered Ruddlesden-Popper phase.  There are some regions of reduced contrast in the $n = 3, 5$ films: rock salt regions which are offset by $a$/2[110]$_{pc}$ will show up as areas of reduced atomic contrast due to the projection through mixed elemental species (neodymium, nickel) into the page.  There are also some regions of locally varying $n$ in the $n = 5$ film, as the Nd$_{6}$Ni$_5$O$_{16}$ compound is not synthesizable through bulk methods alone.  Additional details of the microstructural characterization of these and related Nd$_{n+1}$Ni$_n$O$_{3n+1}$ thin films may be found in Refs. \cite{Pan2022synthesis,ferenc2023limits}.  

\subsection{Electronic transport}\label{sec:transport}

\begin{figure}
    \centering
    \includegraphics[width=0.9\linewidth]{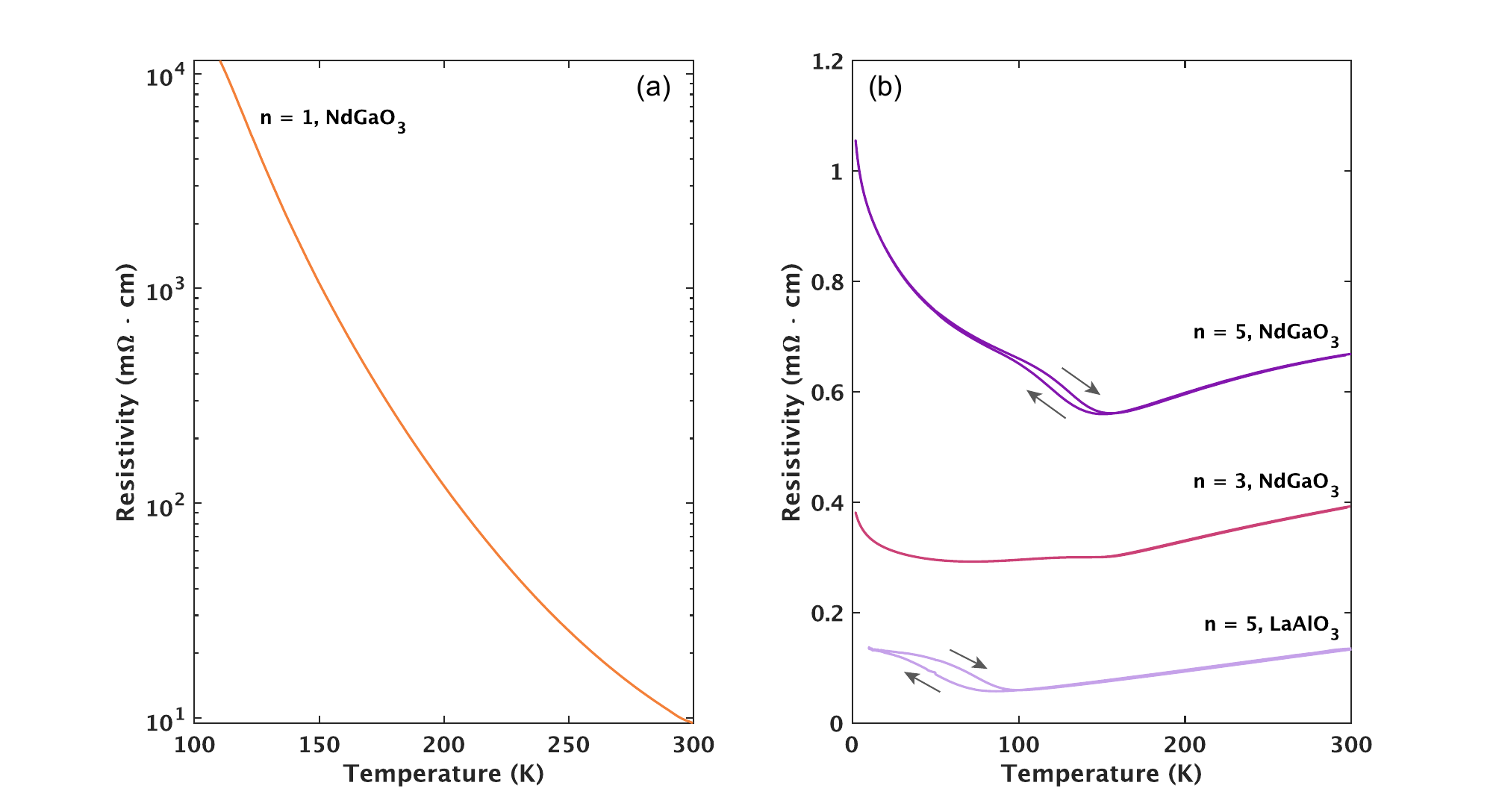}
    \caption{Electrical transport behavior of the $n = 1$ (a) and $n = 3, 5$ (b) films. Grey arrows indicate the direction of the temperature sweep for samples showing hysteretic behavior.  The Nd$_4$Ni$_3$O$_{10}$ and Nd$_6$Ni$_5$O$_{16}$ traces are reproduced from Ref.~\cite{pan2022superconductivity}.}
    \label{fig:transport}
\end{figure}

\noindent Electronic transport measurements as a function of temperature (Fig.~\ref{fig:transport}) were taken down to 1.8~K using a Quantum Design Physical Property Measurement System (PPMS) using standard lock-in techniques at 15~Hz.  Electrical resistivities were determined using devices in van der Pauw or Hall bar geometries. Contacts were patterned with shadow masks and deposited using an electron-beam evaporator (5 nm Cr/100 nm Au). Hall bar channels were defined with a diamond scribe. 

The four films considered in this study have distinct transport properties.  The Nd$_2$NiO$_4$ film exhibits semiconducting behavior, consistent with bulk Nd$_2$NiO$_4$ behavior \cite{takeda1992crystal}.  The Nd$_4$Ni$_3$O$_{10}$ film displays a resistivity kink at $\sim150$ K. Resitivity kinks have been attributed to density wave ordering in $n = 2, 3$ La$_{n+1}$Ni$_n$O$_{3n+1}$ \cite{Chen2024, gupta2024anisotropic, ren2024resolving, zhang2020intertwined}, so the same behavior might be present in these Nd$_3$Ni$_3$O$_{10}$ films. The two Nd$_6$Ni$_5$O$_{16}$ films both exhibit hysteretic metal-insulator transitions.  Under the application of compressive strain,  Nd$_6$Ni$_5$O$_{16}$ synthesized on LaAlO$_3$ has a largely depressed metal-insulator transition temperature $T_\text{MIT}$ and a lower resistivity scale.  This may be potentially attributed to the straightening of Ni-O bonds by compressive strain and improved orbital overlap, which may have implications for superconductivity \cite{catalano2018rare,bhatt2025resolving}, although the precise mechanism is unknown at this stage.

\section{Orbital excitations}\label{sec:orbital}

\subsection{Fluorescence and delocalized ligand states}\label{sec:fluorescence}
\begin{figure}
    \centering
    \includegraphics[width=0.95\linewidth]{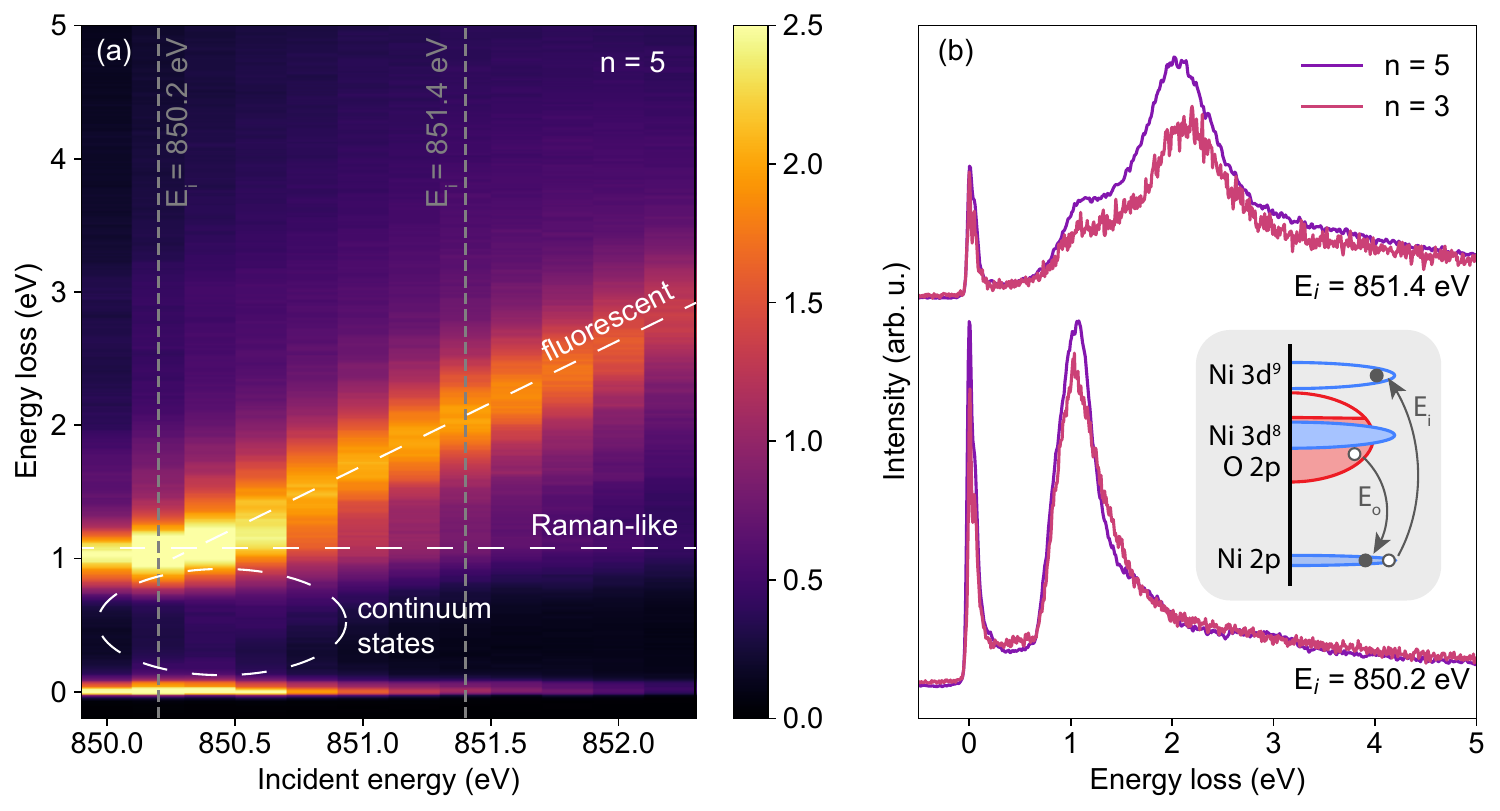}
    \caption{Fluorescent orbital excitations in Nd$_6$Ni$_5$O$_{16}$ ($n=5$) and Nd$_4$Ni$_3$O$_{10}$ ($n=3$). A) Incident-energy dependence of orbital RIXS excitations in Nd$_6$Ni$_5$O$_{16}$. Fluorescent features appear at increasing energy loss as incident energy is increased, while Raman-like features appear at fixed energy loss. B) Comparison of orbital RIXS features in Nd$_6$Ni$_5$O$_{16}$ and Nd$_4$Ni$_3$O$_{10}$ at selected incident energies, highlighting the similar behavior between the two materials. The inset shows the fluorescent RIXS process, in which states from delocalized, hybridized orbitals dominate the decay channel. $E_i$ denotes the incident x-ray photon while $E_o$ denotes the emitted x-ray photon.}
    \label{fig:Emap}
\end{figure}

Figure~\ref{fig:Emap} shows the incident x-ray energy dependence of the \gls*{RIXS} across the Ni $L_3$-edge for Nd$_6$Ni$_5$O$_{16}$ ($n=5$) and Nd$_4$Ni$_3$O$_{10}$ ($n=3$). The spectra include both Raman-like features that occur at constant energy loss for different incident energies and fluorescent-like features which increase in energy with increasing incident x-ray energy. The fluorescence occurs when electrons from delocalized, itinerant states decay to fill the core-hole and implies the presence of ligand states that are hybridized with the resonant element \cite{Mitrano2024exploring}. As discussed in the main text, this indicates that the Ruddlesden-Popper nickelates adopt a $|d^8 \underline{L}\rangle $ configuration, where $\underline{L}$ denotes the hybridized oxygen ligand band \cite{Norman2023, Shen2023electronic}. 

\subsection{Minimal orbital polarization}\label{sec:polarization}
\begin{figure}
    \centering
    \includegraphics[width=0.5\linewidth]{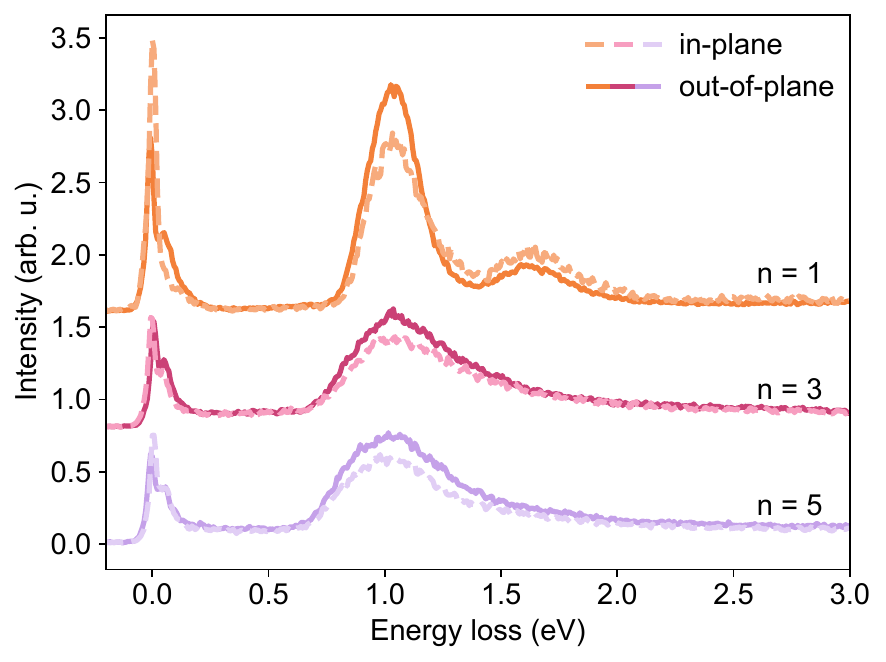}
    \caption{Orbital dichroism of RIXS for Nd$_2$NiO$_4$ ($n=1$) on NdGaO$_3$ (orange), Nd$_4$Ni$_3$O$_{10}$ ($n=3$) on NdGaO$_3$ (pink), and Nd$_6$Ni$_5$O$_{16}$ ($n=5$) on LaAlO$_3$ (purple). RIXS dominantly probing in-plane orbitals is shown with a dashed line and RIXS probing dominantly out-of-plane orbitals is shown with a solid line. For Nd$_2$NiO$_4$ the in-plane orbitals are probed with $\pi$-polarized light near normal incidence ($\theta = 72^\circ$) and out-of-plane orbitals are probed with $\pi$-polarized light near grazing incidence ($\theta = 16^\circ$). For Nd$_4$Ni$_3$O$_{10}$ and Nd$_6$Ni$_5$O$_{16}$, in-plane orbitals are probed with $\sigma$-polarized light near grazing incidence ($\theta=16^\circ$) and out-of-plane orbitals are probed with $\pi$-polarized light near grazing incidence ($\theta=16^\circ$). }
    \label{fig:poldep}
\end{figure}

One key difference between Ruddlesden-Popper nickelates compared with the cuprates and square-planar nickelates is that Ruddlesden-Popper nickelates are expected to have much weaker orbital polarization. We tested for orbital polarization in the higher $n$ Ruddlesden-Popper films by measuring the incident polarization dependence of orbital excitations near grazing incidence ($\theta = 16^\circ$) as shown in Fig.~\ref{fig:poldep}. At this angle, horizontal ($\pi$) polarized light primarily probes the out-of-plane $d_{3z^2-r^2}$ orbital while vertical ($\sigma$) polarized light primarily probes the in-plane $d_{x^2-y^2}$ orbital. The orbital excitations show minimal incident polarization dependence, indicating that holes are distributed roughly equally between both the in-plane and out-of-plane $e_g$ orbitals. For comparison, we also show the orbital dichroism in Nd$_2$NiO$_4$ ($n=1$), which is measured by comparing $pi$-polarized RIXS taken near grazing incidence ($\theta = 16^\circ$) and near normal incidence ($\theta = 72^\circ$), normalized to the integrated orbital intensities. 

\section{Modification to magnetic dispersion from Linear spin-wave theory}\label{sec:LSWT}
\begin{figure}
    \centering
    \includegraphics[width=0.38\linewidth]{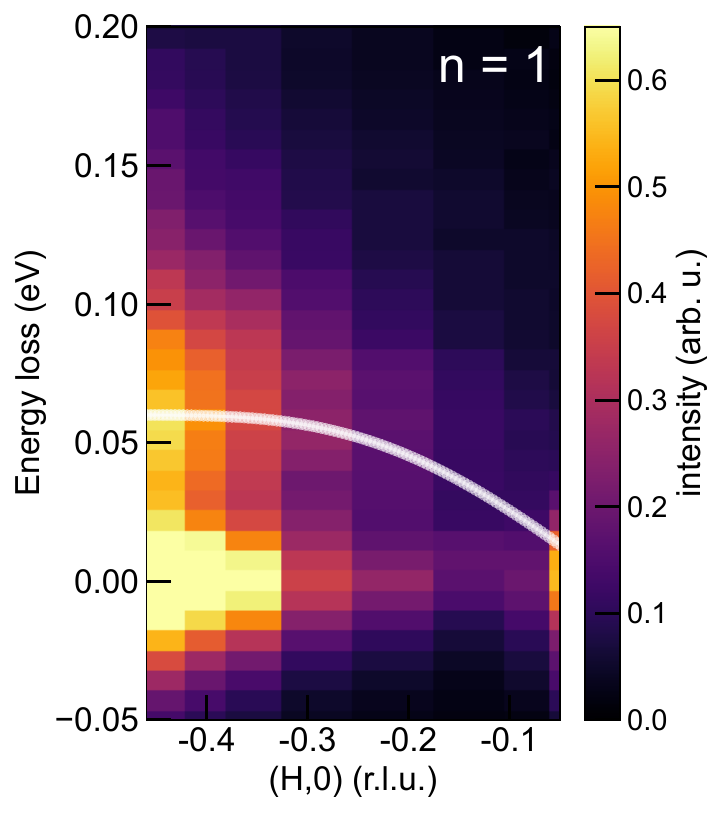}
    \caption{\gls*{RIXS} data for Nd$_2$NiO$_4$ ($n=1$) overlaid with the calculated dispersion from \gls*{LSWT} calculated with $J=16$ meV.}
    \label{fig:n1_LSWT}
\end{figure}

\begin{figure}
    \centering
    \includegraphics[width=0.95\linewidth]{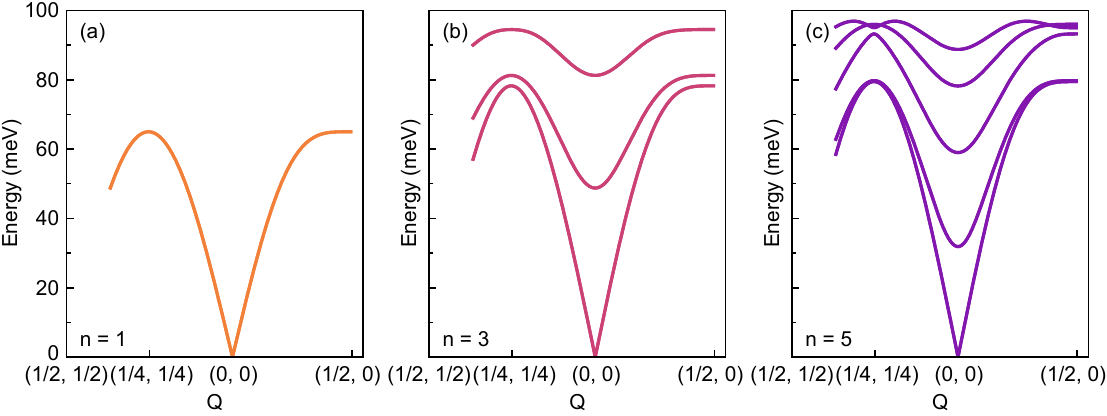}
    \caption{\gls*{LSWT} calculations for magnetic branches for a material with a) a single layer, b) 3 layers, and c) 5 layers in a unit cell, with in-plane exchange $J=16$ meV. For the 3- and 5-layer cases, the out-of-plane exchange $J_z$ is set equal to the in-plane exchange.}
    \label{fig:LSWT}
\end{figure}

We use the package Sunny~\cite{Sunny} to model the expected magnetic dispersions with increasing $n$ within \gls*{LSWT}. For all samples we fix the in-plane magnetic exchange to $J = 16$ meV, matching the magnon bandwidth for our measurements on Nd$_2$NiO$_4$ ($n=1$). We show the calculated magnon dispersion overlaid with the data in Fig. \ref{fig:n1_LSWT}. For higher $n$ we include an out-of-plane magnetic exchange between $n$ adjacent Ni layers, with $J_z = J$, assuming a symmetric octahedral environment. As shown in Fig. \ref{fig:LSWT}, as $n$ is increased, an additional magnon branch emerges for each additional Ni layer. Furthermore, the energies of the magnetic excitations are shifted to higher energies due to the introduction of the additional magnetic exchange pathway. 

\section{Strain dependence of N\lowercase{d}$_6$N\lowercase{i}$_5$O$_{16}$ ($n=5$)}\label{sec:strain}
Here we show the strain imparted by the substrate has minimal impact on the orbital and magnetic features reported in the main text by comparing RIXS spectra for Nd$_6$Ni$_5$O$_{16}$ ($n=5$) synthesized on two different substrates. Films synthesized on LaAlO$_3$ experience a small amount of compressive strain ($\epsilon \approx -0.9\%$), while those on NdGaO$_3$ discussed in the main text experience tensile strain ($\epsilon \approx +1.0\%$). As Nd$_6$Ni$_5$O$_{16}$ ($n=5$) can only be synthesized in thin-film form, strain values are approximated using bulk lattice constants for Nd$_4$Ni$_3$O$_{10}$ ($n=3$) \cite{Li2020contrasting}. Based on the resistivity in Sec.~\ref{sec:transport}, compressive strain suppresses the metal-insulator transition to lower temperatures and increases the conductivity of the samples overall but does not otherwise substantially affect the key findings reported in the main text. 

Strain appears to have a limited impact on the magnetic features in Nd$_6$Ni$_5$O$_{16}$ along $[H,0]$. This suggests that the Ni-O exchange probed by this reciprocal space cut is minimally sensitive to strain, similar to what has been observed in thin films of perovskite NdNiO$_3$~\cite{Fursich2019resonant} and infinite-layer square-planar nickelates~\cite{Gao2024magnetic}. 

\begin{figure}
    \centering
    \includegraphics[width=0.98\linewidth]{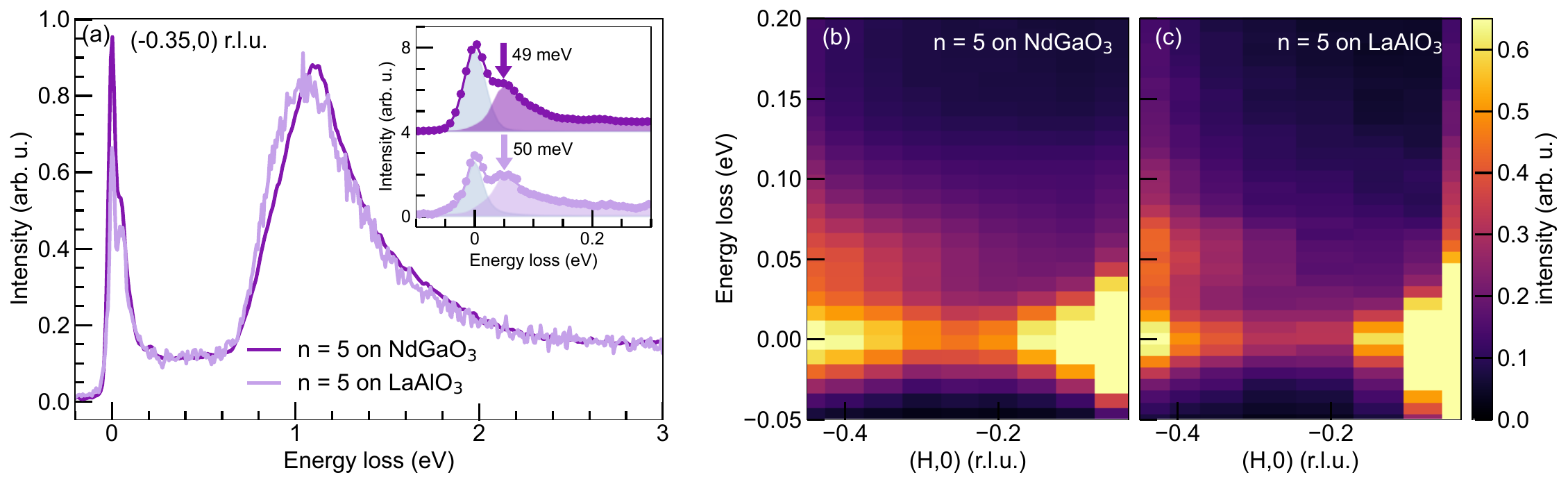}
    \caption{Impact of strain on RIXS spectra. A) RIXS spectra for Nd$_6$Ni$_5$O$_{16}$ on NdGaO$_3$ (tensile) and on LaAlO$_3$ (compressive). The inset shows a zoom-in on the magnetic portion of the RIXS spectra, showing similar magnetic peak positions in both samples. B) Momentum dependence of magnetic excitations in Nd$_6$Ni$_5$O$_{16}$ on NdGaO$_3$ (tensile) along $[H,0]$ (reproduced from main).  C) Momentum dependence of magnetic excitations in Nd$_6$Ni$_5$O$_{16}$ on LaAlO$_3$ (compressive) along $[H,0]$. }
    \label{fig:strain_dep}
\end{figure}

 \FloatBarrier 
\bibliography{refs.bib}